\documentclass[aps,epsfig,twocolumn,multicol,eqsecnum]{revtex4}
\usepackage{amsmath}
\usepackage{amssymb}
\usepackage{graphicx}
\newcommand{\bare}[1]{\mathaccent"7017{#1}}
\def\be{\begin{eqnarray}}
\def\ee{\end{eqnarray}}
\def\be{\begin{equation}}
\def\ee{\end{equation}}
\begin{document}
\title{Diversity of critical behavior
within a universality class}

\author{Volker Dohm}

\affiliation{Institute of Theoretical Physics, RWTH Aachen
University, D-52056 Aachen, Germany}

\date {15 May 2008}

\begin{abstract}

We study spatial anisotropy effects on the bulk and finite-size
critical behavior of the O$(n)$ symmetric anisotropic $\varphi^4$
lattice model with periodic boundary conditions in a
$d$-dimensional hypercubic geometry above, at and below $T_c$. The
absence of two-scale factor universality is discussed for the bulk
order-parameter correlation function, the bulk scattering
intensity, and for several universal bulk amplitude relations. The
anisotropy parameters are observable by scattering experiments at
$T_c$. For the confined system, renormalization-group theory
within the minimal subtraction scheme at fixed dimension $d$ for
$2<d<4$ is employed. In contrast to the $\varepsilon = 4 - d$
expansion, the fixed-$d$ finite-size approach keeps the
exponential form of the order-parameter distribution function
unexpanded. For the case of cubic symmetry and for $n=1$ our
perturbation approach yields excellent agreement with the Monte
Carlo (MC) data for the finite-size amplitude of the free energy
of the three-dimensional Ising model at $T_c$ by Mon [Phys. Rev.
Lett. {\bf 54}, 2671 (1985)]. The $\varepsilon$ expansion result
is in less good agreement. Below $T_c$ a minimum of the scaling
function of the excess free energy is found. We predict a
measurable dependence of this minimum on the anisotropy
parameters. The relative anisotropy effect on the free energy is
predicted to be significantly larger than that on the Binder
cumulant. Our theory agrees quantitatively with the non-monotonic
dependence of the Binder cumulant on the ferromagnetic
next-nearest neighbor (NNN) coupling of the two-dimensional Ising
model found by MC simulations of Selke and Shchur [J. Phys. {\bf A
38}, L739 (2005)]. Our theory also predicts a non-monotonic
dependence for small values of the {\it antiferromagnetic} NNN
coupling and the existence of a Lifshitz point at a larger value
of this coupling. The nonuniversal anisotropy effects in the
finite-size scaling regime are predicted to satisfy a kind of
restricted universality.  The tails of the large-$L$ behavior
at $T \neq T_c$ violate both finite-size scaling and universality even
for isotropic systems as they depend on the bare four-point
coupling of the $\varphi^4$ theory, on the cutoff procedure, and
on subleading long-range interactions.

\end{abstract}
\maketitle

\section{Introduction and summary}

A major achievement of the
renormalization-group (RG) theory is the proof that critical
phenomena can be divided into distinct {\it universality classes}
(for a review see, e.g., \cite{fish-1}). They are characterized by
the spatial dimension $d$ and the symmetry of the ordered state
which, for simplicity, we assume in the following to be $O(n)$
symmetric with an $n$ component order parameter. (For other
universality classes see, e.g., \cite{vi-1}.) Within a given
$(d,n)$ universality class, all bulk systems (with finite-range
interactions and with subleading long-range interactions of the
van der Waals type) have the same critical exponents and the same
thermodynamic functions near criticality in terms of universal
scaling functions that are obtained after a rescaling of two
amplitudes : that of the singular part of the bulk free energy
density $f_{s, b}$ and that of the field $h$ conjugate to the
order parameter. This is summarized in the asymptotic (small
$t=(T-T_c)/T_c$, small $h$) scaling form (below $d = 4$
dimensions)
\begin{equation}
\label{1a} f_{s, b} (t, h) = A_1 |t|^{d \nu} \; W_\pm (A_2 h
|t|^{-
\beta \delta})
\end{equation}
with universal critical exponents $\nu, \beta, \delta$ and the
universal scaling function $W_\pm (z)$ above $(+)$ and below $(-)$
$T_c$. Once the universal
quantities are known one knows the asymptotic thermodynamic
critical behavior of all members of the universality class
provided that only the {\it two} nonuniversal amplitudes $A_1$ and
$A_2$ are specified.  (For the application to real systems,
additional experimental information is necessary to
identify the order parameter and the appropriate thermodynamic
path tangential to the coexistence line.) We refer to this
property as {\it thermodynamic two-scale factor universality}.
Here universality means the independence of all microscopic
details such as lattice structure, lattice spacing, and the
specific form and magnitude of the finite-range or subleading
long-range interaction. This implies that both fluids and
anisotropic solids within the same universality class
have the same scaling function $W_\pm$.

\begin{table*}
  \caption{Subclasses of asymptotic critical behavior within a $(d,n)$
  universality class for $O(n)$ symmetric systems
  in a cubic volume $V=L^d$ with periodic boundary conditions for
  general $n$ above $T_c$ and $n=1$ below $T_c$. All subclasses have the same fixed point
  value $u^*(d,n)$ of the renormalized four-point coupling, the same critical
  exponents, and the same bulk thermodynamic scaling functions.
  This table complements Table IV of \cite{fish-1}.}
\begin{ruledtabular}
\begin{tabular}{|c||c|c|c|c|} \hline\hline
  \parbox[t]{2.5cm}{classes \\ of   interactions \\
  $\delta \widehat K ({\bf k})$ in (\ref{2g})} &
  \parbox[t]{3.7cm}{basic lengths,\\ nonuniversal parameters}
   & \parbox[t]{1.8cm}{bulk \\amplitude \\relations}
   & \parbox[t]{4.9cm}{bulk \\correlation functions}
   & \parbox[t]{4.2cm}{finite-size \\ effects}\\ \hline\hline
  \parbox[t]{2.5cm}{isotropic short \\ range\footnote{ Refs. \cite{fish-1,priv,cd2000-1,cd2000-2,cd2002}}\\ ${\bf k}^2 + O ( k^4)$} &
  \parbox[t]{3.7cm}{ correlation length\footnote{ For isotropic systems, $\xi_+$ and $\xi_-$ denote
  the second-moment bulk correlation lengths above and below
  $T_c$, with a universal ratio $\xi_+ / \xi_-$. For anisotropic
  systems,
  $\xi_\pm^{(\alpha)}$ are the principal bulk correlation lengths
  with universal ratios $\xi_+^{(\alpha)} / \xi_-^{(\alpha)}, \alpha = 1, 2, ...,
  d$.} $\xi_\pm$, \\ two nonuniversal \\ amplitudes  $C_1, C_2$, four-point coupling $u_0$}
  & \parbox[t]{1.8cm}{two-scale \\factor \\universality} &
  \parbox[t]{4.9cm} { $ r /\xi_\pm \lesssim O(1)$ : universal \\isotropic power-law scaling form;\\
  $ r \gg \xi_\pm:$  exponential form with nonuniversal tails}
  & \parbox[t]{4.2cm}{ $ L /\xi_\pm \lesssim O(1)$ : universal \\power-law scaling
  form;\\
  $ L \gg \xi_\pm:$ exponential form with nonuniversal tails}\\ \hline
  \parbox[t]{2.5cm}{anisotropic short\\ range\footnote{Refs. \cite{cd2004,dohm2006}} $\sum_{\alpha,
\beta=1}^d A_{\alpha \beta} \; k_\alpha k_\beta $ \linebreak \linebreak $\det {\bf A} > 0$ \\ $A_{\alpha \beta} = A_{\beta \alpha}$} &
  \parbox[t]{3.7cm}{{$d$ principal correlation lengths  $\xi^{(\alpha)}_\pm$,} \\
  up to {$d(d+1)/2 + 1$ \\nonuniversal parameters\footnote{ The
  reduced anisotropy matrix ${\bf \bar A} = {\bf A} / ({\rm{det}} {\bf A})^{1/d}$ has
  $d (d+1)/2-1$ independent matrix elements $\bar A_{\alpha \beta}$.}
  \\ $C'_1, C'_2, \bar A_{\alpha \beta}$}, \\four-point coupling $u_0$}
  & \parbox[t]{1.8cm}{ multi-parameter \\universality}   &
  \parbox[t]{4.9cm}{  $ r /\xi^{(\alpha)}_\pm \lesssim O(1)$ : universal power-law scaling
  form with  $d(d+1)/2+1 $ nonuniversal para-\\meters in the scaling arguments;\\ $ r \gg \xi^{(\alpha)}_\pm:$
 exponential form with nonuniversal tails} &
   \parbox[t]{4.2cm}{  $ L /\xi^{(\alpha)}_\pm \lesssim O(1)$ :  nonuniversal power-law scaling
   form,
   \\restricted universality;\\ $ L \gg \xi^{(\alpha)}_\pm:$ exponential form with nonuniversal tails}\\
   \hline
  \parbox[t]{2.5cm}{isotropic \\ subleading  long range\footnote{Refs. \cite{cd2002,wi-1,dan-1,dan-2,cd-2003,chamati}}\\
${\bf k}^2 \;- \;b\;
  |{\bf k}|^\sigma$ \\ $ 2 \;< \sigma <\; 4$} &
  \parbox[t]{3.7cm}{ correlation length  $\xi_\pm$,\\ interaction length scale $b^{1/(\sigma-2)}$,
  \\ five nonuniversal \\ parameters $C_1, C_2, b, \sigma, u_0$ } &
  \parbox[t]{1.8cm}{two-scale \\factor universality}
  & \parbox[t]{4.9cm}{$ r /\xi_\pm \lesssim O(1)$ : universal power-law scaling
  form;
  \\ $ r /\xi_\pm > O (1)$ : \\nonuniversal power-law form \\depending on $b$, $\sigma$}
  &\parbox[t]{4.2cm}{ $ L /\xi_\pm \lesssim O(1)$ :  universal power-law scaling
  form;
  \\ $ L / \xi_\pm > O (1) :$ \\nonuniversal power-law form \\depending on $b$, $\sigma$}\\ \hline
\end{tabular}
\end{ruledtabular}
\end{table*}


This important concept of scaling and thermodynamic two-scale
factor universality was extended to the distance $({\bf r})$
dependence of bulk correlation functions \cite{stau} and to the
size $(L)$ dependence of quantities of confined systems
\cite{fish,pri,priv-1} (for  reviews see, e.g., \cite{priv,
priv-2}). It is this {\it extended} hypothesis which is in the
focus of the present paper. We shall present results for the
finite-size critical behavior of the free energy above, at and
below $T_c$ that demonstrate a considerable degree of diversity
within a given $(d,n)$ universality class primarily due to spatial
anisotropy in lattice systems with non-cubic symmetry, but also
due to the lattice spacing $\tilde a$ in systems with cubic
symmetry and due to the bare four-point coupling $u_0$ of the
$\varphi^4$ theory even in the isotropic case. In this context we
also discuss nonuniversal effects related to the cutoff and to
subleading long-range (van der Waals type) interactions. This
diversity suggests to distinguish subclasses of interactions
within a given universality class where the subclasses have
different bulk amplitude relations, different bulk correlation
functions, and, for given geometry and boundary conditions (b.c.),
different finite-size scaling functions. All of these nonuniversal
differences exist in the {\it asymptotic} critical region $ |t|
\ll 1$, $L \gg \tilde a$, and $r \gg \tilde a $ where corrections
to scaling in the sense of Wegner \cite{wegner1972} are
negligible. A summary of these properties is given in Table I. The
basic framework of RG theory is fully compatible with this
diversity of critical behavior.

Spatially anisotropic systems such as magnetic materials, alloys,
superconductors \cite{schn}, and solids with structural phase
transitions \cite{bruce-1,salje} represent an important class of
systems with cooperative phenomena. One may distinguish between
long-range anisotropic interactions (such as dipolar, RKKY, and
effective elastic interactions) and short-range anisotropic
interactions which include the Dzyaloshinskii-Moriya-type
antisymmetric exchange \cite{liu-1} and the spatially anisotropic
Heisenberg exchange interactions which, in the long-wavelength
limit, are described by a $d \times d$ anisotropy matrix ${\bf A}$
\cite{cd2004,dohm2006}. We shall confine ourselves to a detailed
study of the latter type of systems but the general aspects of our
results have an impact also on the former type of anisotropic
systems and on systems of other universality classes \cite{vi-1},
for example on the range of validity of universality for
anisotropic spin glasses \cite{katz-1} or for anisotropic surface
critical phenomena \cite{diehl-1}.

A characteristic feature of spatial anisotropy with non-cubic
symmetry is the fact that there exists no unique bulk
correlation-length amplitude but rather $d$ different amplitudes
$\xi_{0 \pm}^{(\alpha)}$ in the directions $\alpha = 1, ..., d$ of
the $d$ principal axes. Such systems still have a single
correlation-length exponent $\nu$ provided that $\det {\bf A} >
0$. (We do not consider {\it strongly} anisotropic systems with
critical exponents different from those of the usual $(d, n)$
universality classes, see e.g. \cite{tonchev}.) Non-cubic
anisotropy effects in crystals with cubic symmetry can be easily
generated by applying shear forces. In earlier work on two-scale
factor universality \cite{pri,priv,weg-1,aha-74,ger-1,hoh-76},
isotropic systems with a single bulk correlation length
$\xi_\infty$ were considered and important  universal bulk
amplitude relations were derived that depend on only  two
nonuniversal parameters. Recently some of these relations were
reformulated for anisotropic systems within the same universality
class \cite{cd2004,dohm2006}. In Sect. III of the present paper we
give a derivation of these and other relations above and below
$T_c$ and express them in terms of universal scaling functions.
The physical quantities entering these relations depend, in
general, on $d (d+1)/2 + 1$ nonuniversal parameters. We also
present the appropriate formulation of the bulk scattering
intensity of anisotropic systems near criticality in terms of the
eigenvalues of the anisotropy matrix and discuss the nonuniversal
properties of bulk correlation functions.

For {\it  confined} systems with a characteristic length $L$ the
hypothesis of two-scale factor universality is summarized by the
asymptotic (large $L$, small $t$, small $h$) scaling form for the
singular part of the free energy density (divided by $k_B T)$
\cite{pri,priv-2,priv}
\begin{equation}
\label{1b} f_s (t, h, L) = L^{-d} \; {\cal F} (C_1 t L^{1/\nu},
C_2 h L^{\beta \delta/\nu}).
\end{equation}
where ${\cal F}(x, y)$,  for given geometry and b.c., is a
universal scaling function  and where the two constants $C_1$ and
$C_2$  are universally related to the bulk constants $A_1$ and
$A_2$ of  (\ref{1a}). For simplicity we shall confine ourselves to
a hypercubic shape with volume $V = L^d$ and with periodic b.c..
Calculations of $ f_s(t,0,L)$ for this case were carried out
within the spherical model \cite{singh-1}  which supported the
form of (\ref{1b}). For $n=1$ the scaling form (\ref{1b}) was
discussed in the framework of the $\varepsilon = 4 - d$ expansion
\cite{RGJ,GJ}. No theoretical prediction for the function ${\cal
F}(x,y)$ is available up to now for finite $n$ in cubic geometry,
except in the large-$n$ limit \cite{cd2002}. Monte Carlo (MC)
simulations \cite{mon-1,mon-2,mon-3} for three-dimensional Ising
models with nearest-neighbor (NN) couplings on different lattices
of cubic symmetry were consistent with the universality of the
amplitude ${\cal F} (0,0)$. These models belong to the subclass of
(asymptotically) isotropic systems.

It was already noted in \cite{pri,priv,aharony}  that lattice
anisotropy is a marginal perturbation in the RG sense, thus it was
not obvious a priori to what extent two-scale factor universality
is valid in the presence of anisotropic couplings \cite{priv}. It
was  also known that, for most anisotropic systems, (asymptotic)
isotropy can be restored  by an anisotropic scale transformation
\cite{car-1,kam} (for further references see \cite{cd2004}).
Recently it was pointed out \cite{cd2004} that, in systems with
anisotropic interactions of non-cubic symmetry, the scaling
function ${\cal F}$ is indeed affected by anisotropy. In
particular, it was shown \cite{dohm2006} that by means of an
appropriate rescaling of lengths a transformation to an
(asymptotically) isotropic system is always possible provided that
the anisotropy matrix $\bf A$ is positive definite and that the
rescaling is performed along the $d$ {\it nonuniversal} directions
of the  {\it principal} axes which, in general, differ from the
symmetry axes of the system. This rescaling is equivalent to a
shear transformation which distorts the shape, the lattice
structure, and the boundary conditions in a nonuniversal way (e.g.
from a cube to a parallelepiped, from an orthorhombic to a
triclinic lattice, and from periodic b.c. in rectangular
directions to those in non-rectangular directions). This
nonuniversality is reflected in a dependence of the scaling
function ${\cal F}$ on the anisotropy matrix $\bf A$, in addition
to the dependence on $C_1$ and $C_2$.

Specifically, on the basis of the results of renormalized
perturbation theory in Sects. IV - VI, we propose that, for
anisotropic systems with the shape of a cube, (\ref{1b}) is to be
replaced by \cite{scal}
\begin{equation}
\label{1c} f_s (t, h, L) \; = \; L^{-d} \; {\cal F}_{cube} (C_1' t
L'^{1/\nu}, C_2' h' L'^{\beta\delta/\nu}; {\bf \bar A}) \; ,
\end{equation}
with $L' = L (\det {\bf A})^{- 1/(2d)}$, $h'= h (\det {\bf
A})^{1/4}$, and with the reduced anisotropy matrix ${\bf \bar A} =
{\bf A} / (\det {\bf A})^{1/d}$, $\det {\bf A} > 0$. The
nonuniversal constants $C_1'$ and $C_2'$ will be specified in
Sect. VI in terms of the asymptotic amplitudes $\xi_{0+}'$ and
$\xi_c'$ of the second-moment bulk correlation lengths for $T
> T_c$, $h'=0$ and for $T=T_c$, $h'\neq 0$, respectively, of the transformed
isotropic system. The free energy density $f_s' = f_s (\det {\bf
A})^{- 1/2}$ of the parallelepiped with the volume $V' = V (\det
{\bf A})^{- 1/2}$ and with ${\bf \bar A}'={\bf A}' / (\det {\bf
A}')={\bf 1}$ (isotropy) then attains the scaling form
\begin{eqnarray}
\label{1d} f'_s (t, h', L') \; = \; L'^{-d} \; {\cal F}_{iso,{\bf
\bar A}} (C_1' t L'^{1/\nu}, C_2' h' L'^{\beta\delta/\nu}) \;
\end{eqnarray}
where the characteristic length $L'= V'^{1/d}$ determines the overall size of
the parallelepiped and
\begin{eqnarray}
\label{1e}  {\cal F}_{iso, {\bf \bar A}} (x, y) \; = \; {\cal
F}_{cube} (x, y; {\bf \bar A})
 \; .
\end{eqnarray}
Equation (\ref{1d}) has the structure of the isotropic Privman-Fisher
scaling form (\ref{1b}) with a rescaled length $L'$ and with only two nonuniversal constants
$C'_1$ and $C'_2$ which, superficially, appears to be in
agreement with two-scale factor universality. The remaining
$d(d+1)/2 - 1$ nonuniversal parameters, however, are hidden in the
index "iso, ${\bf \bar A}$". This index  is the notation for a
system with the shape of a parallelepiped whose interaction
$\delta \hat K' ({\bf k}') = {\bf k}'^2 + O (k'^4)$ is
(asymptotically) isotropic and whose $d(d-1)/2$ angles and $d-1$
length ratios are determined by the $d (d+1)/2 - 1$ nonuniversal
parameters of the reduced anisotropy matrix  ${\bf \bar A}$. These
parameters appear in the calculation of ${\cal F}_{iso, {\bf \bar
A}}$ via the summation over the discrete ${\bf k'}$ vectors in the
Fourier space of the parallelepiped system since the ${\bf k'}$
vectors depend explicitly on ${\bf \bar A}$, unlike the ${\bf k}$
vectors of the cubic system. Thus for the calculation of ${\cal
F}_{iso, {\bf \bar A}}$ the same nonuniversal information is
required as for the calculation of ${\cal
F}_{cube} (x, y; {\bf \bar A})$.

For general ${\bf A}$ the function ${\cal F}_{cube} (x,0; {\bf
\bar A})$ was presented in \cite{cd2004} for $t \geq 0$ in the
large-$n$ limit. Furthermore, quantitative predictions were made
for the nonuniversal dependence of the critical Binder cumulant
\cite{priv,bin-2}
\begin{equation}
\label{1f} U_{cube} ({\bf \bar A}) \; =
\frac{1}{3}\;\Big[\frac{\partial^4 {\cal F}_{cube}(0,y;{\bf \bar
A})/ \partial y^4}{(\partial^2 {\cal F}_{cube}(0,y;{\bf \bar A})
/\partial y^2)^2}\Big]_{y=0}
 \;
\end{equation}
for $n = 1,2,3$ both in three \cite{cd2004,dohm2006} and two
\cite{dohm2006} dimensions. MC simulations \cite{schulte,stauffer}
for the anisotropic three-dimensional Ising model indeed showed
nonuniversal anisotropy effects which, however, did not agree with
the theoretical prediction. More accurate MC simulations
\cite{selke2005} for the anisotropic two-dimensional Ising model
demonstrated the nonuniversality of the critical Binder cumulant
but no comparison with a quantitative theoretical prediction was
available for this two-dimensional case. Thus the anisotropic
finite-size theory of \cite{cd2004,dohm2006} is as yet
unconfirmed.

In Sects. IV - VI of this paper we derive the finite-size scaling
function ${\cal F}^{ex}_{cube} (x, 0; {\bf \bar A})$ of the
singular part of the excess free energy density $f^{ex}_s = f_s -
f_{s,b}$ at $h=0$ above, at, and below $T_c$ for the $n=1$
universality class in $2<d<4$ dimensions on the basis of the
anisotropic $\varphi^4$ lattice model.  For the isotropic case at
$T_c$ we find excellent agreement with the MC data of Mon
\cite{mon-1,mon-2}. Slightly below $T_c$ we find a minimum of the
scaling function that is similar to the minimum of the scaling
function of the critical Casimir force for the $d = 3$ Ising model
in slab geometry with periodic boundary conditions
\cite{dan-k,vas-1}. For future tests of our theory by MC
simulations we consider both three- and two-dimensional
anisotropies. In both cases we predict a measurable dependence of
the minimum on the anisotropy parameters, thus demonstrating the
nonuniversality of the finite-size scaling function of the excess
free energy density. The magnitude of this anisotropy effect is
predicted to be considerably larger than that on the Binder
cumulant.

We believe that a similar nonuniversal dependence can be derived
for the critical Casimir force by means of our perturbation
approach. For $n \to \infty$ and at $T=T_c$, the nonuniversality
of the Casimir amplitude due to anisotropy was already
demonstrated within the $\varphi^4$ theory in \cite{cd2004}. This
suggests that, for given geometry and b. c., the existing
theoretical
\cite{krech,krech1997,diehl2006,zandi,zandi2004,upton,dantchev2006,dantchev2007}
and MC \cite{krech1997,dan-k,vas-1,hucht} results for the Casimir
force scaling function are not universal within the {\it entire}
universality class but are restricted to the subclass of isotropic
systems. Extensions to the subclass of anisotropic systems  are,
in general, not straightforward  and cannot be obtained just by
transformations but require new nonuniversal input, new analytical
and numerical calculations, and new MC simulations. Experiments,
e.g., in anisotropic superconducting films \cite{schn,wil-1},
could, in principle, demonstrate the nonuniversality in real
systems.

Our present results for the $\varphi^4$ theory cannot be applied
directly to two-dimensional critical phenomena. Nevertheless we
are able to study two-dimensional anisotropy effects within a
three-dimensional model. For the purpose of a comparison with the
two-dimensional MC data \cite{selke2005}, we consider (in Sect.
VIII) a three-dimensional $\varphi^4$ lattice model with the same
two-dimensional anisotropy in the horizontal planes as in the
two-dimensional model Ising model studied by Selke and Shchur
\cite{selke2005}. Our theory agrees quantitatively with the
non-monotonicity of the Binder cumulant as a function of the
anisotropiy {\it ferromagnetic} next-nearest neighbor (NNN)
coupling found in \cite{selke2005}. We also predict a
non-monotonicity for small {\it anti-ferromagnetic} couplings and
the existence of a Lifshitz point at a larger value of this
coupling. Very recent preliminary MC data by Selke
\cite{selke-neu} for the {\it two}-dimensional Ising model indeed
reveal such a non-monotonicity that was not yet detected in
\cite{selke2005}. We predict a similar anisotropy effect for the
excess free energy density of the anisotropic two-dimensional
Ising model. This effect can become quite large if one of the
eigenvalues of ${\bf \bar A}$ approaches zero, in particular if a
Lifshitz point is approached (Sect. VIII).

An important property of the scaling form (\ref{1c}) is that it
depends on ${\bf \bar A}$ but not on other nonuniversal parameters
such as the bare four-point coupling, the lattice spacing, and the
cutoff of $\varphi^4$ field theory.  This is a kind of {\it
restricted universality} since it implies that the same
finite-size scaling functions exist for the large variety of those
systems within a universality class that have the same reduced
anisotropy matrix ${\bf \bar A}$ (and the same geometry and
boundary conditions). In Sect. IX we propose two examples for
testing this hypothesis of restricted finite-size universality by
MC simulations for spin models with {\it anisotropic}
interactions. For recent tests of finite-size universality of
two-dimensional Ising models with (asymptotically) {\it isotropic}
interactions see \cite{selke2006,selke2007} (see also Table 10.1
of \cite{priv}).

Unlike the bulk scaling function $W_\pm(z)$, (\ref{1a}), that is
valid in the entire range $-\infty \leq z \leq \infty$ of the
scaling argument $z$, the finite-size scaling functions such as
${\cal F}_{cube} (x,y; {\bf \bar A})$ are valid only in a limited
range of $x$ and $y$, above the shaded region in Fig. 1. In the
shaded region, nonuniversal nonscaling effects become
nonnegligible and even dominant for sufficiently large $|x|$ and
$|y|$ for both short-range and subleading long-range interactions.
In this region, not only the correlation lengths are relevant but
also additional nonuniversal length scales such as the lattice
spacing $\tilde a$, the inverse cutoff $\Lambda^{-1}$, the length
scale $u_0^{-1/\varepsilon}$ set by the four-point coupling , and
the van-der-Waals interaction-length $b^{1/(\sigma - 2)}$, as
discussed in Sect. X.

\begin{figure}[!h]
\includegraphics[width=80mm]
{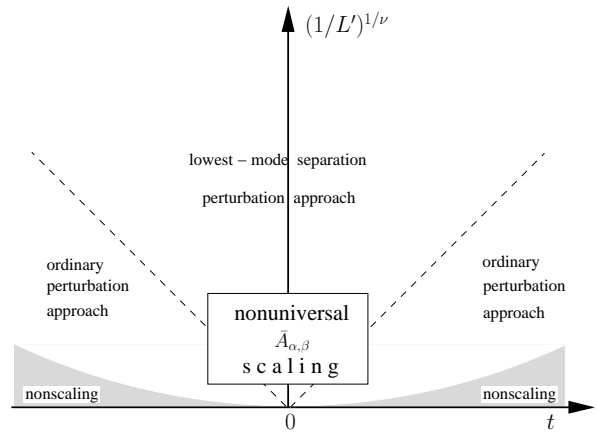} \caption{Asymptotic part of the
$L'^{-1/\nu} - t$ plane at $h=0$ for the anisotropic $\varphi^4 $
theory in a cubic geometry with periodic boundary conditions. In
the central finite-size region (above the dashed lines), the
lowest mode must be separated whereas outside this region ordinary
perturbation theory is applicable. Above the shaded region,
finite-size scaling is valid but with scaling functions that
depend on the anisotropy parameters $\bar A_{\alpha
\beta}$, see (\ref{1c}). In the large - $L'$ regime at $t\neq 0$ (shaded region)
finite-size scaling and universality are violated for both
short-range and subleading long-range interactions and for both
isotropic and anisotropic systems. A similar plot is valid for the
$L'^{-\beta\delta/\nu} - h'$ plane at $T=T_c$.}
\end{figure}

We briefly comment on the methodological aspects of our
finite-size calculations. As far as the field-theoretic
\cite{bre-1} renormalization of the $\varphi^4$ lattice
model is concerned we use the minimal subtraction scheme
\cite{hooft-1} not within the $\varepsilon$ expansion but at fixed
dimension $d$, as introduced in \cite{dohm1985} and further
developed in \cite{schl}. As far as finite-size theory is
concerned we further develop earlier approaches
\cite{RGJ,BZ,EDHW,EDC} that have been successfully used to solve
several finite-size problems in the past
\cite{CDS,CDT,cd-97,cdstau}. After the transformation from the
anisotropic to an isotropic system, the same renormalization
constants ($Z$ factors) and the same fixed-point value $u^*$ of
the renormalized four-point coupling are obtained as for the
standard isotropic $\varphi^4$ Hamiltonian. For this reason, the
same fixed-point Hamiltonian and the same critical exponents
govern isotropic and (weakly) anisotropic systems - they belong to
the same universality class. The crucial point, however, is that
not only the fixed-point value $u^*$ but also the orientation of
the eigenvectors (principal axes) of the fixed-point Hamiltonian
relative to the orientation of the given boundaries of the {\it
confined} anisotropic system determine the finite-size scaling
functions. This is a physical fact that introduces a source of
nonuniversality that cannot be eliminated by transformations and
that makes anisotropic confined systems distinctly different from
isotropic confined systems within the same universality class.

The main result for $f^{ex}_s$ will be obtained in the central
finite-size regime (above the dashed lines in Fig. 1) where the
finite-size effects are most significant and where it is necessary
to separate the lowest-mode from the higher modes. In this regime
finite-size scaling is valid in the form of (\ref{1c}). We compare
the result of our fixed - $d$ perturbation approach
\cite{dohm1985,schl,EDHW,EDC} with that of the $\varepsilon$
expansion approach. The advantage of the former approach is that
it keeps the exponential structure of the order-parameter
distribution function unexpanded. This leads to a result at $T_c$
in excellent agreement with the MC data in the isotropic case
\cite{mon-1,mon-2} and lends credibility also to the quantitative
features of our predictions of anisotropy effects. The
$\varepsilon$ expansion result at $T_c$ turns out to be in less
good agreement.

\begin{figure}[!h]
\includegraphics[width=80mm]
{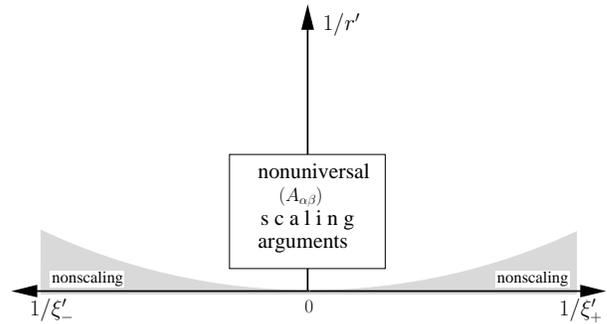} \caption{Asymptotic part of the $r'^{-1}
- \xi_\pm'^{-1} $ plane at $h=0$ for  anisotropic bulk systems.
Above the shaded region, there exists a universal scaling function
$\Phi_\pm(r'/\xi_\pm',0)$ of  the bulk correlation function $G_b$,
(\ref{3n}). The scaling argument, however, contains the spatial
variable $r'\equiv |{\bf x}'|$, (\ref{3x}), that depends on the
anisotropy matrix $ (A_{\alpha
\beta})$ with $d(d+1)/2$ nonuniversal parameters. In the
large - $r'$ regime at $t\neq 0$ (shaded region), scaling and
universality are violated for both short-range and subleading
long-range interactions and for both isotropic and anisotropic
systems. A similar plot is valid in the $r'^{-1} - \xi_h'^{-1} $
plane at $T=T_c, h' \neq 0$, with
$\xi_h'=\xi_c'|h'|^{-\nu/(\beta\delta)}$.}
\end{figure}

The separation of the lowest mode is inadequate in the limit of
large $L'\gg \xi_\pm'$ at fixed $T \neq T_c$. In order to capture
the exponential structure of finite-size effects for large $L'$ we
complement (in Sect. X) our results by ordinary perturbation
theory outside the central finite-size regime (below the dashed
lines in Fig. 1). This includes a small but finite region where
finite-size scaling is violated (shaded region in Fig. 1). There
exists diversity rather than universality of finite-size critical
behavior in this region depending on all microscopic details of
the interactions such as the lattice spacing, the bare four-point
coupling, the cutoff of the $\varphi^4$ theory, and the amplitude
of subleading long-range interactions.  This diversity can be
traced back to a corresponding diversity of the large-distance $(r'
\gg \xi_\pm')$ behavior of the {\it bulk} order parameter correlation
function $G_b$ \cite{cd2000-2} where $r'$ is the distance in the transformed isotropic bulk system,
as discussed in Sect. III. For $G_b$ there exists
a region of  the $r'^{-1} - \xi_\pm'^{-1}$ plane (shaded region in Fig. 2)
that is the analogue of the shaded region of Fig. 1 . In the isotropic case, this region is of
physical relevance for fluids with van der Waals
interactions \cite{cd2002,wi-1,dan-1,dan-2,cd-2003,chamati,dantchev2006,dantchev2007}.

\section{Anisotropic $\varphi^4$ lattice model}
\subsection{Hamiltonian with spatial anisotropy}
We start from the O$(n)$ symmetric $\varphi^4$ lattice Hamiltonian
(divided by $k_B T$)

\begin{eqnarray}
\label{2a} H  &=&   v \Bigg[\sum_{i=1}^N \left(\frac{r_0}{2}
\varphi_i^2 + u_0 (\varphi_i^2)^2 - h \varphi_i \right)
\nonumber\\&+& \sum_{i, j=1}^N \frac{K_{i,j}} {2} (\varphi_i -
\varphi_j)^2 \Bigg], \;
\end{eqnarray}
$r_0(T) = r_{0c} + a_0 t$, $t = (T - T_c) / T_c$ with $a_0>0$,
$u_0>0$. The variables $\varphi_i \equiv \varphi ({\bf x}_i)$ are
$n$-component vectors on $N$ lattice points ${\bf x}_i \equiv
(x_{i1}, x_{i2},\ldots, x_{id})$ of a $d$-dimensional Bravais
lattice with the finite volume $V = Nv$ with the characteristic
length $L= V^{1/d}$ where $v$ is the volume of the primitive cell.
The components $\varphi_i^{(\mu)} \; , \mu = 1, 2, \ldots, n$ of
$\varphi_i$ vary in the continuous range $- \infty \leq
\varphi_i^{(\mu)} \leq \infty$. The couplings $K_{i,j} = K_{j,i}
\equiv K ({\bf x}_i - {\bf x}_j)$ and the temperature variable
$r_0 (T)$ have the dimension of $L^{- 2}$ whereas the variables
$\varphi_i$ have the dimension of $L^{(2-d)/2}$ such that $H$ is
dimensionless. The free energy per unit volume divided by $k_BT$
is
\be
\label{2c} f(t,h,L) =  - V^{-1} \ln Z \; ,
\ee
\be
\label{2d} Z (t, h, L) = \left[\prod_{i = 1}^N \frac{\int d^n
\varphi_i}{v^{n (2-d) / (2d)}} \right] \exp \left(- H \right)
\ee
where $Z$ is the dimensionless partition function. The total
excess free energy density is defined as
\be
\label{2d-1} f^{ex}(t,h,L) =  f(t,h,L) -  f_b(t,h)
\ee
where $f_b(t,h)= \lim_{L \to \infty}f(t,h,L) $ is the bulk free
energy density. Following \cite{priv-1,priv,priv-2} we shall
decompose $f$, for large $L$,  into singular and non-singular
parts
\be
\label{2dd} f(t,h,L) =  f_s(t,h,L) + f_{ns}(t,L) \;
\ee
where  $f_{ns}(t,L)$ has a regular $t$ dependence around $t=0$. In
earlier work on finite-size effects it was supposed
\cite{pri,priv-2,cd2002} that, {\it for periodic boundary
conditions}, one can assume that there exists no $L$ dependence of
the nonsingular part  $f_{ns}$.  Adopting this assumption leads to
a misinterpretation \cite{cd2002} of the singular part $f_s$ of
the free energy density and of the Casimir force in the presence
of a sharp cutoff of $\varphi^4$ field theory. Here we shall not
exclude the possibility of an $L$ dependent nonsingular part
$f_{ns}(t,L) $ even for periodic boundary conditions if long-range
correlations are present . As will be shown in Sect. X, this will
reconcile the earlier results \cite{cd2002} with the concepts of
finite-size scaling.

For periodic b.c., the Fourier representations are $ \varphi({\bf
x}_j) = V^{-1} \sum_{\bf k} e^{i {\bf k} \cdot {\bf x}_j} \hat
\varphi({\bf k})$ and
\begin{equation}
\label{2f} K({\bf x}_i - {\bf x}_j) \; = \; N^{-1} \sum_{\bf k}
e^{i{\bf k} \cdot ({\bf x}_i - {\bf x}_j)} \widehat K ({\bf k})
\;,
\end{equation}
where the summations $\sum_{\bf k}$ run over the $N$ discrete
vectors ${\bf k}$ of the first Brioulluin zone of the reciprocal
lattice. We assume finite-range interactions $K_{i,j}$ with a
finite value $\widehat K ({\bf 0}) = N^{-1} \sum^N_{i,j = 1}
K_{i,j} $. In terms of the Fourier components the Hamiltonian
reads
\begin{eqnarray}
\label{2g} H = V^{-1} \sum_{{\mathbf k}} \frac{1}{2} [r_0 + \delta
\widehat K (\mathbf k)] \hat \varphi({\mathbf k}) \hat
\varphi({-{\mathbf k}})- h \hat \varphi({\mathbf 0}) \nonumber\\ +
\;u_0 V^{-3} \sum_{{\mathbf{kp}}{\mathbf q}} [\hat
\varphi({\mathbf k}) \hat \varphi({{\mathbf p}})] [\hat
\varphi({{\mathbf q}}) \hat \varphi({-{\mathbf k}-{\mathbf
p}-{\mathbf q}})] \qquad
\end{eqnarray}
where $ \delta \widehat K({\bf k}) = 2 [\widehat K({\bf 0}) -
\widehat K ({\bf k})]$. In perturbation theory, $r_0 + \delta
\widehat K ({\bf k})$ plays the role of an inverse propagator.

The Hamiltonian $H$ is isotropic in the vector space of the
$n$-component variables $\varphi_i$ and $\hat \varphi ({\bf k})$
but may be anisotropic in real space and ${\bf k}$ space. A
variety of anisotropies may arise both through the lattice
structure and through the couplings $K_{i,j}$. They manifest
themselves on macroscopic length scales via the $d \times d$
anisotropy matrix ${\bf A} = (A_{\alpha
\beta})$ and the anisotropy tensor ${\bf B} = (B_{\alpha \beta \gamma \delta})$
that appear in the long-wavelength form
\begin{eqnarray}
\label{2h} \delta \widehat K ({\bf k}) &=& \sum_{\alpha,
\beta=1}^d A_{\alpha \beta} \; k_\alpha k_\beta \;\; + \;\;
\sum^d_{\alpha, \beta, \gamma, \delta} B_{\alpha \beta \gamma
\delta} \; k_\alpha k_\beta k_\gamma k_\delta \nonumber\\ &+& \;\;
O (k^6) \;.
\end{eqnarray}
Odd powers of $k_\alpha$ are excluded because of inversion
symmetry of the Bravais lattice. (For the case of non-Bravais
lattices see the discussion in section II.C of Ref.
\cite{Italien}.) For cubic symmetry, ${\bf A}$  has the isotropic
form $A_{\alpha
\beta} = c_0 \; \delta_{\alpha
\beta}$ while the $O(k^4)$ terms of cubic systems differ from those
of isotropic systems. In Sects. III and X we shall consider the
model (\ref{2g}) also in a fully isotropic form with the
short-range interaction $\delta \widehat K ({\bf k}) = {\bf k}^2$
including a finite cutoff $\Lambda$ and, for $n=1$, with the
subleading long-range interaction
\cite{wi-1,cd2002,dan-1,dan-2,chamati,cd-2003,dantchev2006,dantchev2007}
\begin{eqnarray}
\label{2hh} \delta \widehat K ({\bf k}) = {\bf k}^2 \;\; -
b\;|{\bf k}|^\sigma \;\; +\;\; O (k^4) \;, \;\;2<\sigma<4
\end{eqnarray}
with  $b>0$. The second term of the interaction (\ref{2hh})
is usually classified as "irrelevant" in the renormalization-group
sense  \cite{dantchev2006} since it leaves  {\it some} (but not all)
of the universal quantities unchanged: critical exponents and bulk
thermodynamic scaling functions. This terminology is  somewhat
misleading as the term $-b\;|{\bf k}|^\sigma$ changes
not only the leading {\it finite-size} critical behavior
at $T\neq T_c$ (in the shaded region of Fig.1)
but it also destroys the universality of the
leading {\it bulk } critical behavior of the
order-parameter correlation function $G_b$ (and of other
bulk correlation functions): $G_b$  attains an
interaction-dependent power-law structure  \cite{wi-1,dan-2}
in the large-distance regime  at $T\neq T_c$ (in the shaded
region of Fig. 2) whereas  systems with purely short-range
interaction in the same universality class have an {\it exponentially}
decaying $G_b$ in this regime (this decay has, in addition, a
nonuniversal exponential tail, see Sect. X).

The expression for $A_{\alpha
\beta}$ and $B_{\alpha \beta \gamma \delta}$ in terms of the microscopic couplings $K_{i,j}$ is given
by the second moments \cite{dohm2006}
\begin{equation}
\label{2i} A_{\alpha \beta} = A_{\beta \alpha} = N^{-1} \sum^N_{i,
j = 1} (x_{i \alpha} - x_{j \alpha}) (x_{i \beta} - x_{j \beta})
K_{i,j}.
\end{equation}
and the fourth-order moments of $K_{i,j}$, respectively. They have
been classified and studied in the context of the bulk correlation
function in Ref. \cite{Italien}.  The symmetric matrix ${\bf A}$
depends only on the lattice structure and on the pair interactions
$K_{i,j}$ and is independent of the boundary conditions and the
geometry of the system. Its eigenvalues $\lambda_\alpha \; ,
\alpha = 1, 2, ..., d,$ and eigenvectors ${\bf e}^{(\alpha)}$  are
determined by the eigenvalue equation ${\bf A
e}^{(\alpha)}=\lambda_\alpha {\bf e}^{(\alpha)}$ with ${\bf
e}^{(\alpha)} \cdot {\bf e}^{(\beta)} = \delta_{\alpha
\beta}$. In order to have an ordinary critical point of the usual
$(d,n)$ universality classes we assume positive eigenvalues
$\lambda_\alpha, \det {\bf A} = \prod^d_{\alpha = 1}
\lambda_\alpha> 0$, and that the fourth-order moments $B_{\alpha
\beta \gamma \delta}$ enter only the corrections to scaling.
 The critical point occurs at $h = 0$ and at $T = T_c$
corresponding to some critical value $r_0 (T_c) = r_{0 c} $ that
is defined implicitly by $\lim_{t \to 0 +} \chi_b (t, 0)^{-1} = 0$
where $\chi_b (t, h) = - \lim_{L \to \infty}
\partial^2 f (t, h, L) / \partial h^2$ is the bulk susceptibility
for $t > 0$. This implies that $r_0 (T_c) = r_{0 c}(K_{i,j}, v,
u_0)$ depends on the lattice structure,  on $v$, on $u_0$, and on
all couplings $K_{i,j}$.

The matrix ${\bf A}$ affects the observable bulk critical behavior
: the eigenvalues $\lambda_\alpha$ enter the amplitudes of the
bulk correlation lengths $\xi^{(\alpha)}$ in the direction of the
{\it principal axes}; the latter are determined by the
eigenvectors ${\bf e}^{(\alpha)}$ of ${\bf A}$ which provide the
reference axes for the spatial dependence of the anisotropic bulk
order-parameter correlation function
\begin{equation}
\label{2l} G_b ({\bf x_i-x_j}; t, h) = \lim_{V \to \infty} \left\{
< \varphi_i \varphi_j> - < \varphi
>^2\right\}
\end{equation}
where $< \varphi > (t, h, L) = - \partial f (t, h, L) / \partial
h$. Correspondingly, the matrix ${\bf A}$ determines the
anisotropy of the ${\bf k}$ dependence of the Fourier transform
\begin{equation}
\label{2ll} \widehat G_b ({\bf k}; t, h) = v \sum_{\bf x} e^{- i
{\bf k} \cdot {\bf x}} G_b ({\bf x}; t, h)
\end{equation}
which is proportional to the observable scattering intensity. The
{\it principal} axes must be distinguished from the {\it symmetry}
axes of the Bravais lattice. The latter depend only on the lattice
points ${\bf x}_i$ but not on the couplings $K_{i,j}$. Below an
example is given where the principal axes differ from the symmetry
axes.

The long-wavelength approximation takes into account only the
leading $O (k_\alpha k_\beta)$ term of $\delta \widehat K ({\bf
k})$. In real space this is equivalent to using the $\varphi^4$
continuum Hamiltonian for the vector field $\varphi ({\bf x})$
\begin{eqnarray}
\label{2n} H_{field} &=& \int\limits_V d^d x \Bigg[\frac{r_0}{2}
\varphi^2 + \sum_{\alpha,
\beta=1}^d \frac{A_{\alpha \beta}}{2} \frac{\partial \varphi}
{\partial x_\alpha} \frac{\partial \varphi} {\partial x_\beta}
 \nonumber\\ &+& u_0 (\varphi^2)^2 - h \varphi \Bigg]
\end{eqnarray}
with some cutoff $\Lambda$.

Various types of anisotropies may result not only from pair
interactions on rectangular lattice structures but also from
nonrectangular lattice structures, from effective many-body
interactions as well as from  distortions of the lattice
structure, e.g., due to external shear forces. The
semi-macroscopic continuum model (\ref{2n}) is expected to be of
general significance in that it provides a complete
long-wavelength description of a large class of real systems near
criticality whose nonuniversal properties can be condensed into
the $d (d+1)/2$ parameters of the anisotropy matrix ${\bf A}$, in
addition to the nonuniversal parameters $r_0, u_0, h, \Lambda$.
The quantities $A_{\alpha
\beta}$ depend on all microscopic details (lattice structure,
electronic structure, many-body interactions) which, in general,
are not known a priori for a given material. Thus the matrix
elements $A_{\alpha
\beta}$ represent phenomenological parameters of a truly
nonuniversal character. Consequently, physical quantities
depending on $A_{\alpha
\beta}$ ( such as ${\cal
F}_{cube} (0, 0; {\bf \bar A})$, the Binder cumulant $ U_{cube}
({\bf \bar A})$, and the critical Casimir amplitude) are
nonuniversal as well.

For an appropriate formulation of the bulk order-parameter
correlation function (see Sect. III) and of finite-size scaling
functions (see Sect. VI) it will be important to employ the
reduced anisotropy matrix ${\bf \bar A} = {\bf A} / (\det {\bf
A})^{1/d}$ , $ {\bf \bar A} {\bf e}^{(\alpha)} = \bar
\lambda_\alpha {\bf e}^{(\alpha)}$ with the eigenvalues $\bar
\lambda_\alpha = \lambda_\alpha / (\det {\bf A})^{1/d} > 0$ and
with $\det {\bf \bar A} = \prod^d_{\alpha = 1} \bar \lambda_\alpha
= 1$. The matrix ${\bf \bar A}$ is independent of the kind of
variables $\varphi_i$ on the lattice points and independent of the
number $n$ of components of $\varphi_i$. It is well defined, e.g.,
also for models with fixed-length spin variables $ {\bf S_i }$
with $ |{\bf S_i }| = 1$ and for Ising models with discrete spin
variables $\sigma_i = \pm 1$ instead of the continuous vector
variables $\varphi_i$. Thus the XY and Ising Hamiltonians  $H_{XY}
= - \sum_{i,j} J_{i,j} {\bf S_i \cdot S_j}$ and $H_{Ising} = -
\sum_{i,j} J_{i,j} \sigma_i \sigma_j$ have the same reduced
anisotropy matrix ${\bf \bar A}$ and the same reduced eigenvalues
$\bar \lambda_\alpha$ as the $\varphi^4$ lattice Hamiltonian if
these models are defined on the same lattice points ${\bf x}_i$
and if the couplings $J_{i,j}$ are proportional to $K_{i,j}$.

As an illustration we consider an $L \times L \times L$
simple-cubic lattice model (Fig. 3) with a lattice constant
$\tilde a$ and with the following couplings: nearest-neighbor (NN)
couplings $K_x, K_y, K_z$ along the cubic symmetry axes,
next-nearest-neighbor (NNN) couplings $J_1, J_2, J_3$ only in the
$\pm (1,1,0)$, $ \pm (0,1,1)$,  $ \pm (1,0,1)$ directions [but not
in the $\pm (-1,1,0), \pm (0,-1,1), \pm (-1,0,1)$ directions] ,
and a third-NN coupling $\overline K $ only in the diagonal $\pm
(1, 1, 1)$ direction (Fig. 3).
\begin{figure}[!h]
\includegraphics[width=80mm]
{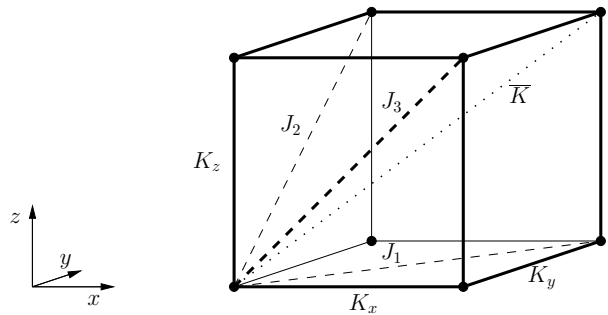} \caption{Lattice points $ {\bf x}_j$ of
the primitive cell (cube) of the anisotropic simple-cubic lattice
model (\ref{2a}) and (\ref{2g}) with ${\bf \bar A} \neq {\bf 1}$.
Solid, dashed, and dotted lines indicate the NN couplings
$K_\alpha$, the NNN couplings $J_i$, and the third-NN coupling
$\overline K$.}
\end{figure}
The corresponding anisotropy matrix is obtained from (\ref{2i}) as
\begin{equation}
 \label{32}
 {\bf  A} = 2\tilde a^2 \left(\begin{array}{ccc}
  D_x & J_1+\overline K & J_3+\overline K \\
  J_1+\overline K & D_y & J_2+\overline K \\
  J_3+\overline K & J_2+\overline K & D_z \\
\end{array}\right) \; ,
\end{equation}
with the diagonal elements $ D_x  =  K_x  +   J_1 + J_3  +
\overline K $ , $ D_y =  K_y  +  J_2 + J_1  +  \overline K $ , $
D_z =  K_z  +  J_3 + J_2  +  \overline K $ . For quantitative
analytical and numerical studies this model with seven different
couplings would, of course, be much too complicated. We shall
present explicit quantitative results only for two nontrivial
cases:

(i) Model with {\it three-dimensional anisotropy}: isotropic
ferromagnetic NN couplings $K_x=K_y=K_z\equiv K > 0$ and three
equal anisotropic NNN couplings $J_1=J_2=J_3\equiv J$. MC
simulations for three-dimensional Ising models with this type of
anisotropy (with $\overline K = 0$) have been performed by Schulte
and Drope \cite{schulte} and by Sumour et al. \cite{stauffer}. The
corresponding reduced anisotropy matrix (with $\overline K \neq
0$) is
\begin{equation}
 \label{32a}
 {\bf \bar A}  = (1-3w^2+2w^3)^{-1/3} \; \left(\begin{array}{ccc}
  1 & w & w \\
  w & 1 & w \\
  w & w & 1 \\
\end{array}\right) \; ,
\end{equation}
which depends only on the single anisotropy parameter
\begin{equation}
\label{32b} w \; = \; \frac{J \; + \; \overline K}{K \; + \; 2J \;
+ \; \overline K} \; .
\end{equation}
The eigenvalues of ${\bf A}$ and ${\bf \bar A}$ are $\lambda_1 = 2
\tilde a^2 (K  +  4 J  +  3 \overline K) $ , $ \lambda_2  =
\lambda_3  =  2 \tilde a^2 (K  +  J  )$  , and $ \bar \lambda_1 =
(1-3w^2+2w^3)^{-1/3}  (1  +  2 w) $ , $ \bar \lambda_2  =  \bar
\lambda_3  =  (1-3w^2+2w^3)^{-1/3}   (1  - w) $ , respectively.
The eigenvectors
\begin{equation}
 \label{2pp}
 {\bf e}^{(1)} = \frac{1}{\sqrt{3}}\left(\begin{array}{c}
  1 \\
  1 \\
  1 \\
\end{array}\right) \; , {\bf e}^{(2)} = \frac{1}{\sqrt{2}}\left(\begin{array}{c}
  -1 \\
  1 \\
  0 \\
\end{array}\right) \; ,{\bf e}^{(3)} = \frac{1}{\sqrt{6}}\left(\begin{array}{c}
  1 \\
  1 \\
  -2 \\
\end{array}\right)
\end{equation}
defining the principal axes are not parallel to the cubic symmetry
axes. The possible range of $w$ consistent with $\det {\bf \bar
A}(w) > 0$ is $ - \frac{1}{2} < w < 1$.  In the limit $K
\rightarrow 0, J \rightarrow 0$ at fixed $\overline K \neq 0$
corresponding to $w \rightarrow 1$ the model describes a system of
variables $\varphi_i$ on decoupled one-dimensional chains with NN
interactions $\overline K$. In the previous version \cite{cd2004}
of this model with $\overline K = 0$  the range of $w$ was
restricted to $- \frac{1}{2} < w \leq \frac{1}{2}$. A vanishing of
$\lambda_2$ and $\lambda_3$ occurs for $J \to - K$. At some value
$w = w_{LP}$ near $- 1/2$ (corresponding to $\lambda_1 = 0$) our
model is predicted to have a Lifshitz point with a wave-vector
instability in the direction of ${\bf e}^{(1)}$, i.e., in the
$(1,1,1)$ direction (see also Sect. VIII. E).

(ii) Model with {\it two-dimensional anisotropy}: An anisotropic
NNN coupling $J_1 \equiv J$ is taken into account only in the
$x-y$ planes of the three-dimensional sc lattice whereas all other
anisotropic couplings $J_2, J_3$ and $\overline K$ vanish. This
model is of interest for comparison with the MC data by Selke and
Shchur \cite{selke2005} for the {\it two-dimensional} anisotropic
Ising model as will be discussed in Sect. VIII. For further recent
studies of the anisotropic two-dimensional Ising model see also
\cite{zandvliet}.

The bulk critical behavior of both models (i) and (ii) belongs to
the same $d = 3$ universality class as that of the isotropic model
with $K_x=K_y=K_z=K > 0$ and $J_1=J_2 =J_3= \overline K = 0$
provided that $\lambda_\alpha > 0 \; , \alpha = 1,2,3$.

\subsection{Rotation and rescaling: shear transformation}

In order to derive an appropriate representation of the
anisotropic bulk order-parameter correlation function (see Sect.
III), to develop an appropriate formulation of finite-size
perturbation theory (see Sect. IV), and to treat the anisotropic
Hamiltonian $H$ by RG theory (see Sect. V) it is necessary to
first transform $H$ such that the $O(k_\alpha k_\beta)$ terms of
$\delta \widehat K ({\bf k})$ attain an isotropic form. This is a shear
transformation that consists of a rotation and rescaling of lengths in
the direction of the principal axes \cite{dohm2006}. The rotation
is provided by the orthogonal matrix $\bf U$ with matrix elements
$U_{\alpha
\beta} = e_\beta^{(\alpha)}$, $({\bf U}^{-1})_{\alpha \beta} =
e_\alpha^{(\beta)}$ where $e^{(\alpha)}_\beta $ denote the
Cartesian components of the eigenvectors ${\bf e}^{(\alpha)}$. The
rescaling is provided by the diagonal matrix $ {{\mbox
{\boldmath$\lambda$}} = \bf U AU}^{-1}$ with diagonal elements $
\lambda_\alpha >0$. In ${\bf k}$ space the transformation is ${\bf
k}'={\mbox {\boldmath$\lambda$}}^{1/2} {\bf U k}$ such that the $O
(k'_\alpha k'_\beta)$ term of $\delta \widehat K$ is brought into
an isotropic form with ${\bf A'} = {\bf 1}$,
\begin{equation}
\label{2o} \delta \widehat K ({\bf k})  = \delta \widehat K ({\bf
U}^{-1}{\mbox {\boldmath$\lambda$}}^{-1/2} {\bf k'}) \equiv \delta
\widehat K' ({\bf k'}) = \sum^d_{\alpha = 1} k'^2_\alpha + O
(k'^4)\;.
\end{equation}
In real space the transformed lattice points are ${\bf x'_j} =
{\mbox {\boldmath$\lambda$}}^{-1/2} {\bf U}{\bf x_j}$. This
transformation leaves the scalar product ${\bf k'\cdot x'_j} =
{\bf k \cdot x_j}$ invariant. Thereby the volume of the primitive
cells is changed to $v' = (\det {\bf A})^{- 1/2} v$.
Correspondingly the total volume of the transformed system is $V'
= N v' = (\det {\bf A})^{- 1/2} V$ with the characteristic length
$L'= V'^{1/d}$ .

Our transformation is defined such that the values of the
couplings $ K_{i,j}$ on the transformed lattice as well as the
temperature variable $r_0 (T)$ including the values
of $r_{0c}, a_0$ and $t$ are invariant [see also Eq. (\ref{4zz})
below]. This requires us to perform the additional transformations
$\varphi'_j = (\det {\bf A})^{1/4} \varphi_j , u'_0=(\det{\bf
A})^{-1/2}u_0$ and
\begin{equation}
 \label{2oo}h' = (\det {\bf A})^{1/4} h .
\end{equation}
In terms of
the Fourier transform $\hat \varphi'({\bf k'}) = \; v' \sum^N_{j=
1} e^{-i{\bf k'} \cdot {\bf x'}_j} \varphi'_j$ the transformed
lattice Hamiltonian reads
\begin{eqnarray}
\label{2p} \lefteqn{H' = V'^{-1} \sum_{{\mathbf k'}} \frac{1}{2}
[r_0 + \delta \widehat K' ({\bf k'})] \hat \varphi'({\mathbf k'})
\hat \varphi'({-{\mathbf k'}})}  \nonumber
\\ &+ \;u'_0 V'^{-3} \sum_{{\mathbf{k'p'}}{\mathbf q'}} [\hat
\varphi'({\mathbf k'}) \hat \varphi'({{\mathbf p'}})] [\hat
\varphi'({{\mathbf q'}}) \hat \varphi'({-{\mathbf k'}-{\mathbf
p'}-{\mathbf q'}})]  \nonumber \\ &- h' \hat \varphi'({\mathbf
0})\;.
\end{eqnarray}
\begin{figure}[!h]
\includegraphics[width=80mm]
{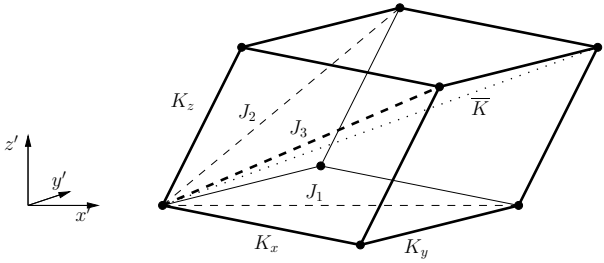} \caption{Lattice points ${\bf x}'_j$ of
the primitive cell (parallelepiped) of the transformed lattice
model (\ref{2p}) and (\ref{2q}) with the volume $v' = (\lambda_1
\lambda_2 \lambda_3)^{-1/2} \tilde a^3$. Solid, dashed, and dotted
lines indicate the NN couplings $K_\alpha$, the NNN couplings
$J_i$, and the third-NN coupling $\overline K$. The couplings are
the same as in Fig. 3 but ${\bf A'} = {\bf 1}$ (compare Figs. 1
and 2 of \cite{dohm2006}). }
\end{figure}

We illustrate this transformation by the example of the
simple-cubic lattice model shown in Fig. 3. The primitive cell
with the volume $v' = (\det {\bf A})^{- 1/2} \tilde a^3$ of the
transformed system is shown in Fig. 4. It has the shape of a
parallelepiped whose lengths and angles are determined such that
the transformed second-moment matrix ${\bf A}' = {\bf 1}$ is the
unity matrix although there are still the {\it same} NN couplings
$K_\alpha$, NNN couplings $J_i$, and third-NN coupling $\overline
K$ as in the simple-cubic lattice model of Fig. 3.

Working with $H'$ rather than $H$ will be of advantage in the
context of {\it bulk} properties and bulk renormalizations in
Sect. V. This is not the case for the confined system. Although
the $O ({\bf k'}^2)$ term of $\delta \widehat K' ({\bf k'})$ in
(\ref{2o}) and (\ref{2p}) looks quite simple, namely ${\bf k'}
\cdot {\bf k'}$ with a trivial anisotropy matrix ${\bf A'} = {\bf
1}$, the summations $\sum_{{\bf k'}}$ are nontrivial.

For concreteness consider the simplified example (i) with the
matrix (\ref{32a}) and the eigenvectors (\ref{2pp}). While the
${\bf k}$ vectors of the $sc$ lattice have the simple form
\begin{equation}
 \label{2-p}
 {\bf k} \;= \; \left(\begin{array}{c}
  k_1 \\
  k_2 \\
  k_3 \\
\end{array}\right)  \; = \; \frac{2 \pi}{L}\left(\begin{array}{c}
  m_1 \\
  m_2 \\
  m_3 \\
\end{array}\right)
\end{equation}
with the integer numbers $m_i = 0, \pm 1, \pm 2$, \ldots, the
${\bf k'}$ vectors are considerably more complicated,
\begin{equation}
 \label{2qq}
 {\bf k'} \; = \; \left(\begin{array}{c}
  k'_1 \\
  k'_2 \\
  k'_3 \\
\end{array}\right)  \; = \; \frac{2 \pi}{L \sqrt{6}}\left(\begin{array}{c}
  (m_1 + m_2 + m_3) \sqrt{2 \lambda_1} \\
  (m_2 - m_1) \sqrt{3 \lambda_2} \\
  (m_1 + m_2 - 2 m_3) \sqrt{\lambda_3} \\
\end{array}\right) \; .
\end{equation}
These ${\bf k}'$ vectors reflect the shape and lattice structure
of the transformed system. Thus the price paid for transforming
${\bf A} \neq {\bf 1}$ to ${\bf A}' = {\bf 1}$ is to work with
more complicated ${\bf k}'$ vectors. This example demonstrates
that the effect of anisotropy cannot be eliminated for {\it
confined} systems. In our applications the summations in
finite-size perturbation theory will be performed in the simpler
${\bf k}$ space whereas {\it bulk}  integrals (with infinite
cutoff) are simplified in ${\bf k'}$ space.

In real space the Hamiltonian $H'$ reads
\begin{eqnarray}
\label{2q} H'  &=&   v' \Bigg[\sum_{i=1}^N \left(\frac{r_0}{2}
{\varphi'_i}^2 + u'_0 ({\varphi'_i}^2)^2 - h' \varphi'_i \right)
\nonumber\\ &+& \sum_{i, j=1}^N \frac{K_{i,j}} {2} ({\varphi'_i} -
{\varphi'_j})^2 \Bigg].
\end{eqnarray}
By substituting the transformations defined above one easily
verifies
\begin{equation}
\label{2r} H (r_0, h, u_0, K_{i,j},v,L) = H' (r_0, h', \
u_0',K_{i,j},v',L').
\end{equation}
The measure for the temperature distance from criticality $r_0 -
r_{0c} = a_0 t$ is the same for both $H$ and $H'$. Defining the
free energy density $f'$ (divided by $k_B T$) as
\begin{eqnarray}
\label{2s} f'(t,h',L')  =  - {V'}^{-1} \ln Z' (t, h', L') \; ,
\end{eqnarray}
\begin{eqnarray}
\label{2t} Z' (t, h', L') = \left[\prod^N_{j = 1} \frac{\int d^n
\varphi'_j} {(v')^{n (2-d) / (2d)}} \right] \exp \left( - H'
\right),
\end{eqnarray}
one obtains the exact relations
\be
\label{2tt} Z (t, h, L) \; = \; (\det {\bf A})^{- n N / (2 d)} \;
Z'(t, h', L') \; ,
\ee
\begin{equation}
\label{2u} f(t,h,L) = (\det {\bf A})^{- 1/2} f'(t,h',L')\; + \;
[n/(2d v)] \ln (\det {\bf A}) \; .
\end{equation}
The last term is a bulk contribution, i.e., independent of $L$.
Furthermore it is independent of $t$, i.e., a non-singular bulk
contribution, thus the singular bulk parts of
$f_b(t,h)=f(t,h,\infty)$ and of $f'_b(t,h')=f'(t,h',\infty)$ as
well as the total singular parts of $f (t, h, L)$ and of $f' (t,
h', L')$ are related by
\begin{equation}
\label{2v} f_{s,b} (t, h) = (\det {\bf A})^{- 1/2} f'_{s,b} (t,
h') \; ,
\end{equation}
\begin{equation}
\label{2vv} f_s(t,h,L) = (\det {\bf A})^{- 1/2} f_s'(t,h',L') \; .
\end{equation}
The bulk correlation function of the transformed system is
\begin{equation}
\label{2w} G'_b ({\bf x_i' - x_j'}; t, h') = \lim_{V' \to \infty}
\left\{ < \varphi_i' \varphi_j'>' - (< \varphi'
>')^2\right\}
\end{equation}
where $<...>'$ denotes the average with the weight $\sim $ exp
$(-H')$. It is related to $G_b$ by
\begin{equation}
\label{2y} G_b ({\bf x}; t, h) = (\det {\bf A})^{- 1/2} G'_b
({\mbox {\boldmath$\lambda$}}^{-1/2} {\bf U} {\bf x}; t, (\det{\bf
A})^{1/4} h).
\end{equation}
The corresponding relation between the Fourier transforms is
\begin{equation}
\label{2x} \widehat G_b ({\bf k}; t, h) = \widehat G'_b ({\mbox
{\boldmath$\lambda$}}^{1/2} {\bf U} {\bf k}; t, (\det{\bf
A})^{1/4} h).
\end{equation}
(In the arguments of Eqs. (\ref{2tt}) - (\ref{2x}) we have, for
simplicity, not indicated explicitly the additional
transformations of $u_0 = (\det{\bf A})^{1/2} u_0'$ and of $v' =
(\det {\bf A})^{- 1/2} v$.) In terms of the transformed field $
\varphi'({\bf x}') = (\det{\bf A})^{1/4}\varphi({\bf U}^{-1}
{\mbox{\boldmath$\lambda$}}^{1/2} {\bf x'})$ the Hamiltonian
(\ref{2n}) attains the form of the standard isotropic
Landau-Ginzburg-Wilson Hamiltonian
\begin{eqnarray}
\label{2z}H_{field} = H'_{field} &=& \int\limits_{V'} d^d x'
\big[\frac{r_0} {2} \varphi'({\bf x}')^2 +  \frac{1} {2} (\nabla'
\varphi')^2 \nonumber\\ &+&
 u'_0 (\varphi'^2)^2 - h' \varphi' \big]
\end{eqnarray}
where $\nabla'  \varphi' \equiv (\partial  \varphi' /
\partial x_1', \ldots, \partial  \varphi' / \partial x_d')$ with a
transformed cutoff.

\section{Bulk critical behavior of anisotropic systems}

Before turning to finite-size theory of anisotropic systems it is
necessary to first discuss the bulk critical behavior of
anisotropic systems and its relation to that of isotropic systems.
We are not aware of such a discussion in the literature. It is
well known that anisotropic systems and isotropic systems have the
same critical exponents (in a limited range of the anisotropy, see, e.g., \cite{bruce}, and references therein).
Within the $\varphi^4$ theory this is immediately seen from the
relation of dimensionally regularized bulk integrals (at infinite
cutoff) such as
\begin{eqnarray}
\label{3-a} u_0 \int\limits_{\bf k}^\infty (r_0 + {\bf k} \cdot
{\bf Ak})^{-1} &=& u'_0 \int\limits_{{\bf k}'}^\infty (r_0 + {\bf
k}' \cdot {\bf k}')^{-1} \nonumber\\ &=& - \;
\frac{A_d}{\varepsilon} r_0^{(d-2)/2} \; u_0',
\end{eqnarray}
\begin{eqnarray}
\label{3dd} \int\limits_{\bf k'}^\infty \; = \; (\det {\bf
A})^{1/2} \int\limits_{\bf k}^\infty \; \equiv \; (\det {\bf
A})^{1/2} \prod^d_{\alpha = 1} \; \int\limits^\infty_{- \infty}
\; \frac{d k_\alpha}{2\pi} \;,  \nonumber\\
\end{eqnarray}
provided that $\det {\bf A} > 0 $. We see that the ${\bf A}$
dependence is completely absorbed by the coupling $u'_0$ and that
the $d=4$ pole term $\sim \varepsilon^{-1}$ does not depend on the
matrix ${\bf A}$. This leads to identical field-theoretic
functions for anisotropic and isotropic systems (as functions of
the renormalized couplings $u'$ and $u$, respectively) and yields
the same critical exponents and fixed point value $u'^* = u^*$ for
anisotropic and isotropic systems (see Sect. V). (The $d=2$ pole
of the integral (\ref{3-a})  which has nothing to do with the
critical behavior in $ d > 2$ dimensions can be incorporated in
the geometric factor $A_d$ \cite{dohm1985} which is finite in $2 <
d \leq 4 $ dimensions [see (\ref{5c1}) below].)

For this reason not much attention has been paid to the role
played by anisotropy in bulk critical phenomena. This is not
justified, however, in the context of the important feature of
two-scale factor universality \cite{stau, priv}. Its validity has
been established by the RG theory only for {\it isotropic} systems
at $h=0$ with short-range interactions \cite{weg-1,aha-74,ger-1,hoh-76}. A
brief derivation was also given by Privman and Fisher \cite{pri}
and by Privman et al. \cite{priv} using scaling assumptions at $h
\neq 0$. Their ansatz for the order-parameter correlation
function, however, is not valid for anisotropic systems since they
assumed the existence of a {\it single} bulk correlation length
$\xi_\infty$. Recently it has been pointed out that two-scale
factor universality is absent in anisotropic systems \cite{cd2004}
and that anisotropy has an effect on several universal bulk
amplitude combinations \cite{dohm2006} but no derivation was
given. In particular, two important universal amplitude relations
derived by Privman and Fisher \cite{pri} [Eqs. (\ref{3g}) and
(\ref{3h}) below] have not been discussed in the context of
anisotropic systems. Furthermore, the bulk order-parameter
correlation function of anisotropic systems was discussed only for
$h=0$ and $ T \geq T_c$ \cite{cd2004}. Here we extend this
discussion to $ h \neq 0$ and $ T<T_c $ and provide an appropriate
formulation of the order-parameter correlation function and of the
scattering intensity in terms of both the eigenvalues
$\lambda_\alpha$ and the reduced eigenvalues $\bar \lambda_\alpha$
of the anisotropic system. We also present the derivation of
several amplitude combinations for anisotropic systems in terms of
universal scaling functions.

All of the {\it thermodynamic} bulk relations given in the
following subsections A and B remain valid also in the presence of
subleading long-range interactions of the type (\ref{2hh}). This
is not the case, however, for bulk correlation functions at $T
\neq T_c$, $h=0$ and $h \neq 0$, $T=T_c$ in the large-distance
regime corresponding to the shaded region in Fig. 2.

\subsection{Two-scale factor universality in
isotropic bulk systems}

First we summarize the bulk critical behavior of systems described
by the (asymptotically) isotropic lattice Hamiltonian $H'$ and the
continuum Hamiltonian $H'_{field}$. Near $T_c$ the bulk free
energy density can be decomposed uniquely into singular and
non-singular parts as $f_b' (t, h') = f_{s, b}' (t, h') + f'_{ns,
b} (t)$ where $f_{ns,b}' (t)$ has a regular $t$ dependence. It is
well established that $f'_{s, b}$ has the asymptotic (small $t$,
small $h'$) scaling form below $d = 4$ dimensions
\begin{equation}
\label{3a} f'_{s, b} (t, h') = A_1' |t|^{d \nu} \; W_\pm (A'_2 h'
|t|^{- \beta \delta})
\end{equation}
with  the universal scaling function $W_\pm (z)$, $\infty \leq z
\leq \infty$. This function is independent of the cutoff procedure
of $H'_{field}$. We use the normalization $W_+ (0) = 1$. The two
amplitudes $A'_1$ and $A'_2$ are nonuniversal.

Because of isotropy it is justified to define a {\it single}
(second-moment) bulk correlation length
\begin{equation}
\label{3b} \xi_\pm'(t, h') = \left(\frac{1} {2d} \frac{\sum_{\bf
x'}\; {\bf x}'^2 \;G_b' ({\bf x}'; t, h')}{\sum_{\bf x'}\; G_b'
({\bf x}'; t, h')} \right)^{1/2} \;
\end{equation}
above and below $T_c$, respectively.  In (\ref{3b}) we have
assumed sufficiently rapidly decaying correlations, i.e., general
$n$ for $T \geq T_c$ but $n = 1$ for $T < T_c$. We assume, in the
asymptotic region $|{\bf x'}| \gg (v')^{1/d} \; , \xi'_\pm \gg
(v')^{1/d}$ and for $| {\bf x'} | / \xi'_\pm \lesssim O (1)$ and
small $h'$, the asymptotic {\it isotropic} scaling form
\begin{equation}
\label{3c} G'_b ({\bf x'}; t, h') = D'_1 | {\bf x'} |^{- d + 2 -
\eta} \Phi_\pm (|{\bf x'}| / \xi'_{\pm}, D'_2 h'
|t|^{-\beta\delta} ) \; ,
\end{equation}
\begin{equation}
\label{3d} \xi'_\pm(t, h') = \xi'_{0+} |t|^{- \nu} X_\pm (D'_2 h'
|t|^{-\beta\delta}) \; ,
\end{equation}
with universal scaling functions $\Phi_\pm(x,y)$ and $X_\pm(y)$.
We use the normalization $X_+ (0) = 1$, thus $\xi_+'(t,0) =
\xi'_{0 +} t^{- \nu}$ above $T_c$. The length  $\xi'_{0 +}$  will
be needed as a reference length in the formulation of renormalized
finite-size theory in Sect. V. The corresponding scaling form of
the Fourier transform $\widehat G'_b$ of (\ref{3c}) is
\begin{equation}
\label{3da} \widehat G'_b ({\bf k}'; t, h') = D'_1 | {\bf k}'|^{-
2 + \eta} \widehat \Phi_\pm (|{\bf k}'| \xi_\pm', D_2' h' |t|^{-
\beta \delta}) \; ,
\end{equation}
\begin{eqnarray}
\label{3db}  &&\widehat \Phi_\pm (x', y') = 2 \pi^{(d-1)/2}
\;\Gamma ((d-1)/2)^{-1}
\int\limits_0^\infty ds \; s^{1 - \eta} \nonumber\\
&& \times \int\limits_{-1}^1 d (\cos \vartheta) (\sin
\vartheta)^{d-3}\; e^{- i s \cos \vartheta} \Phi_\pm (s/x', y') \;
.
\end{eqnarray}
The three amplitudes $D_1'$, $D_2'$, and $\xi'_{0 +}$ in Eqs.
(\ref{3c}) - (\ref{3da}) are nonuniversal. The basic content of
two-scale factor universality is that all of these amplitudes are
universally related to the two thermodynamic amplitudes $A'_1$ and
$A_2'$. The relations read
\begin{equation}
\label{3f}  \left(\xi'_{0 +}\right)^d A_1'  = Q_1(d,n) = universal
\;,
\end{equation}
\begin{equation}
\label{3g} A_2' / D_2' \; = \; P_2 (d, n) = universal \; ,
\end{equation}
\begin{equation}
\label{3h} D_1' (A_2')^{-2} (A_1')^{- 1 - \gamma / (d \nu)} \; =
\; P_3 (d, n) = universal \; .
\end{equation}
In Ref. \cite{pri} the universal constants $P_2$ and $P_3$ were
denoted by $Q_2$ and $Q_3$. In order to conform with Refs.
\cite{tarko,priv} and to avoid confusion we reserve the notation
$Q_2$ and $Q_3$ for the {\it different} universal constants in
Eqs. (\ref{3k}) and (\ref{3l}) below. For the sake of clarity we
present the explicit expressions for $Q_i$ and $P_i$ in terms of
universal scaling functions in Appendix A. An equivalent
formulation of Eq. (\ref{3f}) is
\begin{equation}
\label{3f1}\lim_{t \to 0 +} \left[f_{s,b}' (t, 0) \; \xi_+' (t,
0)^d \right] \; = \; Q_1 (d, n) = universal \; .
\end{equation}
The validity of (\ref{3f}) and (\ref{3f1}) has been established by
the RG theory \cite{weg-1,hoh-76}.

Furthermore, the following amplitude ratios
\begin{equation}
\label{3k} (\Gamma_+' / \Gamma_c') (\xi_c' / \xi_{0 +}')^{2-\eta}
= Q_2(d,n)= universal \; ,
\end{equation}
\begin{equation}
\label{3l} \widehat D_\infty' (\xi_{0 +}')^{2 - \eta} / \Gamma_+'
\; = Q_3(d,n) = universal\; ,
\end{equation}
have been proposed  \cite{tarko} to be universal. The constants
$\Gamma_+' \; ,  \Gamma_c'$ and $\xi_c'$ are defined as follows:
$\chi_b' (t, 0) = \Gamma_+' t^{- \gamma} \; {\rm for} \; t > 0,
\chi_b' (0, h') = \Gamma_c' | h' |^{- \gamma/(\beta \delta)}$
where $\chi_b' (t, h') = - \partial^2 f_b (t, h') /
\partial h'^2$, and $\xi'_\pm (0, h') \equiv \xi'_h = \xi_c' \;| h' |^{- \nu /(
\beta \delta)}$. The length $\xi'_h$ with the amplitude $\xi_c'$ is a natural reference length
of finite-size theory at $t=0$, $h' \neq 0$ [see (\ref{6-l})
below]. $\widehat D_\infty'$ is the asymptotic (small ${\bf k'}$)
amplitude of $\widehat G'_b ({\bf k'}; 0, 0) \approx \widehat
D_\infty'  |{\bf k'}|^{ -2 + \eta }$. Alternatively, Eq.
(\ref{3l}) can be formulated as $D_\infty' (\xi_{0 +}')^{2 - \eta}
/ \Gamma_+' = \widetilde Q_3 (d, n)$, or, equivalently,
\begin{eqnarray}
\label{3l1} \lim_{|{\bf x}'| \to \infty} \left\{G'_b ({\bf x}' ;
0, 0) \left(|{\bf x}'| / \xi_{0 +}' \right)^{d - 2 + \eta}\right\}
\left(\xi_{0 +}' \right)^d / \Gamma_+' \nonumber\\ = \widetilde
Q_3 (d, n) = \left(D_\infty' / \widehat D_\infty' \right) Q_3 (d,
n) = universal \quad
\end{eqnarray}
where $D_\infty'$ is the asymptotic (large-${\bf x}'$) amplitude
of $G_b' ({\bf x}' ; 0,0) \approx D'_\infty | {\bf x}'|^{- d + 2 -
\eta}$. The derivation of Eqs. (\ref{3k}) - (\ref{3l1}) is
sketched in App. A. Again, all of the constants on the left-hand sides of
(\ref{3k}) - (\ref{3l1}) are universally related to $A'_1$ and
$A'_2$.

Below $T_c$ at $h' = 0$ we have, for $n = 1$, $\xi'_- = \xi'_{0 -}
| t |^{- \nu}$ with the universal ratio
\begin{equation}
\label{3e} \xi'_{0 -} / \xi'_{0 +} = X_- (0) = universal \; .
\end{equation}

Previously the bulk relations (\ref{3f}) - (\ref{3e}) were
expected to be universal for all systems {\it within a given
universality class} \cite{priv}. Consistency with the universality
of (\ref{3k}) and (\ref{3l}) was found \cite{tarko} for
two-dimensional (square and triangular) Ising lattices and
three-dimensional (sc and bcc) Ising lattices with {\it isotropic}
nearest-neighbor interactions (see also \cite{liu}). All of these
systems, however, belong to the subclass of asymptotically
isotropic systems with an anisotropy matrix ${\bf A} = c_0 {\bf
1}$ or ${\bf \bar A} = {\bf 1}$ and with an isotropic scattering
intensity \cite{tarko}. Also the honeycomb-lattice Ising model
considered in \cite{pri} belongs to that subclass, with a constant $c_0$ different from that for the triangular lattice or the square lattice. As will be shown in Subsect. B  below,
Eqs. (\ref{3f}) and (\ref{3f1}) - (\ref{3e}) must be reformulated
for anisotropic systems with noncubic anisotropy at $O(k_\alpha
k_\beta)$ whereas (\ref{3g}) and (\ref{3h}) remain valid also for
anisotropic systems provided that $A_1', A_2', D_1'$ and $D_2'$
are transformed appropriately.

It has been shown \cite{cd2000-1,cd2000-2,cd2002,dan-2} that the
universal scaling form (\ref{3c}), (\ref{3d}) is not valid in the
regime $r'\equiv|{\bf x}'| \gg \xi'_+$ above $T_c$. The same can be shown
for $n=1$ in the regime $r' \gg \xi'_-$ below
$T_c$. Note that this regime is part of the {\it asymptotic}
critical region $r' \gg \tilde a$ and $ \xi'_\pm \gg
\tilde a$ corresponding to the shaded area in the
${r'}^{-1}$ - $\xi_\pm'^{-1}$ plane in Fig. 2. In this
regime, corrections to scaling in the sense of Wegner
\cite{wegner1972} are still negligible. One must distinguish at
least four cases: (i) For the $\varphi^4$ lattice model with
short-range interactions, the exponential decay above $T_c$
depends explicitly on the lattice spacing $\tilde a$ via the {\it
exponential correlation length} $\xi_{{\bf e}}$
\cite{cd2002,fish-2}; in Sect. X we shall show that it also
depends on the bare four-point coupling $u_0$. (ii) For the
$\varphi^4$ continuum theory with a {\it smooth} cutoff $\Lambda$,
the exponential decay depends explicitly on $\Lambda$ via
$\xi_{{\bf e}}(\Lambda)$ \cite{cd2000-1,cd2000-2} (see also Sect.
X). (iii) For the $\varphi^4$ continuum theory with a {\it sharp}
cutoff $\Lambda$, a nonuniversal oscillatory power-law decay
dominates the exponential decay \cite{cd2002,cd2003}. (iv) In the
presence of subleading long-range interactions of the type
(\ref{2hh}), the power law $\sim b |{\bf x}'|^{- d - \sigma}$
\cite{wi-1,dan-2} dominates the exponential short-range behavior.
For $T>T_c$ this has been shown explicitly for the mean spherical model where the
asymptotic structure for $|{\bf x}'|/\xi_+ \gg 1$  at $h=0$ is \cite{dan-2}
\begin{eqnarray}
\label{3xx}  G'_b ({\bf x'}; t, 0)
= \frac{D'_1}{ |{\bf x'|}^{d-2}} \Big[\Phi_+\Big(\frac{|{\bf x}'|}{\xi_+ }\Big)
+ \frac{b}{ |{\bf x}'|^{\sigma-2}} {\cal D} \Big(\frac{|{\bf x}'|}{\xi_+ }\Big)\Big] \nonumber\\
\end{eqnarray}
with $ {\cal D}(|{\bf x}'|/\xi_+) \sim (|{\bf x}'|/\xi_+)^{-4}$.
In all cases (i) - (iv), scaling in the sense  of (\ref{3c}),
(\ref{3d}) and two-scale factor universality are violated in the
regime $|{\bf x}'| \gg \xi'_\pm$ (shaded area in Fig. 2) because, in addition to the
reference length $\xi'_{0+}$, the nonuniversal lengths $\tilde a$,
$u_0^{-1/\varepsilon} $, $\Lambda^{-1}$, and $b^{1/(\sigma - 2)}$
govern the leading large $ |{\bf x}'| $ behavior.

\subsection{Absence of two-scale factor universality in
anisotropic bulk systems}

Now we turn to the anisotropic system. According to (\ref{2oo}),
(\ref{2v}) and (\ref{3a}), the asymptotic scaling form of
$f_{s,b}$ is given by (\ref{1a}) with the nonuniversal amplitudes
\begin{equation}
\label{3mm} A_1 = A'_1 (\det {\bf A})^{- 1/2}, A_2 = A'_2 (\det
{\bf A})^{ 1/4} \;.
\end{equation}
In order to represent the order-parameter correlation function
(\ref{2l}) in an appropriate asymptotic scaling form it is
necessary to employ both of the diagonal matrices ${\mbox
{\boldmath$\lambda$}}$ and ${\bf \bar{\mbox
{\boldmath$\lambda$}}}$ with diagonal elements $\lambda_\alpha$
and $ \bar \lambda_\alpha$. Using (\ref{2y}), (\ref{2x}),
(\ref{3c}) and (\ref{3da}) we write $G_b$ and $\widehat G_b$ as
\begin{eqnarray}
\label{3n} G_b ({\bf x}; t, h) &=& D_1 |{\bf \bar{\mbox
{\boldmath$\lambda$}}}^{-1/2} {\bf U}{\bf x}|^{- d + 2 - \eta}
\nonumber\\ &\times& \Phi_\pm (|{\mbox
{\boldmath$\lambda$}}^{-1/2} {\bf U}{\bf x}| / \xi'_\pm, D_2 h
|t|^{-\beta\delta}) \; ,
\end{eqnarray}
\begin{eqnarray}
\label{3na} \widehat G_b ({\bf k}; t, h) &=& D_1 |{\bf \bar{\mbox
{\boldmath$\lambda$}}}^{1/2} {\bf U}{\bf k}|^{- 2 + \eta}
\nonumber\\ &\times& \widehat \Phi_\pm (|{\mbox
{\boldmath$\lambda$}}^{1/2} {\bf U}{\bf k}| \xi'_\pm, D_2 h
|t|^{-\beta\delta})
\end{eqnarray}
with the nonuniversal amplitudes
\begin{equation}
\label{3nn}D_1 = D'_1 (\det {\bf A})^{(- 2 + \eta) / (2 d)} \; ,
\end{equation}
\begin{equation}
\label{3n1} D_2 = D'_2 (\det {\bf A})^{1/4} \;.
\end{equation}
Here we identify the spatial variable $r'$ in the scaling argument
of $\Phi_\pm(r'/ \xi'_\pm,0)$ used in Fig. 2 as
\begin{equation}
\label{3x}  r'\equiv |{\bf x'}|=|{\mbox
{\boldmath$\lambda$}}^{-1/2} {\bf U}{\bf x}|\;,
\end{equation}
which, for given ${\bf x}$, depends on all of the $d(d+1)/2$ parameters contained in ${\bf A}$.

While the simple transformations (\ref{3mm}) and (\ref{3n1})
follow immediately from the transformations of $h, \varphi_i$, and
$V$, the transformation of $D_1$ is less trivial. Using Eqs.
(\ref{3mm}), (\ref{3nn}), and (\ref{3n1}) we find that the
universal amplitude relations (\ref{3g}) and (\ref{3h}) of
isotropic systems remain valid also for anisotropic systems :
\begin{equation}
\label{3n2} A_2 / D_2 \; = \; P_2 (d, n) = universal \; ,
\end{equation}
\begin{equation}
\label{3n3} D_1 A_2^{-2} A_1^{- 1 - \gamma / (d \nu)} \; = \; P_3
(d, n) = universal
\end{equation}
with the same universal constants $P_2$ and $P_3$ as in (\ref{3g})
and (\ref{3h}). Eq. (\ref{3n}) differs from the representation of
$G_b$ of \cite{cd2004} at $h=0$ where, instead of $D_1$, an
overall amplitude $A_G' = D_1'(\det {\bf A})^{-1 /2} $ was
employed. The latter representation is inappropriate as $A'_G$ is
not universally related to $A_1$ and $A_2$. The relations
(\ref{3g}) and (\ref{3n2})  follow from the sum rule (see App. A)
\begin{eqnarray}
\label{3-n} \chi_b' (t, h')= -\partial^2 f_b' (t, h') / \partial
h'^2 = v'\sum_{\bf x'}\; G_b' ({\bf x}'; t, h') \nonumber \\=
\chi_b (t, h)= -\partial^2 f_b (t, h) / \partial h^2= v \sum_{\bf
x}\; G_b ({\bf x}; t, h).
\end{eqnarray}
Less obvious are the relations (\ref{3h}) and (\ref{3n3}). Their
derivation is, in fact, based on an additional assumption about
the {\it unsubtracted} order-parameter correlation function (see
App. A). The physical significance of (\ref{3n3}) is that, at
criticality, the bulk correlation function and its Fourier
transform, if expressed in term of ${\bf \bar{\mbox
{\boldmath$\lambda$}}}$,
\begin{equation}
\label{3n4} G_b ({\bf x}; 0, 0) = D_1 \; \Phi_\pm (0,0) \; | {\bf
\bar{\mbox {\boldmath$\lambda$}}}^{-1/2} {\bf U}{\bf x}|^{- d + 2
- \eta} \; ,
\end{equation}
\begin{equation}
\label{3n5} \widehat G_b ({\bf k}; 0, 0) = D_1 \; \widehat
\Phi_\pm (0, 0) |{\bf \bar{\mbox {\boldmath$\lambda$}}}^{1/2} {\bf
Uk}|^{- 2 + \eta}\; ,
\end{equation}
have an overall amplitude $D_1$ that is universally determined by
the {\it thermodynamic} amplitudes $A_1$ and $A_2$ of the bulk
free energy $f_{s,b}$. Unlike in isotropic systems, however, the
spatial dependence of $G_b$ and the ${\bf k}$ dependence of
$\widehat G_b$ at criticality are governed by the $d$ reduced
nonuniversal eigenvalues $\bar \lambda_\alpha$ (with $d - 1$
independent parameters). In addition, the knowledge of $d (d - 1)
/ 2$ nonuniversal parameter is needed in order to specify the
orthogonal matrix ${\bf U}$, i.e., to specify the directions ${\bf
e}^{(\alpha)}$ of the principal axes relative to the symmetry axes
of the system. Thus $1 + (d-1) + d (d - 1) / 2 = d (d + 1) / 2$
nonuniversal parameters are needed at $T = T_c$ and $h = 0$, and
$d (d+1) / 2 + 1$ nonuniversal parameters at finite $h$. These
parameters can be measured by elastic-scattering experiments at
bulk criticality of anisotropic solids.

Now we discuss the temperature and $h$ dependence of $G_b$ away
from criticality. Along the direction ${\bf e}^{(\alpha)}$ of the
principal axis $\alpha$ the spatial dependence of (\ref{3n}) is,
for $|\tilde x^{(\alpha)}| / \xi_\pm^{(\alpha)} \lesssim O(1)$,
\begin{eqnarray}
\label{3o} G_b ({\bf \tilde x}^{(\alpha)}; t, h) &=& D_1 (| \tilde
x^{(\alpha)} |/ \bar \lambda_\alpha^{1/2})^{- d + 2 - \eta} \nonumber\\
&\times& \Phi_\pm \left(|\tilde x^{(\alpha)}| /
\xi_\pm^{(\alpha)}, D_2 h |t |^{- \beta \delta}\right)
\end{eqnarray}
with ${\bf \tilde x}^{(\alpha)} = \tilde x^{(\alpha)} {\bf
e}^{(\alpha)}$ and, because of (\ref{2oo}) and (\ref{3d}),
\begin{eqnarray}
\label{3p} \xi_\pm^{(\alpha)} (t, h) &=& \lambda_\alpha^{1/2}
\xi_\pm'
(t,(\det {\bf A})^{1/4} h) \nonumber\\
&=& \xi^{(\alpha)}_{0+} |t|^{- \nu} X_\pm (D_2 h
|t|^{-\beta\delta}).
\end{eqnarray}
Thus along the different principal axes (see Fig. 1 (b) of Ref.
\cite{dohm2006}) there exist $d$ different principal correlation
lengths $\xi_\pm^{(\alpha)} (t, h)$ which constitute $d$ different
nonuniversal reference lengths with $d$ nonuniversal amplitudes
$\xi_{0 +}^{(\alpha)} = \lambda_\alpha^{1/2} \xi_{0 +}'$. Their
ratios
\begin{equation}
\label{3q} \xi_{0 +}^{(\alpha)} / \xi_{0 +}^{(\beta)} =
(\lambda_\alpha / \lambda_\beta)^{1/2}
\end{equation}
are also nonuniversal. Below $T_c$ at $h = 0$ we have
$\xi_-^{(\alpha)} = \xi_{0 -}^{(\alpha)} | t |^{- \nu}$ with the
universal ratio for each $\alpha$
\begin{equation}
\label{3r} \xi_{0 -}^{(\alpha)} / \xi_{0 +}^{(\alpha)} = X_- (0) =
universal
\end{equation}
but for $\alpha \neq \beta$ the ratios $\xi_{0 -}^{(\alpha)} /
\xi_{0 -}^{(\beta)} = (\lambda_\alpha / \lambda_\beta)^{1/2}$ and
$ \xi_{0 -}^{(\alpha)} / \xi_{0 +}^{(\beta)} = (\lambda_\alpha /
\lambda_\beta)^{1/2} X_- (0)$ are nonuniversal. Because of $A'_1 =
A_1 \prod_{\alpha = 1}^d \lambda_\alpha^{1/2} $, Eqs. (\ref{3f})
and (\ref{3f1}) imply
\begin{eqnarray}
\label{3u} A_1 \prod^d_{\alpha = 1} \xi_{0 +}^{(\alpha)} &=&
\lim_{t \to 0 +} \left[f_{s,b} (t, 0) \; \prod^d_{\alpha = 1}
\xi_+^{(\alpha)}(t,0) \right]\nonumber\\ &=& \;Q_1(d,n) =
universal \;.
\end{eqnarray}
The susceptibility $\chi_b (t, h)$ with $\chi_b (t, 0) = \Gamma_+
t^{- \gamma}$ above $T_c$ and $\chi_b (0, h) = \Gamma_c | h |^{-
\gamma /
\beta \delta}$ have the amplitudes $\Gamma_+ = \Gamma_+'$ and
$\Gamma_c = \Gamma_c' (\det {\bf A})^{- \gamma / (4 \beta
\delta)}$. Here we have used (\ref{2oo}) and (\ref{3-n}). From Eq.
(\ref{3p}) we have $\xi_\pm^{(\alpha)}(0,h) =
\xi_c^{(\alpha)}|h|^{- \nu/(\beta \delta)}$ with
\begin{equation}
\label{3w} \xi_c^{(\alpha)} = \lambda_\alpha^{1/2} (\det {\bf
A})^{- \nu / (4 \beta \delta)} \xi_c' \;.
\end{equation}
Eq. (\ref{3k}) then implies for each $\alpha = 1, ..., d$
\begin{eqnarray}
\label{3y} \left(\Gamma_+ / \Gamma_c \right)
\left(\xi_c^{(\alpha)} / \xi_{0 +}^{(\alpha)} \right)^{2 - \eta}
 = Q_2 (d, n) = universal \; , \quad
\end{eqnarray}
and from Eqs. (\ref{3l1}) and (\ref{3o}) we obtain for each $\beta
 $
\begin{eqnarray}
\label{3z} \lim_{| {\bf \tilde x}^{(\beta)} | \to \infty} \;
\left\{ G_b ({\bf \tilde x}^{(\beta)}; 0,0) \left( \frac{| {\bf
\tilde x}^{(\beta)} | } {\xi_{0 +}^{(\beta)}} \right)^{d - 2 +
\eta} \right\} \frac{\prod_{\alpha = 1}^d \limits \;  \xi_{0
+}^{(\alpha)}} {\Gamma_+}  \nonumber\\ = \; \widetilde Q_3 (d, n)
= universal \;. \qquad \qquad
\end{eqnarray}
$Q_1 \; , Q_2$ and $\widetilde Q_3 = (D_\infty / \widehat
D_\infty) \; Q_3$ are the same universal numbers for both
isotropic and anisotropic systems within the same $(d, n)$
universality class.

Similar reformulations of universal amplitude relations are
necessary for  $R_{\sigma \xi} $ and $R^T_\xi$ involving the
surface tension, Eq. (2.58) of Ref. \cite{priv}, and the stiffness
constant (superfluid density) $\rho_s = \xi_T^{2-d}$, Eqs. (2.17)
and (3.54) of Ref. \cite{priv}, respectively. Corresponding
nonuniversal anisotropy effects must be taken into account in the
formulation of universal relations involving correction-to-scaling
amplitudes (Wegner \cite{wegner1972} amplitudes) as well as of
universal {\it dynamic} bulk amplitude combinations
\cite{hohenberg} such as $R_\lambda, R_2, \tilde R^-_m$, and $
R_\Gamma$ defined in Ref. \cite{priv}.

For completeness, we briefly mention also those universal bulk
amplitude relations that do not involve the correlation length, as
listed in Eqs. (2.45) - (2.48), (2.51), and (2.52) of Ref.
\cite{priv}. It is straightforward to show that, as a consequence
of the scaling structure of $f'_{s,b}$, Eq. (\ref{3a}), and of the
universality of the scaling function $W_\pm (z)$ that these
relations remain valid also for anisotropic systems, i.e., they
are independent of the anisotropy parameters $A_{\alpha
\beta}$. Consider, for example, the asymptotic amplitudes $A'_\pm$
and $\Gamma'_\pm$ of the bulk specific heat $C'_b = \partial^2
f'_{b,s} / \partial t^2 = (A'_\pm / \alpha) | t |^{- \alpha}$ and
of the bulk susceptibility $\chi'_b = - \partial^2 f'_{s,b} /
\partial h'^2 = \Gamma'_\pm | t |^{- \gamma}$ of the isotropic system above and below
$T_c$ at $h' = 0$, respectively, and, correspondingly,  $C_b =
\partial^2 f_{b,s} / \partial t^2 = (A_\pm / \alpha) | t |^{-
\alpha}$ , $\chi_b = - \partial^2 f_{s,b} /
\partial h^2 = \Gamma_\pm | t |^{- \gamma}$ of the
anisotropic system. (For $\chi'_b$ and $\chi_b$ below $T_c$ we
consider, for simplicity, only $n = 1$.) Their amplitude ratios
are given by $A'_+ / A'_- = A_+ / A_- = W_+ (0) / W_- (0)$ and by
\begin{equation}
\label{3bb}\frac{\Gamma'_+} {\Gamma'_-} \; =  \frac{\Gamma_+}
{\Gamma_-} \; = \; \left.\frac{\partial^2 W_+ (y) / \partial y^2}
{\partial^2 W_- (y) / \partial y^2}\right|_{y = 0}.
\end{equation}
Thus the nonuniversal parameters $A_{\alpha \beta}$  drop out
completely. Corresponding statements hold for the amplitude
combinations denoted by $R_\chi, R_C, R_A$ in Ref. \cite{priv}.

A Monte Carlo (MC) study \cite{schulte} of the {\it anisotropic}
three-dimensional Ising model appeared to be at variance with the
universality of the bulk susceptibility ratio (\ref{3bb}).
Subsequent MC simulations \cite{stauffer} of the same anisotropic
model on larger lattices, however, are consistent with the
universality of (\ref{3bb}).

The analysis of anisotropy effects near criticality can of course
be extended also to the scaling form of other bulk correlation
functions such as $< \varphi' ({\bf x}'_i)^2 \varphi' ({\bf
x}'_j)^2>$. It can also be extended to the case of general $n$
below $T_c$ where one must distinguish between longitudinal and
transverse correlations. Furthermore, extensions of this analysis
should be applied also to critical dynamics \cite{hohenberg} and
to boundary critical phenomena \cite{diehl86,dietrich88}.

In conclusion, all critical exponents and bulk scaling functions
$W_\pm(z) \; , \Phi_\pm(x',y')$ with $|x'|\lesssim O(1)$, and
$X_\pm(y')$ of anisotropic systems are universal, i.e., they are
the same as those of isotropic systems in the same universality
class. However, as far as the bulk correlation function $G_b ({\bf
x}; t, h)$ is concerned, the knowledge only of the scaling
{\it function} $\Phi_\pm(x',y')$ would be empty unless one knows how the
{\it arguments} $x',y'$ of $\Phi_\pm$  are related to observable
properties. In particular right at criticality, the spatial
dependence of $G_b ({\bf x}; 0,0)$ is not at all contained in the
scaling function but only in the factor $| {\bf \bar{\mbox
{\boldmath$\lambda$}}}^{-1/2} {\bf U}{\bf x}|^{- d + 2 - \eta}$
[see (\ref{3n4})]. This requires the knowledge of up to $d (d + 1)
/ 2 + 1$ nonuniversal parameters. As far as the universal bulk
amplitude relations are concerned, two-scale factor universality
(with only {\it two} independent nonuniversal amplitudes) is valid
only for a subset of such relations, namely for those that do not
involve the correlation length [such as (\ref{3g}), (\ref{3h}),
(\ref{3n2}), (\ref{3n3}), (\ref{3bb}) and those of Ref.
\cite{priv} mentioned above]. The other relations [such as
(\ref{3u}), (\ref{3y}), (\ref{3z})] provide universal relations
between quantities depending on up to $d (d + 1) / 2 + 1$
independent nonuniversal parameters, thus seven parameters in
three dimensions. This is the property of {\it multi-parameter
universality} referred to in Table I.

Furthermore, for anisotropic systems there exist nonuniversal
anisotropy effects of the large - distance regime of $G_b ({\bf
x}; t, h)$ (corresponding to the shaded region in Fig. 2)
at $t \neq 0, h=0$ and  $h\neq 0, t=0$ in combination
with the nonuniversal nonscaling features of the isotropic cases
(i) -(iv) mentioned at the end of subsection A above.

\setcounter{equation}{0}
\section{Perturbation approach in the central
finite-size regime}

\subsection{General remarks}

Consider the transformed Hamiltonian $H'$  with a
one-component order parameter at $h' = 0$ in a finite geometry
with a characteristic length $L'$ in the presence of periodic
boundary conditions. It is expected that, for short-range
interactions, there exist three different types of finite-size
critical behavior of $f'_s (t, L') - f'_{s,b} (t)$ where $f'_{s,b}
(t)$ is the singular bulk part : (a) the exponential $L'$
dependence $\sim \exp (- L' / \xi'_{{\bf e}+})$ for large
$L'/\xi'_{{\bf e}+} \gg 1$ at fixed temperature $T > T_c$ with
$\xi'_{{\bf e}+}$ being the exponential bulk correlation length
above $T_c$, (b) the power-law behavior $\sim L'^{-d}$ for large
$L'$ at fixed $L'/\xi'_\pm \; , 0 \leq L'/\xi'_\pm \leq O (1)$, above, at
and below $T_c$ where $\xi'_\pm$ is the second-moment bulk correlation
length (\ref{3b}), (c) the exponential $L'$ dependence $\sim \exp (- L' /
\xi'_{{\bf e}-})$ for large $L'/\xi'_{{\bf e}-} \gg 1$ at fixed
temperature $T < T_c$ with $\xi'_{{\bf e}-}$ being the exponential
bulk correlation length below $T_c$. For a description of the
cases (a) and (c), ordinary perturbation theory with respect to
$u'_0$ of the isotropic $\varphi^4$ theory is sufficient. For the
case (b), a separation of the lowest mode and a perturbation
treatment of the higher modes is necessary \cite{BZ, RGJ, EDHW,
EDC}.

For anisotropic systems, the distinction between the regimes (a),
(b) and (c) remains relevant except that there exist no single
correlation lengths $\xi_{{\bf e}+}, \xi_+$,  $\xi_-$, and
$\xi_{{\bf e}-}$. In this and the subsequent sections we treat the
case (b) on the basis of the lattice Hamiltonian (\ref{2a}) for $n
= 1, h = 0$ and defer the cases (a) and (c) to Section X. The case
(b) corresponds to the central finite-size region above the dashed
lines in Fig. 1. For simplicity we assume a cubic shape with
volume $V = L^d$ and a simple-cubic lattice with lattice constant
$\tilde a$. Now the summations $\sum_{\bf k}$ run over $N$
discrete vectors ${\bf k} \equiv (k_1, k_2, \ldots, k_d)$ with
Cartesian components $k_\alpha = 2 \pi m_\alpha / L, m_\alpha = 0,
\pm 1, \pm 2, \cdots, \alpha = 1,2, \cdots, d$ in the range $ -
\pi / \tilde a \leq k_\alpha < \pi / \tilde a$.

The goal is to derive the finite-size scaling form of the singular
finite-size part $f_s $ of the free energy density of the
anisotropic system (\ref{2a}) for $n=1$ at $h=0$ with an
anisotropy matrix ${\bf A}$. We shall show that, for small $|t|$
and large $L$ in the regime (b), $f_s $ has the scaling form
\begin{equation}
\label{4a1} f_s (t, L; {\bf A}) \; = \; L^{-d} \;{\cal F} (t (L' /
\xi_{0+}')^{1/\nu}; {\bf \bar A})
\end{equation}
where the scaling argument is expressed in terms of the
transformed length $L' \; = \; (\det {\bf A})^{- 1 / (2d)} L \;$,
rather than in terms of $L$, and where $\xi_{0+}'$ is the
asymptotic amplitude of the bulk correlation length  defined in
(\ref{3b}) on the basis of the transformed Hamiltonian $H'$.
Because of  (\ref{2vv}), ${\cal F}$ is identical with the
finite-size scaling function of the free energy density of the
transformed system
\begin{equation}
\label{4a2} f'_s (t, L'; {\bf \bar A}) \; = \; L'^{-d} \;{\cal F}
(t (L' / \xi_{0+}')^{1/\nu}; {\bf \bar A}).
\end{equation}
The advantage of the transformed system is that its bulk
renormalizations (see Sect. V) are well known from the standard
isotropic $\varphi^4$ field theory. Thus, in order to derive the
scaling function ${\cal F}$, it is most appropriate to develop
perturbation theory first within the transformed system with the
Hamiltonian $H'$, (\ref{2p}) and (\ref{2q}), with $v' = (\det {\bf
A})^{- 1/2} \tilde a^d$.

\subsection{Perturbation approach}

It is necessary to reformulate in detail the field-theoretic
perturbation approach of \cite{EDC} in the context of our
anisotropic lattice model in order to correctly identify the total
finite-size part  of the free energy density $f'_s$ including all
temperature independent contributions $\propto L'^{-d}$ and to
identify the new parts of the theory that are affected by the
anisotropy. The decomposition into the lowest mode and higher
modes reads $\varphi'_j = \Phi' + \sigma'_j $,
\begin{eqnarray}
\label{4b}\Phi' = L'^{-d}
\hat \varphi'({\bf 0}) = N^{-1} \sum_j \varphi'_j,\\ \sigma'_j =
L'^{-d} {\sum_{\bf k'\neq0}} e^{i{\bf k'} \cdot {\bf x'}_j} \hat
\varphi'({\bf k'}),
\end{eqnarray}
where $L'^d = (\det {\bf A})^{- 1/2} L^d$.
Correspondingly, the lattice Hamiltonian $H'$ and the partition
function $Z'$, (\ref{2t}), are decomposed as $ H' = H'_0 +
\widetilde H' (\Phi', \sigma')$ ,
\be
\label{4e} H'_0 (r_0 \; , u'_0 \;, L' \; , \Phi'^2) = L'^d
\left(\frac{1}{2} r_0 \Phi'^2 + u'_0 \Phi'^4 \right) \,,
\ee
\begin{eqnarray}
\label{4f} \widetilde H'(\Phi',\sigma') &=& v' \Bigg\{\sum_{j=1}^N
\left[( \frac{r_0}{2}+6u'_0\Phi'^2) \sigma'^2_j +
4u'_0\Phi'\sigma_j'^3 + u'_0 \sigma'^4_j \right] \nonumber\\ &+&
\sum_{i,j=1}^N \frac{K_{i,j}}{2} (\sigma'_i - \sigma'_j)^2
\Bigg\},\;
\end{eqnarray}
\begin{eqnarray}
\label{4g} Z' = \frac{L'^{d/2}}{(v')^{1/d}} \int\limits^\infty_{-
\infty} d \Phi' \exp \left\{- \left[H'_0 +
{\bare{\Gamma'}}(\Phi'^2)\right]\right\} \;,
\end{eqnarray}
\begin{eqnarray}
\label{4h} {\bare{\Gamma'}} (\Phi'^2) &=& -\; \ln
\left[\prod_{{\bf k'\neq0}}\frac{1}{(v')^{1/d} L'^{d/2}} \int \; d
\hat \sigma'({\bf k'}) \right] \nonumber\\ &\times& \exp \left[ -
\widetilde H' (\Phi', \sigma') \right]
\end{eqnarray}
where $\hat \sigma'({\bf k'}) \equiv \hat \varphi'({\bf k'})$ for
${\bf k'} \neq{\bf 0}$ and $\widetilde H'$ is expressed in terms
of $\hat \sigma' ({\bf k'})$. No terms $ \sim \Phi' \;\sigma'_j$
and $\sim \Phi'^3 \sigma'_j$ appear in (\ref{4f}) because of
$\sum_j \sigma'_j = 0$. The integration measure $\int d \;\hat
\sigma' ({\bf k'})$ is defined in Eq. (\ref{bb6}) of Appendix B.
[The corresponding (but different) functional integration $\int
D\sigma$ of the (cut-off dependent) continuum model  was not
explicitly defined in Eq. (2.11) of \cite{EDC}.] The quantity $
{\bare{\Gamma'}} (\Phi'^2)$ can be interpreted as a constraint
free energy, with the constraint being that the zero-mode
amplitude $\Phi'$ is fixed. The quantity $\exp [- {\bare{\Gamma'}}
(\Phi'^2)]$ is proportional to the order-parameter distribution
function of isotropic systems \cite{CDS} which is a physical
quantity in its own right. Therefore, in contrast to the
$\varepsilon$ expansion approach of \cite{RGJ} and \cite{BZ}, we
shall not expand the exponential form $\exp [- {\bare{\Gamma'}}
(\Phi'^2)]$ but only ${\bare{\Gamma'}} (\Phi'^2)$. The advantage
of our approach has been demonstrated for the specific heat below
$T_c$ in Refs. \cite{EDC,CDT}.

Following \cite{EDHW} and \cite{EDC} we decompose $\widetilde H'
(\Phi', \sigma')=  H'_1 + H'_2$ into an unperturbed Gaussian part
\be
\label{4j} H'_1=  v' \left[\sum_{j=1}^N \frac{r'_{0L}}{2}
\sigma'^2_j + \sum_{i,j=1}^N \frac{K_{i,j}}{2} (\sigma'_i -
\sigma'_j)^2 \right]
\ee
and a perturbation part
\be
\label{4k} H'_2 = v' \left\{\sum_{j=1}^N
\left[6u'_0(\Phi'^2-M'^2_0) \sigma'^2_j + 4u'_0\Phi'\sigma_j'^3 +
u'_0 \sigma_j'^4 \right] \right\}.
\ee
The crucial point is to incorporate the lowest-mode average
\be
\label{4l} M'^2_0 (r_0, u'_0, L') = \frac{
\int\limits_{\infty}^\infty d \Phi' \;\Phi'^2 \exp (-
H'_0)}{\int\limits_{\infty}^\infty d \Phi' \;\exp (- H'_0)}
\ee
into the parameter
\be
\label{4m} r'_{0{\rm L}} (r_0, u'_0, L') = r_0 + 12 u'_0 M_0'^2
\ee
of the unperturbed part $H'_1$ and to treat the term
$6u'_0(\Phi'^2-M'^2_0)\sigma_j'^2$ of $H'_2$ as a perturbation.
The treatment of the Gaussian fluctuations $\sigma'^2_j$ as a
perturbation is similar in spirit to an earlier perturbation
approach for Dirichlet boundary conditions \cite{dohm1989} where
part of the Gaussian fluctuations of the higher modes were
included in the perturbation part of the Hamiltonian. The
positivity of $r'_{0{\rm L}}
> 0$ for all $r_0$ permits us to extend the theory to the region
below $T_c$. For finite $L'$, $M'^2_0$ and $r'_{0{\rm L}}$
interpolate smoothly between the mean-field bulk limits above and
below $T_c$
\be
\label{4n} \lim_{L' \rightarrow \infty} M'^2_0 \equiv
 M_{mf}'^2 = \left\{ \begin{array}{r@{\quad\quad}l}
                 0 \hspace{1.0cm} & \mbox{for} \;\;\; r_0\geq0\;, \\ -r_0/(4u'_0) & \mbox{for} \;\;\;
                 r_0\leq0\;,
                \end{array} \right.
\ee
\be
\label{4o} \lim_{L' \rightarrow \infty} r'_{0{\rm L}} \equiv
r_{mf} = \left\{
\begin{array}{r@{\quad\quad}l}
                 r_0 & \mbox{for} \;\;\; r_0\geq0\;, \\  -2r_0 & \mbox{for} \;\;\;
                 r_0\leq0\;.
                \end{array} \right.
\ee
The contribution of $H'_1$ to $L'^{-d}
\;{\bare{\Gamma'}}(\Phi'^2)$ is (compare Eq. (\ref{bb11}) in
Appendix B)
\begin{eqnarray}
\label{4p} - \frac{1}{L'^d} \ln \left[\prod_{\bf k'\neq0}\;  \int
\frac{d \hat \varphi'({\bf k'})}{(v')^{1/d} L'^{d/2}}
\right] \exp (- H'_1) = - \frac{N-1}{2 L'^d} \ln (2 \pi) \nonumber\\
+ \frac{1}{2L'^d} {\sum_{\bf k'\neq0}} \ln \{[r'_{0{\rm L}} +
\delta \widehat K' (\mathbf k')] (v')^{2/d}\}\;. \qquad \qquad
\end{eqnarray}
The leading contributions of the perturbation term $6 u'_0
(\Phi'^2 - M_0'^2) \sigma_j'^2$ of $H'_2$ to $L'^{- d}
{\bare{\Gamma'}} (\Phi'^2)$ read
\begin{eqnarray}
\label{4q} 6 u'_0 (\Phi'^2 - M_0'^2) S_1 (r'_{0{\rm L}}) - 36
u_0'^2 (\Phi'^2 - M_0'^2)^2 S_2 (r'_{0{\rm L}}) \nonumber\\ + \; O
(u_0'^3 (\Phi'^2 - M_0'^2)^3) \qquad \qquad
\end{eqnarray}
where
\be
\label{4r} S_m (r'_{0{\rm L}}) = L'^{-d} {\sum_{\bf k'\neq0}}
\left\{\left[r'_{0{\rm L}} + \delta \widehat K' ({\bf k'})\right]
\right\}^{-m}.
\ee
The terms $\sim u'_0 \Phi' \sigma_j'^3$ and $u'_0 \sigma_j'^4$ of
$H'_2$ yield higher-order contributions of $O (u_0'^2 \Phi'^2 \;,
u'_0)$ which will be neglected in the following. We emphasize,
however, that leading finite-size effects caused by the four-point
coupling $u'_0$ are taken into account in Eq.(\ref{4q}) as it
contains the coupling between the fluctuations $\Phi'^2 - M_0'^2$
of the lowest mode and those of the higher modes $\hat \sigma'
({\bf k'})$. For a discussion of the order of the neglected terms
see also Refs. \cite{dohm1989,EDC}.

The starting point for our perturbation expression of the bare
free energy density (\ref{2s}) is
\begin{eqnarray}
\label{4s} f' &=&   L'^{-d} \bare\Gamma' (0) - L'^{-d} \ln
\left\{\frac{L'^{d/2}}{(v')^{1/d}} \; \int\limits_{-
\infty}^\infty d \Phi' \exp \left[- H'^{eff} \right] \right\}
\nonumber\\ &=& - \; \frac{N-1}{2 L'^d} \ln(2 \pi) \nonumber\\ &+&
\frac{1}{2L'^d} {\sum_{\bf k'\neq0}} \ln \left\{\left[r'_{0{\rm
L}}
+ \delta \widehat K' (\mathbf k')\right] (v')^{2/d}\right\} \nonumber \\
&-& L'^{-d}\ln \left\{\frac{L'^{d/2}}{(v')^{1/d}} \int
\limits^\infty_{-
\infty} d \Phi' \; \exp \left[- H'^{eff} \right]\right\} \nonumber\\
&-& 6 u'_0 M_0'^2 S_1 (r'_{0{\rm L}}) - 36 u_0'^2 M_0'^4 S_2
(r'_{0{\rm L}})\;,
\end{eqnarray}
\be
\label{4t} H'^{eff} = L'^d \left(\frac{1}{2} r'^{eff}_0 \Phi'^2 +
u_0'^{eff} \Phi'^4 \right),
\ee
\be
\label{4u} r_0'^{eff} = r_0 + 12 u'_0 S_1 (r'_{0{\rm L}}) + 144
u_0'^2 M_0'^2 S_2 (r'_{0{\rm L}}),
\ee
\be
\label{4v} u_0'^{eff}  = u'_0 - 36 u_0'^2 S_2 (r'_{0{\rm L}}).
\ee
Apart from the different form of the lattice interaction $\delta
\widehat K' ({\bf k}')$ and the different vectors ${\bf k}'$,  Eq.
(\ref{4s}) differs from the previous Eqs. (4.3), (4.11) and (4.12)
of the isotropic field theory of Ref. \cite{EDC} in two respects:
(i) In (\ref{4s}) there are additive logarithmic finite-size terms
proportional to $L'^{-d}\ln [(v')^{1/d}]$; in the regime (b)
mentioned above, they will cancel each other, and a dependence on
$\ln [(v')^{1/d}]$ will remain only in the bulk part [see
(\ref{c1}) in App. C and (\ref{4z5})-(\ref{4z7}) ]. (ii) In
(\ref{4s}) there are the additive logarithmic  finite-size terms
\begin{eqnarray}
\label{4v-1} - \; \frac{N-1}{2 L'^d} \ln(2 \pi) - \frac{1}{ L'^d}
\ln L'^{d/2} \nonumber\\ = - \frac{1}{2 v'} \ln (2\pi)
  + \frac{1}{2 L'^d} \ln \frac{2\pi}{L'^d}
\end{eqnarray}
where $v'=L'^d/N$. These terms are independent of $t$ and $h$,
therefore such terms do not affect the physical quantities
considered in Ref. \cite{EDC} which are {\it derivatives} of the
free energy with respect to  $t$ and $h$. These terms, however,
must not be omitted in the calculation of the free energy itself.
While the first term on the r.h.s. of (\ref{4v-1}) is an
unimportant nonsingular {\it bulk} part, the second term yields a
nonnegligible contribution to the universal value of the
finite-size scaling function ${\cal F}^{ex}$ at $T_c$ which is a
measurable quantity. (The second term affects the argument of the
first logarithmic term of the scaling function at $T_c$ given in
(\ref{6-1}) below.) Omission of this term would cause a
misidentification of the finite-size scaling function of the
excess free energy density. This would yield an incorrect result
in a comparison with Monte Carlo data \cite{mon-1,mon-2,mon-3}
that measure the {\it total} amplitude of the $L^{-d}$ term of the
excess free energy of two- and three-dimensional spin models.

Our approach incorporates, in an approximate form, the effect of
the finite-size fluctuations $\Phi'^2 - M_0'^2$ of the lowest mode
amplitude around its average $M_0'^2$ that are present in the
central finite-size critical region. This is not taken into
account in the effective Hamiltonian of \cite{BZ} which contains
fluctuations of $\Phi'^2$ around zero. Setting $M_0'^2 = 0$ and
$r'_{0{\rm L}} = r_0$ in Eqs. (\ref{4s}) - (\ref{4v}) would yield
the bare free energy density corresponding to perturbation theory
based on the effective Hamiltonian of \cite{BZ}. This would
restrict the theory to the regime $r_0 \geq 0$. A foundation of
Eqs. (\ref{4s}) - (\ref{4v}) can also be given on the basis of the
order-parameter distribution function \cite{CDS}.

\subsection{Improved perturbation expression}

In its present form the saddle point contribution of the
lowest-mode integral in (\ref{4s}) for large $L'$ below $T_c$ is
\begin{eqnarray}
\label{4w} \lim_{L' \to \infty} &-& L'^{-d} \ln
\left\{\frac{L'^{d/2}}{(v')^{1/d}} \int \limits^\infty_{- \infty}
d \Phi' \; \exp \left[- H'^{eff} \right]\right\}\nonumber\\ &=& \;
- \; \frac{(r_0'^{eff})^2}{16 u_0'^{eff}}
\end{eqnarray}
which, after expansion of $(u_0'^{eff})^{-1}$ with respect to
$u'_0$, would produce arbitrary large powers of $u'_0$. On the
other hand it is clear at the outset that, because of neglecting
the terms $\sim u'_0 \Phi'^2 \sigma_j'^3$ and $\sim u'_0
\sigma_j'^4$ of $H'_2$, the neglected terms in (\ref{4s}) are bulk
terms of $O (u'_0)$ corresponding to two-loop terms. Therefore it
is necessary to further improve the perturbation expression
(\ref{4s}). Here our reformulation of the $\ln \int d \Phi' e^{-
H'^{eff}}$ term will be guided by the requirement that
higher-order powers of $u_0$ are neglected already at the level of
$H'^{eff}$, {\it before} integrating over $\Phi'$. For this
purpose we rewrite the logarithm of the integral over the lowest
mode as
\begin{eqnarray}
\label{4x} \ln \left\{\frac{L'^{d/2}}{(v')^{1/d}} \int
\limits^\infty_{-
\infty} d \Phi' \; \exp \left[- H'^{eff} \right]\right\} \nonumber\\
=  \ln \int\limits_{-\infty}^\infty d s \exp \left[-
\frac{r_0'^{eff} L'^{d/2}}{2 (u_0'^{eff})^{1/2}} s^2 - s^4\right]
\nonumber\\  + \frac{1}{2} \; \ln \; \left[\frac{L'^{d/2}}
{(v')^{2/d} (u_0'^{eff})^{1/2}}\right] \; .
\end{eqnarray}
For the reason given above it is appropriate to expand the factors
$(u_0'^{eff})^{- 1/2}$ in both terms of Eq. (\ref{4x}) in powers
of $u_0$ and to neglect terms of $O (u_0'^{3/2})$ corresponding to
a truncation of the expansion
\be
\label{4y} (u_0'^{eff})^{- 1/2}\; = \; u_0'^{- 1/2} \; + \; 18
u_0'^{1/2} S_2 (r'_{0{\rm L}}) \; + \; O (u_0'^{3/2}) \; .
\ee
  In summary our improved perturbation
expression for the bare free energy density reads
\begin{eqnarray}
\label{4z} f' \;=\; &-& \frac{N-1}{2 L'^d} \ln(2 \pi) \nonumber\\
&+& \frac{1}{2L'^d} {\sum_{\bf k'\neq0}} \ln
\left\{\left[r'_{0{\rm
L}} + \delta \widehat K' (\mathbf k')\right] (v')^{1/2 d}\right\} \nonumber \\
&-& \frac{1}{L'^d} \ln \int\limits_{-\infty}^\infty d s \;\exp (-
\frac{1}{2} y_0'^{eff} s^2 - s^4)   \nonumber\\
&-& \frac{1}{2 L'^d} \ln\; \left[\frac{L'^{d/2} w_0'^{eff}}
{v'^{2/d}}\right] - 6 u'_0 M_0'^2 S_1 (r'_{0{\rm L}}) \nonumber\\
&-& 36 u_0'^2 M_0'^4 S_2 (r'_{0{\rm L}})
\end{eqnarray}
with
\begin{eqnarray}
\label{4z1} y_0'^{eff} & = & L'^{d/2} u_0'^{- 1/2}
\Big\{r_0[1+18u'_0S_2(r'_{0{\rm L}})]+ 12
u'_0S_1(r'_{0{\rm L}}) \nonumber\\
& + & 144 u_0'^2M_0'^2S_2(r'_{0{\rm L}}) \Big\},
\end{eqnarray}
\be
\label{4z2} w_0'^{eff} \; = \; u_0'^{- 1/2}
\left[1+18u'_0S_2(r'_{0{\rm L}}) \right] \; .
\ee
Now, because of $u'_0 M'^2_0 \sim O (u'^{1/2}_0)$ at $T_c$,
$y_0'^{eff}$ and $w_0'^{eff}$ and  the last two terms in (\ref{4z})
contain terms only up to $O(u_0'^{1/2})$ at $T_c$. One can verify
that in the bulk limit below $T_c$ the last two terms $- 6 u'_0
M'^2_0 S_1$ and $- 36 u_0'^2 M'^4_0 S_2$ of (\ref{4z}) which are of $O (1)$ are exactly
cancelled by the $O (1)$ terms of the saddle-point contribution $-(y_0'^{eff})^2 / (16
L'^d)$ of  the integral term of (\ref{4z}). Thus
Eq.(\ref{4z}) correctly contains the bare bulk free energy density
$f_b'^\pm \equiv \lim_{L' \rightarrow \infty}f'$ in one-loop order
[i.e., up to $O (1)$]
\be
\label{4z3} f_b'^+ \; = \; - \; \frac{\ln (2 \pi)} {2 v'} \; + \;
\frac{1}{2} \int\limits_{\bf k'} \ln \{[r_0 + \delta \widehat K'
(\mathbf k')] (v')^{2/d} \} + O (u'_0),
\ee
\begin{eqnarray}
\label{4z4}f_b'^- = \frac{1}{2} r_0 M_{mf}'^2 + u'_0 M_{mf}'^4 -
\; \frac{\ln (2 \pi)} {2 v'} \nonumber\\  + \; \frac{1}{2}
\int\limits_{\bf k'} \ln \{[- 2 r_0 + \delta \widehat K' (\mathbf
k')] (v')^{2/d} \} + O(u'_0)\;\;\;\;
\end{eqnarray}
above and below $T_c$, respectively, where
\begin{eqnarray}
\label{4z5a} \int\limits_{\bf k'} \; = \; (\det {\bf A})^{1/2}
\int\limits_{\bf k} \; \equiv \; (\det {\bf A})^{1/2}
\prod^d_{\alpha = 1} \; \int\limits^{\pi / \tilde a}_{- \pi /
\tilde a} \; \frac{d k_\alpha}{2\pi} \; . \nonumber\\
\end{eqnarray}
We shall rewrite (\ref{4z}) in terms of $r_0 - r_{0c}$ where
\begin{eqnarray}
\label{4zz} r_{0c} = - \; 12 u'_0 \int\limits_{\bf k'}
\frac{1}{\delta \widehat K' ({\bf k'})}  = - \; 12 u_0
\int\limits_{\bf k} \frac{1}{\delta \widehat K ({\bf k})}
\end{eqnarray}
is the critical value of $r_0$ up to $O(u_0)$.  On the level of
bare perturbation theory, the  application of Eq. (\ref{4z}) will
be in the central finite-size regime $|r_0 - r_{0c} | \lesssim
O(u_0'^{1/2} \; L'^{- d/2})$. On the level of the asymptotic
renormalized theory this will correspond to the finite-size regime
$0 \leq | t (L' / \xi_0')^{1/\nu} | \lesssim O (1)$ above, at, and
below $T_c$, i.e., the regime (b) mentioned above. If applied to
the regime $L' / \xi'_- \gg 1$ below $T_c$, $f'$ also contains
bulk and finite-size terms of $O (u'_0)$ which would need to be
complemented by two-loop calculations.

The right-hand side of Eq. (\ref{4z}) can be decomposed as
\begin{eqnarray}
\label{4z5}f' (r_0 &-& r_{0c}, u'_0, L', K_{i,j}, v') =
f'^{(1)}_{ns,b} (r_0 - r_{0c}, u'_0, K_{i,j}, v') \nonumber\\
&+& \delta f' (r_0 - r_{0c}, u'_0, L',K_{i,j}, v' ), \qquad
\end{eqnarray}
where $f'^{(1)}_{ns,b}$ is a non-singular {\it bulk} part up to
linear order in $r_0 - r_{0c}$,
\begin{eqnarray}
\label{4z6}&&f'^{(1)}_{ns,b} (r_0 - r_{0c}, u'_0, K_{i,j}, v') = -
\frac{\ln (2 \pi)} {2 v'} \nonumber\\ &&+ \; \frac{1}{2}
\int\limits_{\bf k'} \ln \{[ \delta \widehat K' ({\bf k'})]
(v')^{2/d} \} + \frac{ r_0 - r_{0c}}{2} \int\limits_{\bf
k'}[\delta \widehat K' ({\bf k'})]^{-1} \;. \nonumber\\
\end{eqnarray}
As expected from bulk theory \cite{str}, the remaining finite-size
part $\delta f'$ has a finite limit for $v' \to 0$ at fixed $r_0 -
r_{0c}$ in $2 < d < 4$ dimensions. It turns out that the resulting
function depends only on ${\bf \bar A}$ rather than on ${\bf A}$,
\be
\label{4z7}\lim_{v' \to 0} \; \delta f' (r_0 - r_{0c}, u'_0, L',
K_{i,j}, v') = \delta f' (r_0 - r_{0c}, u'_0, L', {\bf \bar A})\;.
\ee
The r.h.s. of (\ref{4z7}) can be further decomposed as
\begin{eqnarray}
\label{4z8}\delta f'(r_0 - r_{0c}, u'_0, L', {\bf \bar A}) =
f'^{(2)}_{ns,b} (r_0 - r_{0c}, u'_0) \nonumber\\+ \; f'_s(r_0 -
r_{0c}, u'_0, L', {\bf \bar A}) \; , \quad
\end{eqnarray}
where $f'^{(2)}_{ns,b}$ is a non-singular {\it bulk} part
proportional to $(r_0 - r_{0c})^2$ \cite{str}.  We are interested
in the asymptotic singular finite-size part $f'_s$. In the limit
$v' \to 0$ our result for $\delta f'$ does not contain an $L$
{\it dependent non-singular} part. The limit $v' \to 0$ is
justified in the power-law regime (b) mentioned above where the
$v'$ dependent terms of our perturbation expression (\ref{4z})
give rise only to corrections to scaling. However, although the
limit (\ref{4z7}) does exist in the exponential regimes (a) and
(c), it is not justified to neglect the $v'$ dependencies in the
exponential arguments, as will be discussed in Section X.

\subsection{Bare perturbation result}

The calculation of $\delta f'$ is outlined in App. B and C for the
power-law regime $|r_0 - r_{0c}| \lesssim O (u'_0 L'^{- d/2}) \; ,
L' \gg (v')^{1/d} \; , |r_0 - r_{0c}|^{1/2} (v')^{1/d}\ll 1$. The
result reads for $2<d<4$
\begin{eqnarray}
\label{4aa}&&\delta f'(r_0 - r_{0c}, u'_0, L', {\bf \bar A}) =
\nonumber\\&& \; - A_d \; (r'_{0{\rm L}})^{- \varepsilon/2}
\left[\frac {(r'_{0{\rm L}})^2}{4d}  \; + \; \frac{(r_0 -
r_{0c})^2} {4 \varepsilon} \; - \; 18 u_0'^2 M_0'^4\right]
\nonumber\\  && \;+ \; \frac{1}{L'^d}\Bigg\{-\;  \ln
\int\limits_{-\infty}^\infty d s \;\exp (- \frac{1}{2}
y_0'^{eff}({\bf \bar A}) s^2 - s^4) \nonumber\\ &&
\phantom{\frac{1}{L'^d}\Bigg\{-\;}\;- \frac{1}{2}\ln \left[\frac{2
\pi w_0'^{eff} ({\bf \bar A})}{(L')^{\varepsilon/2} \;
 } \; \right]
 + \frac{1}{2}J_0(r'_{0{\rm L}}L'^2, {\bf \bar A}) \nonumber\\ && - \frac{3u'_0
M_0'^2 L'^2}{2\pi^2 }I_1(r'_{0{\rm L}}L'^2, {\bf \bar A}) -
\frac{9u_0'^2 M_0'^4L'^4}{4\pi^4 }I_2(r'_{0{\rm L}}L'^2, {\bf \bar
A} )\Bigg\}\nonumber\\
\end{eqnarray}
with
\begin{eqnarray}
\label{4bb}&&y_0'^{eff}({\bf \bar A}) =
\left(\frac{V'}{u_0'}\right)^{1/2} \Bigg\{(r_0 -
r_{0c})\nonumber\\ && \Big[1 + 18 u_0' \Big(\frac{A_d(d-2)}{2
\varepsilon} (r'_{0{\rm L}})^{- \varepsilon/2} +
\frac{L'^\varepsilon}{16\pi^4}I_2(r'_{0{\rm L}} L'^2, {\bf \bar
A})\Big)\Big] \nonumber\\ &&+ \;12u'_0\Big(-\; \frac{A_d}{
\varepsilon}(r'_{0{\rm L}})^{(d-2)/2} +
\frac{(L')^{2-d}}{4\pi^2}I_1 (r'_{0{\rm L}} L'^2, {\bf \bar A} )\Big)\nonumber\\
&&+ \; 144 u_0'^2 M_0'^2\Big(\; \frac{A_d(d-2)}{2
\varepsilon}(r'_{0{\rm L}})^{- \varepsilon/2} +
\frac{L'^\varepsilon}{16\pi^4}I_2(r'_{0{\rm L}} L'^2, {\bf \bar A}
)\Big)\Bigg\} \;,\nonumber\\
\end{eqnarray}
\begin{eqnarray}
\label{4cc}w_0'^{eff} ({\bf \bar A})  &=&  u_0'^{- 1/2} \Big[1 +
18 u_0' \Big(\; \frac{A_d(d-2)}{2 \varepsilon} (r'_{0{\rm L}})^{-
\varepsilon/2} \nonumber\\ &+&
\frac{L'^\varepsilon}{16\pi^4}I_2(r'_{0{\rm L}} L'^2, {\bf \bar A}
)\Big)\Big] \; ,
\end{eqnarray}
\be
\label{4dd} r'_{0{\rm L}}(r_0 - r_{0c}, u_0',L')=r_0 - r_{0c} + 12
u_0' M_0'^2,
\ee
\be
\label{4ee} M_0'^2 =  (V'u_0')^{- 1/2} \; \vartheta_2 (y_0'),
\ee
\be
\label{4ff} y_0' = (r_0 - r_{0c}) (V'/u_0')^{1/2} \; ,
\ee
\be
\label{4f1} \vartheta_m(y_0') = \frac{\int\limits_0^\infty d s
\;s^m \exp (- \frac{1}{2} y_0' s^2 - s^4)} {\int\limits_0^\infty d
s \; \exp (- \frac{1}{2} y_0' s^2 - s^4)},
\ee
\begin{eqnarray}
\label{4gg} J_0(r'_{0{\rm L}} L'^2, {\bf \bar A})&=&
\int\limits_0^\infty \frac{dy}{y}
  \Bigg[\exp {\left(-\frac{r'_{0{\rm L}} L'^2y}{4\pi^2}\right)} \nonumber\\ &\times& \left\{
  (\pi / y)^{d/2}
  - K_d (y, {\bf \bar A}) + 1 \right\}  - \exp (-y)\Bigg]
  \;,\nonumber\\
\end{eqnarray}
\begin{eqnarray}
\label{4hh} I_m (r'_{0{\rm L}} L'^2, {\bf \bar A}) &=&
\int\limits_0^\infty
{\rm{d}}y \;y^{m-1} \exp[- r'_{0{\rm L}} L'^2 y / (4 \pi^2)] \nonumber\\
&\times& \{K_d
(y, {\bf \bar A}) - (\pi / y)^{d/2} - 1\} \;, \nonumber\\
\end{eqnarray}
\begin{eqnarray}
\label{4ii} &&K_d (y, {\bf \bar A}) = \sum_{\bf n} \;\exp (- y
{\bf n} \cdot {\bf \bar A n})
\end{eqnarray}
with ${\bf n} = (n_1, n_2, ..., n_d) \; , n_\alpha = 0, \pm 1,
..., \pm \infty$. The behavior of the functions $J_0 \; , I_1$,
and $I_2$ for small and large arguments $r'_{0{\rm L}} L'^2$ is
given in App. C.

The crucial information on the anisotropy is contained in the sum
(\ref{4ii}). By means of the Poisson identity \cite{morse} [see
also (\ref{b12})] one can show that this function satisfies
\be
\label{4iii} K_d (y; {\bf A})  \;  = (\det {\bf A})^{-1/2}
\left(\frac{\pi}{y}\right)^{ d/2} K_d \left(\frac{\pi^2}{y} , {\bf
A}^{-1} \right)\; .
\ee
The sum (\ref{4ii}) could formally be rewritten in ${\bf k'}$
space as $\sum_{\bf k'} \;\exp (- y' {\bf k'} \cdot {\bf k'})$
with $y' = y L'^2/(4\pi^2)$ but in practice there is no advantage
of using the more complicated ${\bf k'}$ vectors (see the example
(\ref{2qq}) in Sect. II). For this reason the sums in the
three-dimensional calculations in Sect. VIII will be performed in
${\bf k}$ space.

As expected, the bare perturbation result (\ref{4aa}) for $\delta
f'$ does not yet correctly describe the critical behavior: (i) In
the bulk limit at $t \neq 0$ the small-$t$ behavior is $\delta
f'_b \sim |t|^{d/2}$ rather than $ \sim |t|^{d \nu}$. (ii) At
$t=0$ the leading large-$L'$ behavior is $\delta f'  \sim
L'^{-d^2/4}$ rather than $\sim L'^{-d} $. These defects will be
removed by turning to the renormalized theory.

\setcounter{equation}{0}
\section{Minimal Renormalization at fixed dimension}

The bare perturbation form of $\delta f'$ requires additive and
multiplicative renormalizations, followed by a mapping of the
renormalized free energy $\delta f'_R$ from the critical to the
noncritical region where perturbation theory is applicable. It is
well known that, for the multiplicative renormalizations, the
usual bulk $Z$ factors are sufficient \cite{brezin}. For both
multiplicative and additive renormalizations, the absence of
$L$-dependent pole terms has been checked explicitly up to
$O({u'_0}^2)$ for the case of periodic boundary conditions
\cite{dohm1989,GJ}. In particular, there is no need for an
$L$-dependent shift of the temperature variable. We employ the
minimal subtraction scheme at fixed dimension $2<d<4$ without
using the $\varepsilon$ expansion \cite{dohm1985}. This approach
has already been successfully employed in previous finite-size
studies \cite{EDHW,EDC,CDS,CDT,cd-97,cdstau} and is applicable
above, at, and below $T_c$ with the same renormalization
constants. Thus it permits us to derive a {\it single} finite-size
scaling function of the free energy in the central finite-size
critical region above, at, and below $T_c$.

The multiplicatively renormalized quantities are
\be
\label{5a} u' = \mu^{- \varepsilon} A_d Z_{u'}^{-1} Z_{\varphi'}^2
u_0'
\ee
and $r = Z_r^{-1} (r_0 - r_{0c}) = a t$, $\varphi'_R =
Z_{\varphi'}^{- 1/2} \varphi'$ with an arbitrary inverse reference
length $\mu$. $L'$ is not renormalized. Furthermore, the reduced
anisotropy matrix ${\bf \bar A}$ is not renormalized either as it
does not change the ultraviolet behavior at $d = 4$. If our
calculation is extended to a finite external field $h'$,
(\ref{2oo}), the additional renormalization $h'_R =
Z_{\varphi'}^{1/2} h'$ is necessary \cite{cd-97}.

The geometric factor of bulk theory [see (\ref{3-a})]
\be
\label{5c1} A_d \;=\; \frac{\Gamma(3-d/2)}{2^{d-2} \pi^{d/2}
(d-2)} = \; S_d \; \Gamma (1 + \frac{\varepsilon}{2}) \Gamma (1 -
\frac{\varepsilon}{2})
\ee
appears naturally in Eqs. (\ref{4aa}) - (\ref{4cc}) rather than
the more commonly used factor $S_d =
2^{1-d}\pi^{-d/2}[\Gamma(d/2)]^{-1}$. The perturbation results of
amplitudes and scaling functions depend on the choice of the
geometric factor in (\ref{5a})
\cite{dohm1985,schl,str,kleinert,EDC} (see, e.g., the universal
ratio $Q_1$ in (\ref{6o}) below, see also the comment after
(\ref{5y}) below). The advantage of the factor (\ref{5c1}) is that
it describes the full $d$ dependence of single-loop integrals in
$2 < d < 4$ dimensions such as (\ref{3-a}), in contrast to the
factor $S_d$. For this reason we have incorporated $A_d$ in the
definition of the renormalized coupling $u'$, (\ref{5a}).  Any
other choice, such as $S_d$ instead of $A_d$, would introduce
artificial $d$ dependencies into the perturbation results. For the
same reason we employ $A_d$ in the definition of the
multiplicatively and additively renormalized free energy density
\begin{eqnarray}
\label{5d}   f'_R(r, u', L',\mu, {\bf \bar A}) &=& \delta f'(Z_rr,
       \mu^{\varepsilon}Z_{u'}Z_{\varphi'}^{-2}A_d^{-1}u',L',
       {\bf \bar A})\nonumber\\  &-& \frac{1}{8}\mu^{-\varepsilon} r^2 A_d
       A(u',\varepsilon)\; .
\end{eqnarray}
Because of relations such as (\ref{3-a}) the $Z$
factors $Z_r (u', \varepsilon)$, $Z_{u'} (u', \varepsilon)$, and
$Z_{\varphi'} (u', \varepsilon)$ depend on $u'$ in the same way as
the usual $Z$ factors depend on $u$ in the standard isotropic
$\varphi^4$ theory. The same statement holds for the additive
renormalization constant $A(u',\varepsilon)$ because of
\begin{eqnarray}
\label{3-b} \int\limits_{\bf k}^\infty \ln (r_0 + {\bf k} \cdot
{\bf Ak}) &=&  (\det {\bf A})^{-1/2}\int\limits_{{\bf k}'}^\infty
\ln (r_0 + {\bf k}' \cdot {\bf k}') \nonumber\\ &=& - \;(\det {\bf A})^{-1/2} \frac{2
A_d}{d \varepsilon} \; r_0^{d/2} \; .
\end{eqnarray}
Thus the renormalization constants read up to one-loop order $ Z_r
(u', \varepsilon) = 1 + 12 u' / \varepsilon $, $ Z_{u'} (u',
\varepsilon) = 1 + 36 u' / \varepsilon$, $Z_{\varphi'} (u',
\varepsilon) = 1$, $ A(u',\varepsilon)= - 2/\varepsilon $. The $Z$
factors $Z_{u'}$ and $Z_r$ are sufficient to renormalize
$y'^{eff}_0$, (\ref{4z1}), and $w'^{eff}_0$, (\ref{4z2}), whereas
the additive renormalization constant $ A(u',\varepsilon)$ is
needed to absorb the pole term $\sim -A_d (r_0 - r_{0c})^2 /(4
\varepsilon$) in the square brackets of (\ref{4aa}). After
substituting these renormalization constants one verifies that the
resulting renormalized free energy density $ f'_R$ has a finite
limit for $\varepsilon\rightarrow0$ at fixed $u'>0$.

We define the dimensionless amplitude function
\begin{eqnarray}
\label{5i} F'_R(r/\mu^2,u',L' \mu, {\bf \bar A}) =
 \mu^{-d} A_d^{-1} f'_R(r,u',\mu,L', {\bf \bar
A})\;. \qquad \nonumber\\
\end{eqnarray}
From the $\mu$ independence of $\delta f'(r_0 - r_{0c}, u_0', L',
{\bf \bar A})$ one can derive the renormalization-group equation
(RGE) for the amplitude function
\begin{eqnarray}
\label{5j}
&&(\mu\,\partial_\mu+r\zeta_r\partial_r+\beta_{u'}\partial_{u'}+d)
  F'_R(r/\mu^2,u',L' \mu, {\bf \bar A}) \nonumber\\ &&= -
  [r^2/(2 \mu^4)] B(u')\;,
\end{eqnarray}
where the field-theoretic functions $\beta_{u'}(u', \varepsilon),
\zeta_r(u')$, and $B(u')$ are defined as usual
\cite{dohm1985,schl}. Eq. (\ref{5j}), however, differs from the
corresponding bulk RGE (119) of \cite{str} since here we are using
$r$ rather than the bulk correlation lengths $\xi'_\pm$ as the
appropriate measure of the temperature variable. Using $r$ rather
than $\xi'_\pm$ is of advantage in finite-size theories where a
{\it single} finite-size scaling function is derived for both $r
\geq 0$ and $r < 0$. The functions $\zeta_r (u')\; , \; \beta_{u'}
(u',1)$ and $B(u')$ as well as the fixed point value $u'^* = u^*$
are accurately known \cite{schl,larin} from Borel resummations.
Integration of the RGE yields
\begin{eqnarray}
\label{5m} &&F'_R\Big(\frac{r}{\mu^2},u',L'\mu, {\bf \bar A}\Big)=
\;l^d \; \Bigg\{F'_R\Big(\frac{r(l)}{l^2\mu^2},u'(l),L'l\mu, {\bf
\bar A} \Big) \quad \nonumber\\ &&+ \frac{r(l)^2}{2l^4\mu^4}
\int\limits_1^l
B(u'(l))\Big\{\exp\int\limits_l^{l'}\Big[2\zeta_r(u'(l'')) -
\varepsilon\Big]\frac{dl''}{l''}\Big\}\frac{dl'}{l'}\Bigg\} \;,
\nonumber\\
\end{eqnarray}
with an as yet arbitrary flow parameter $l$ and $u'(1)\equiv u'$.
The effective parameters  $r(l)$ and $u'(l)$ are defined as usual
\cite{dohm1985}.

Eqs. (\ref{4dd}) and (\ref{5m}) show that in the arguments of the
functions $J_0,I_1,I_2$ and of the pole terms
$\sim\varepsilon^{-1}$ of Eqs.(\ref{4aa}) - (\ref{4cc}) the
parameter $r'_{0{\rm L}}$ will appear in the form of the effective
renormalized counterpart
\begin{eqnarray}
\label{5p} r'_{{\rm L}} (l) &\equiv& r'_{0{\rm L}} (r (l),
l^\varepsilon \mu^\varepsilon
A_d^{-1} u' (l), L') \nonumber\\
&=& r (l) + 12 (\mu l)^{\varepsilon/2} A_d^{-1/2} u' (l)^{1/2}
(L')^{- d/2} \vartheta_2 (y' (l)) \; ,\nonumber\\
\end{eqnarray}
\be
\label{5q} y' (l) = r (l) \mu^{-2} l^{-2} (L' \mu l)^{d/2}
A_d^{1/2} u' (l)^{- 1/2}.
\ee
Correspondingly, the effective renormalized counterparts of $
y_0'^{eff}$ and of $w_0'^{eff}$ appearing in the renormalized form
of the logarithmic part of $\delta f'$ are given by
\begin{eqnarray}
\label{5r} &&y'^{eff}(l, {\bf \bar A}) = (l \mu L')^{ d/2}A_d^{
1/2} u'(l)^{-1/2}\nonumber\\ && \times \Bigg\{\frac{r
(l)}{\mu^2l^2}\Big[1 + 18 u'(l)R_2 \Big(\frac{r'_{{\rm L}} (l)}{
\mu^2l^2 }, l \mu L', {\bf \bar A} \Big)\Big] \nonumber\\&&+ 12 u'
(l) R_1 \Big(\frac{r'_{{\rm L}} (l)}{ \mu^2l^2 }, l \mu L', {\bf
\bar A}\Big)
\nonumber\\
&&+ 144  (l \mu L')^{- d/2} A_d^{- 1/2} u'(l)^{3/2} \vartheta_2
(y' (l)) \nonumber\\ &&\times R_2  \Big(\frac{r'_{{\rm L}} (l)}{
\mu^2l^2 }, l \mu L', {\bf \bar A} \Big) \Bigg\} \; ,
\end{eqnarray}
\begin{eqnarray}
\label{5s} w'^{eff} (l, {\bf \bar A}) \; = \; u' (l)^{- 1/2}
\Big[1 + 18 u'(l)R_2 \Big(\frac{r'_{{\rm L}} (l)}{ \mu^2l^2 }, l
\mu L', {\bf \bar A} \Big)\Big] \; ,\nonumber\\
\end{eqnarray}

\be
\label{5t} R_1 (q, p, {\bf \bar A}) = \varepsilon^{-1} q [1 - q^{-
\varepsilon/2}] + p^{\varepsilon - 2} (4 \pi^2 A_d)^{-1} I_1 (q
\;p^2, {\bf \bar A}) \; ,
\ee

\begin{eqnarray}
\label{5u} R_2 (q, p, {\bf \bar A}) &=& - \; \varepsilon^{-1} [1 -
q^{- \varepsilon/2} ] - \frac{1}{2} q^{- \varepsilon / 2}
\nonumber\\ &+& p^\varepsilon (16 \pi^4 A_d)^{-1} I_2 (q \;p^2,
{\bf \bar A}) \; .
\end{eqnarray}
This suggests that the most natural choice of the flow parameter
$l$ is made by
\be
\label{5v} r'_{{\rm L}} (l) = \mu^2l^2.
\ee
It ensures the standard choice in the bulk limit both above and
below $T_c$ \cite{dohm1985}
\begin{eqnarray}
\label{5w} \lim_{L \rightarrow \infty} \mu^2l^2 = \left\{
\begin{array}{r@{\quad \quad}l}
                         \mu^2 l_+^2 = r(l_+)& \mbox{for} \;T > T_c ,\\
                         \mu^2 l_-^2 = -2r(l_-) & \mbox{for} \;T <
                 T_c ,
                \end{array} \right.
\end{eqnarray}
and appropriately implies $\mu l \propto L'^{-1}$ for large finite
$L'$ at $T=T_c$. As a natural choice for the reference length
$\mu^{-1}$ we take $\mu^{-1} = \xi_{0 +}'$ where \cite{schl}
\be
\label{5y} \xi_{0 +}' = \left[Z_r (u', \varepsilon) a_0^{-1} Q^*
\exp \left(\int_{u'}^{u'^*} \frac{\zeta_r (u'^*) - \zeta_r
(u'')}{\beta_{u'} (u'', \varepsilon)}\, d u'' \right)\right]^{1/2}
\ee
is the asymptotic amplitude of the second-moment bulk correlation
length of the isotropic system above $T_c$, as defined in
(\ref{3d}). The dimensionless amplitude $ Q^* = 1 + O (u^{*2}) = 1
+ O (u'^{*2})$ is the fixed point value of the amplitude function
$Q (1, u', d)$ of the second-moment bulk correlation length above
$T_c$ \cite{schl}.  Owing to the choice of the factor $A_d$,
(\ref{5c1}), the $O(u')$ term of $Q (1, u', d)$ and the $O(u^*)$
term of $Q^*$ vanish \cite{schl,krause,str}, similar to the
vanishing of the order-parameter amplitude function at $O(u')$
\cite{schl}. The same observation was recently made for the
correlation-length amplitude within the $\epsilon$ expansion
\cite{diehl2006} where the same geometric factor $A_d$ was
employed (apart from a harmless factor of 2). In three dimensions
the amplitude $Q^*$ it is accurately known from Borel resummations
\cite{krause}.

Eqs. (\ref{5p}) and (\ref{5v}) determine $l=l(t,L')$ as a function
of the reduced temperature $t$ and the size $L'$. With this choice
of $l$, Eqs.(\ref{5m}) and (\ref{5i}) provide a mapping of the
functions $F'_R$ and $ f'_R$ from the critical to the noncritical
region.

In summary, the singular part of the contribution $\delta f'$ to
the free energy density of the isotropic system is contained in
\begin{eqnarray}
\label{5aa} &&f'_R(r,u',L',\mu, {\bf \bar A}) = \delta f'(Z_rr,
\mu^{\varepsilon}Z_{u'}Z_{\varphi'}^{-2}A_d^{-1}u',L', {\bf \bar
A})\nonumber\\ && -\frac{1}{8}\mu^{-\varepsilon} r^2 A_d
A(u',\varepsilon)  = f'_R\big(r(l),u'(l),l\mu,L', {\bf \bar
A}\big) \nonumber\\ && + \;\frac{A_d r(l)^2}{2(l\mu)^\varepsilon}
\int\limits_1^l
B(u'(l'))\Big\{\exp\int\limits_l^{l'}\Big[2\zeta_r(u'(l'')) -
\varepsilon\Big]\frac{dl''}{l''}\Big\}\frac{dl'}{l'}, \nonumber\\
\end{eqnarray}
\begin{eqnarray}
\label{5bb} && f'_R \big(r(l),u'(l),l\mu,L',{\bf \bar A}\big) =
\nonumber\\ && -\; A_d (l\mu)^d / (4d) + 18 u' (l) L'^{-d}
\;[\vartheta_2(y'(l))]^2 \nonumber\\ &&+ \frac{1}{L'^d}\Bigg\{-\;
\ln \int\limits_{-\infty}^\infty d s \exp \big[- \frac{1}{2}
y'^{eff}(l, {\bf \bar A})(l) s^2 - s^4\big] \nonumber\\ && -
\frac{1}{2}\ln \left[\frac{2 \pi A_d^{1/2} w'^{eff} (l, {\bf \bar
A} )}{ (l \mu L')^{\varepsilon/2}}\right]\; + \frac{1}{2}J_0(l^2
\mu^2 L'^2,{\bf \bar A}) \nonumber\\ &&- \frac{3(l\mu
L')^{\varepsilon/2} u'(l)^{1/2}}{2\pi^2 A_d^{1/2}}
\;\vartheta_2(y'(l))\; I_1(l^2 \mu^2 L'^2,{\bf \bar A}) \nonumber\\
&&- \frac{9(l\mu L')^\varepsilon u'(l)}{4\pi^4 A_d }
\;[\vartheta_2(y'(l))]^2\; I_2(l^2 \mu^2 L'^2,{\bf \bar
A})\Bigg\}\; .
\end{eqnarray}
In the bulk limit the function  $f_{R,b}'^\pm (l_\pm \mu, u'
(l_\pm)) = \lim_{L \rightarrow \infty}f'_R (r (l), u' (l), l \mu,
L', {\bf \bar A})$ becomes independent of ${\bf \bar A}$,
\be
\label{5cc} f'^+_{R,b} \big(l_+ \mu, u' (l_+) \big) \; = \; - \;
A_d (l_+ \mu)^d / (4 d) \; ,
\ee
\begin{eqnarray}
\label{5dd} && f'^-_{R,b} \big(l_- \mu, u' (l_-)\big) \; = \nonumber\\
&&- A_d (l_- \mu)^d  \left\{ \frac{1}{64 u' (l_-)} \; + \;
\frac{1} {4d} \; + \; \frac{81}{64} \; u' (l_-) \right\} \;,
\nonumber\\
\end{eqnarray}
above and below $T_c$, respectively, where $l_+$ and $l_-$ are
determined by Eq. (\ref{5w}). The last integral term in
(\ref{5aa}) contains both a contribution $\propto  t^2 l
^{-\alpha/\nu} $ to the singular finite-size part $f_s'$ and a
contribution $\propto t^2$ to the nonsingular bulk part
$f'^{(2)}_{ns,b} $ of $\delta f'$ [see Eq. (\ref{4z8}), compare
also Eqs. (\ref{6j}) and (\ref{6l})].

\setcounter{equation}{0}
\section{Finite-size scaling function of the free energy density}

\subsection{Result in ${\bf 2<d<4}$ dimensions}

In order to derive the finite-size scaling function ${\cal F}$ we
consider (\ref{5aa}) - (\ref{5dd}) the limit of small $l \ll 1$ or
$ l\rightarrow 0$. In this limit we have $u'(l) \rightarrow
\;u'(0)\equiv u'^* = u^*$, $ r(l)/(\mu^2l^2)\rightarrow \;Q^*\; t
\;l^{-1/\nu}$,
\be
\label{6c} y'(l) \rightarrow \; \tilde y = \tilde x \;Q^*\; (\mu l
L')^{- \alpha / (2 \nu)} A_d^{1/2} {u^*}^{-1/2},
\ee
\be
\label{6d} \tilde x = t (\mu L')^{1/\nu} = t (L'/\xi_{0 +}
')^{1/\nu}\; .
\ee
In (\ref{6c}) we have used the hyperscaling relation
\be
\label{6-d} 2 - \alpha = d \nu \;.
\ee
Because of the choice (\ref{5v}), Eq. (\ref{5p}) implies $ \mu l
L' \rightarrow \; \tilde l = \tilde l(\tilde x)$ where  $\tilde l
(\tilde x)$ is determined implicitly by
\be
\label{6f} \tilde y + 12 \vartheta_2(\tilde y) = \tilde l^{d/2}
A_d^{1/2} {u^*}^{-1/2},
\ee
\be
\label{6g} \tilde y =\; \tilde x\;Q^*\;\tilde l^{- \alpha / (2
\nu)} A_d^{1/2} {u^*}^{-1/2}.
\ee
Simultaneously, these two equations determine $\tilde y = \tilde
y(\tilde x)$. Furthermore we have
\be
\label{6h} w'^{eff}(l, {\bf \bar A})\rightarrow \; W (\tilde x,
{\bf \bar A}) = {u^*}^{- 1/2} \left[1 + 18 \; u^* R_2(1, \tilde l,
{\bf \bar A}) \right] \;,
\ee
\begin{eqnarray}
\label{6i} && y'^{eff}(l, {\bf \bar A})\rightarrow \; Y(\tilde x,
{\bf \bar A}) = {\tilde l}^{ d/2} A_d^{ 1/2}
{u^*}^{-1/2}\nonumber\\ &&\times \Bigg\{Q^* \tilde x \;{\tilde
l}^{-1/\nu}\Big[1 + 18 u^*R_2(1,\tilde l, {\bf \bar A}) \Big] +
\;12 u^* R_1(1, \tilde l, {\bf \bar A}) \nonumber\\ && + \; 144
{u^*}^{3/2} {\tilde l}^{- d/2} A_d^{- 1/2}
 \vartheta_2 (\tilde y )R_2(1,\tilde l, {\bf \bar A})  \Bigg\}.
\end{eqnarray}
The asymptotic ($l\rightarrow0$) behavior of the integral in
(\ref{5aa})
\begin{eqnarray}
\label{6j} && \int\limits_1^l
B(u'(l'))\Big\{\exp\int\limits_l^{l'}\Big[2\zeta_r(u'(l'')) -
\varepsilon\Big]\frac{dl''}{l''}\Big\}\frac{dl'}{l'} \nonumber\\
&& \rightarrow\; - \;\frac{ \nu}{\alpha } \;B(u^*) +
O(l^{\alpha/\nu}) \;\;
\end{eqnarray}
is known from bulk theory \cite{str}. In (\ref{6j}) the subleading
term $O(l^{\alpha/\nu})$, together with the prefactor
$r(l)^2/(l\mu)^{\varepsilon}$ in (\ref{5aa}), contributes to the
regular bulk term $f'^{(2)}_{ns, b}$ proportional to $t^2$ of Eq.
(\ref{4z8}).

In summary, the asymptotic form of the singular part $f'_s$
(\ref{4a2}) of the reduced free energy density of the isotropic
system at $h' = 0$ is obtained from $f'_R$, (\ref{5aa}), in the
limit of small $l$ as
\be
\label{6k} f'_R \rightarrow f'_s(t,L') = L'^{-d}
\mathcal{F}(\tilde x, {\bf \bar A})
\ee
where the finite-size scaling function is given by
\begin{eqnarray}
\label{6l} &&\mathcal{F}(\tilde x, {\bf \bar A})= -\; A_d \;
\left[ \frac{\tilde l^d}{4d} \; + \;\frac{\nu\;{Q^*}^2 \tilde x^2
\tilde l^{- \alpha/\nu}}{2\alpha} \;B(u^*)\right] \nonumber\\ && +
\; 18 u^* \left[\vartheta_2 (\tilde y)\right]^2 - \frac{1}{2}\ln
\left(\frac{2 \pi A_d^{1/2} W (\tilde x, {\bf \bar A})}{
\tilde l^{\varepsilon/2}}\right) \nonumber\\
&&-\;  \ln \int\limits_{-\infty}^\infty d s \;\exp \big[-
\frac{1}{2}  Y(\tilde x, {\bf \bar A}) s^2 - s^4\big] \nonumber\\
&&+ \frac{1}{2}J_0( {\tilde l}^2, {\bf \bar A}) - \frac{3\;{\tilde
l}^{\varepsilon/2} {u^*}^{1/2}}{2\pi^2 A_d^{1/2}}
\;\vartheta_2(\tilde y)\; I_1( {\tilde l}^2, {\bf \bar A})
\nonumber\\ &&- \frac{9\;{\tilde l}^\varepsilon u^*}{4\pi^4 A_d }
\;[\vartheta_2(\tilde y)]^2\; I_2( {\tilde l}^2, {\bf \bar A}) \;.
\end{eqnarray}
This result is valid for $2<d<4$ in the range $L'\gg \tilde a$ and
$0 \leq |\tilde x| \lesssim O (1)$ above, at, and below $T_c$ (but
not for the exponential regime $|\tilde x| \gg 1$, see Sect. X).
It incorporates the correct bulk critical exponents $\alpha$ and $
\nu$ and the complete bulk function $B(u^*)$ (not only in one-loop
order). There is  only one adjustable parameter that is contained
in the nonuniversal bulk amplitude $\xi'_{0+}$ of the scaling
variable $\tilde x$, (\ref{6d}). For finite $L', f_s' (t, L')$ is
an analytic function of $t$ near $t = 0$, in agreement with
general analyticity requirements. From previous studies at finite
external field \cite{cd-97,cd-2000} we infer that the extension of
(\ref{6k}) to $h' \neq 0$ has the structure
\begin{equation}
\label{6-l} f_s' (t, h', L') = L'^{-d} {\cal F} (\tilde x, h' (L'
/ \xi_c')^{\beta \delta / \nu}; {\bf \bar A})
\end{equation}
where $\xi_c'$ is defined after (\ref{3l}). Thus the constants
$C_1'$ and $C_2'$ in (\ref{1c}) and (\ref{1d}) can be chosen most
naturally as $C_1' = (\xi_{0 +}')^{- 1/\nu}$ and $C_2' =
(\xi_c')^{- \beta \delta/\nu}$.

Of particular interest is the finite-size amplitude ${\cal F} (0,
{\bf \bar A}) \equiv {\cal F}_c ({\bf \bar A})$ at $T_c \;,$
\begin{eqnarray}
\label{6-1} \mathcal{F}_c({\bf \bar A}) &=& \left(18 -
\frac{36}{d}\right) u^* \left[\vartheta_2 (0)\right]^2 -
\frac{1}{2}\ln \left(\frac{2 \pi A_d^{1/2} W_c
({\bf \bar A})}{ \tilde l_c^{\varepsilon/2}}\right) \nonumber\\
&-&  \ln \int\limits_{-\infty}^\infty d s \;\exp
\big[- \frac{1}{2}  Y_c({\bf \bar A}) s^2 - s^4\big] \nonumber\\
&+& \frac{1}{2} \; J_0( {\tilde l_c}^2, {\bf \bar A})  -
\frac{\tilde l_c^2}{8 \pi^2} \; I_1( {\tilde l_c}^2, {\bf \bar A})
- \frac{\tilde l_c^4}{64 \pi^4} \; I_2( {\tilde l_c}^2, {\bf
\bar A}) \nonumber\\
\end{eqnarray}
where $\tilde l_c^{d/2} =12 {u^*}^{1/2}  A_d^{-1/2}
\vartheta_2(0)$ and
\be
\label{6-3} W_c ({\bf \bar A}) = {u^*}^{- 1/2} \left[1 + 18 \; u^*
R_2(1, \tilde l_c, {\bf \bar A}) \right],
\ee
\be
\label{6-4} Y_c({\bf \bar A}) = 144 u^* \vartheta_2 (0) \Bigg\{
R_1(1, \tilde l_c, {\bf \bar A}) + \; R_2(1,\tilde l_c, {\bf \bar
A})  \Bigg\}
\ee
with $\vartheta_2 (0) \;=\; \Gamma (3/4) / \Gamma (1/4) $ and
\be
\label{6-6}R_1(1, \tilde l_c, {\bf \bar A} )= \tilde l_c^{2-d} (4
\pi^2 A_d)^{-1} I_1( {\tilde l_c}^2, {\bf \bar A}) \; ,
\ee
\be
\label{6-7}R_2(1,\tilde l_c, {\bf \bar A})= - \frac{1}{2} + \tilde
l_c^\varepsilon (16 \pi^4 A_d)^{-1} I_2( {\tilde l_c}^2, {\bf \bar
A}) \; .
\ee
In the bulk (large $|\tilde x|$) limit Eqs.(\ref{6k}) and
(\ref{6l}) yield $\lim_{L' \to \infty} f_s' (t, L') = f_{s,b}'^\pm
(t)$ where
\be
\label{6m} f_{s,b}'^+ (t) \; = \; - \; A_d \; Q^{*d\nu} \;
\left[\frac{1}{4d} \; + \; \frac{\nu}{2 \alpha} \; B (u^*) \right]
\xi_{0+}'^{-d} t^{d \nu},
\ee
\begin{eqnarray}
\label{6n} && f_{s,b}'^- (t) \; = \; - \; A_d \; Q^{*d\nu} 2^{d
\nu}
 \nonumber \\ && \times \left[\frac{1}{64 u^*} \; +
\; \frac{1}{4d} \; + \; \frac{81}{64} \; u^* \; + \; \frac{\nu}{8
\alpha} \; B (u^*) \right] \; \xi_{0+}'^{-d} |t|^{d \nu} \nonumber\\
\end{eqnarray}
above and below $T_c$, respectively, with the universal bulk
amplitude ratios
\be
\label{6o} f_{s,b}'^+ \; (t) \xi_+'^d \; \equiv  Q_1 = \; - \; A_d
Q^{* d \nu} \left[\frac{1}{4d} \; + \; \frac{\nu}{2 \alpha} \; B
(u^*) \right] \; ,
\ee
\begin{eqnarray}
\label{6p} &&\frac{f_{s,b}'^- (t)} {f_{s,b}'^+(t)} \; = \;
\frac{A^-}{A^+} \nonumber\\ && = \; 2^{d \nu} \; \frac {1 / (64
u^*) \; + 1/(4d) \; + 81 u^*/64 \; + \; \nu B (u^*) / (8 \alpha) }
{1/(4d) \; + \; \nu B (u^*) / (2 \alpha)} \;. \nonumber\\
\end{eqnarray}
(For $Q_1$ compare (\ref{3f1}).) Here we have used the bulk
identifications
\be
\label{6q} \mu l = \left\{
\begin{array}{r@{\quad\quad}l}
                \mu l_+ = Q^{* \nu} \; \xi_{0+}'^{-1} t^\nu \;   & \mbox{for} \;\;\;  T > T_c\;, \\
\mu l_- = Q^{* \nu} \; \xi_{0+}'^{-1} (2 | t |)^\nu & \mbox{for}
\;\;\;
                 T < T_c \;,
                \end{array} \right.
\ee
as implied by the choice (\ref{5w}). As noted in Sect. III, a
complete two-loop calculation would yield further bulk
contributions of $O (u^*)$ in $f_{s,b}'^\pm$. Owing to the
truncation (\ref{4y}), no terms of  $O({u^*}^{2})$ and higher
order appear in (\ref{6n}) and (\ref{6p}).

In order to present the scaling function
\be
\label{6t} {\cal F}^{ex} (\tilde x; {\bf \bar A})\; = {\cal
F}(\tilde x; {\bf \bar A})\;- {\cal F}_b^\pm (\tilde x).
\ee
of the
excess free energy density $f'^{ex}_s (t, L'; {\bf \bar A})  = f'_s (t, L'; {\bf \bar A}) -
f'_{s,b}(t)$  we shall also need the
${\bf \bar A}$ independent bulk part
\begin{eqnarray}
\label{6s}  {\cal F}_b^\pm (\tilde x) \; = \left\{
\begin{array}{r@{\quad \quad}l}
                         \; L'^d f_{s,b}'^+ \; =  \; Q_1 \tilde x^{d \nu}& \mbox{for} \;T > T_c ,\\
                         \; L'^d f_{s,b}'^- \;=  \; Q_1^- |\tilde x|^{d \nu} & \mbox{for} \;T <
                 T_c ,
                \end{array} \right.
\end{eqnarray}
with $Q_1^- = (A^-/A^+)  Q_1$, representing the large  $|\tilde
x|$ behavior of ${\cal F}(\tilde x, {\bf \bar A})$.
It should be noted that it is not obvious how to interpret the $d
\times d$ matrix ${\bf \bar A}$ for the case of non-integer
dimensions $d$.

In the spirit of the fixed - $d$ minimal subtraction approach
\cite{dohm1985} we shall evaluate ${\cal F} (\tilde x, {\bf\bar
A})$ and ${\cal F}^{ex} (\tilde x, {\bf\bar A})$ in $d = 3$
dimensions without any further expansion with respect to $u^*$.
This is in contrast to the $\varepsilon$ expansion which is a
double expansion with respect to $u^*$ and $\varepsilon = 4 - d$.

\subsection{Epsilon expansion}

Considering $u^*$ as a smallness parameter and using the results
of App. C we obtain from (\ref{6-1}) at fixed $2 < d < 4$
\begin{eqnarray}
\label{7a} &&{\cal F}_c ({\bf \bar A}) = \frac{1}{2} \; \ln
\left\{ \frac{(12)^{4/d} [\Gamma (3/4)]^{\varepsilon/d}
{u^*}^{2/d}}{24
\pi A_d^{2/d} [\Gamma (1/4)]^{\varepsilon/d}}\right\} \nonumber\\
&&- \ln \;\Big [\frac{1}{2} \Gamma (1/4) \Big] \; + \; \frac{1}{2}
J_0 (0, {\bf \bar A}) \nonumber\\ &&+ \frac{1}{8 \pi^2}
\left[\frac{12 \; \Gamma (1/4)}{A_d^{1/2} \Gamma
(3/4)}\right]^{4/d} I_1 (0, {\bf
\bar A}) {u^*}^{2/d} + O (u^*, {u^*}^{4/d}) \; . \nonumber\\
\end{eqnarray}
Substituting $u^* \; = \; \varepsilon/36 \; + \; O
(\varepsilon^2)$ and expanding all $d$ dependent quantities with
respect to $\varepsilon = 4-d$ yields the $\varepsilon$-expansion
result at $T_c$
\begin{eqnarray}
\label{9b}{\cal F}_c ({\bf \bar A}) \; = \; \frac{1}{4} \; \ln
\varepsilon \; + \; f_0 ({\bf \bar A}) \; + \; f_1 ({\bf \bar A})
\; \varepsilon^{1/2} \; + \; O (\varepsilon) \;, \nonumber\\
\end{eqnarray}
\begin{eqnarray}
\label{9c} f_0 ({\bf \bar A}) \; = \; - \; \frac{1}{4} \; \ln 18
\; - \; \ln \; \left[\frac{1}{2} \; \Gamma (1/4)\right]
\nonumber\\ + \; \frac{1}{2} \int\limits_0^\infty \frac{dy}{y}
\left[\left(\frac{\pi}{y}\right)^2 \; - \; K_4 (y, {\bf \bar A}) +
1 - e^{- y} \right] \; ,
\end{eqnarray}
\begin{eqnarray}
\label{9d}f_1 ({\bf \bar A}) \; = \; \frac{\Gamma (1/4)}{\pi
\Gamma (3/4) \sqrt{2} } \int\limits_0^\infty dy \left[K_4 (y, {\bf
\bar A})
- \left(\frac{\pi}{y}\right)^2 - 1 \right] \; , \nonumber\\
\end{eqnarray}
where now ${\bf \bar A}$ denotes a $4 \times 4$ matrix. The
$\varepsilon$ expansion result (\ref{9b}) - (\ref{9d}) is
independent of which renormalization scheme and which kind of
perturbation approach is used. The same result is obtained if one
starts with the effective Hamiltonian of Br\'ezin and Zinn-Justin
\cite{BZ} or with the cumulant expansion of Rudnick et al.
\cite{RGJ}. Because of the strict expansion with respect to $u^*$
and $\varepsilon$, the exponential structure of the distribution
$\sim \exp [- H'^{eff}]$ is destroyed. As expected, the
$\varepsilon$ - expansion term $\sim \ln \varepsilon$ is not well
behaved for $\varepsilon \to 0$ since at $d = 4$ the finite
lattice constant $\tilde a $ must not be neglected.

A nontrivial question arises if the $\varepsilon$-expansion result
is applied to three-dimensional anisotropic systems with a matrix
${\bf \bar A} \neq {\bf 1}$. It appears that, to some extent, it
is ambiguous how the physical $3 \times 3$ matrix ${\bf \bar A}$
(which, in general, has 5 independent nonuniversal matrix
elements) should be continued to $d = 4$ in order to evaluate the
coefficients $f_0 ({\bf \bar A})$ and $f_1 ({\bf \bar A})$. This
matrix ${\bf \bar A}$ in (\ref{9b}) is necessarily a $4 \times 4$
matrix which, in general, has 9 independent nonuniversal matrix
elements, i.e., four additional nonuniversal parameters. It is not
unique how to choose the magnitude of these four additional matrix
elements. The results for $f_0 ({\bf \bar A})$ and $f_1 ({\bf \bar
A})$ in four dimensions will significantly depend on this choice.

As a possible choice we propose the following. In order to
describe the physical system with the symmetric three-dimensional
matrix
\begin{equation}
 \label{6-33}
 {\bf \bar A}  \; =  \; \left(\begin{array}{ccc}
  a & b & c \\
  b & d & e \\
  c & e & f \\
\end{array}\right)
\end{equation}
with $\det {\bf \bar A} = {\bf 1}$ it seems reasonable to extend
this matrix to the four-dimensional counterpart
\begin{equation}
 \label{6-34}
 {\bf \bar A}  \; = \; \left(\begin{array}{cccc}
  a & b & c & 0 \\
  b & d & e & 0 \\
  c & e & f & 0 \\
  0 & 0 & 0 & 1
\end{array}\right) \; .
\end{equation}
This choice guarantees that no arbitrary anisotropy is introduced
in the fourth dimension and that $\det {\bf \bar A} = {\bf 1}$.

A corresponding problem would arise in an $\varepsilon$ expansion
in $d = 2 + \varepsilon$ dimensions. In Sect. VIII. D below we
shall present an example where we compare the anisotropy effects
of two- and three-dimensional models. The result of this
comparison supports the suggestion given above for the dimensional
extension of the matrix ${\bf \bar A}$ in an $\varepsilon$
expansion.

\subsection{Large - $n$ limit}

For comparison with the case $n=1$ we also present the exact
result for the finite-size scaling function ${\cal F}_\infty$ of
the free energy density per component $f_\infty(t,L)$ of the
$\varphi^4$ lattice model (\ref{2a}) with $v= \tilde a^d$ and
$V=L^d$ in the limit $n \to \infty$ at fixed $u_0 n$. From Eqs.
(45) and (46) of Ref. \cite{cd1998} we have
\begin{eqnarray}
\label{10a} f_\infty(t,L) = \lim_{n \to \infty}\{-(n V)^{-1} \ln
Z(t,0,L)\} \;\;\;\nonumber\\ = \hat f_0 -
\frac{(r_0-\chi_\infty^{-1})^2}{16 u_0 n} + \frac{1}{2V}
{\sum_{\bf k}} \ln \{[\chi_\infty^{-1} + \delta \widehat K
(\mathbf k)] \tilde a^2\}\;\;\;
\end{eqnarray}
where $Z(t,0,L)$ is defined by (\ref{2d})  and $\chi_\infty^{-1}$
is determined implicitly by $ \chi_\infty^{-1}= r_0 +  4 u_0n
V^{-1} {\sum_{\bf k}} [\chi_\infty^{-1} + \delta \widehat K
(\mathbf k)]^{-1}$. The additive constant in (\ref{10a}) is $\hat
f_0 = - [\ln (2 \pi)] / (2 \tilde a^d)$. Using the results of App.
C  leads to the singular part of $f_\infty$ in the regime $L' \gg
\tilde a$ and $0 \leq | \tilde x | \lesssim O (1)$ at $h = 0$ for
$2 < d < 4$
\begin{eqnarray}
\label{8a}f_{\infty,s} (t, L; {\bf  A}) \; = \; L^{-d} \; {\cal
F}_\infty (\tilde x; {\bf \bar A}) \; ,
\end{eqnarray}
\begin{eqnarray}
\label{8b}{\cal F}_\infty (\tilde x; {\bf \bar A}) = \frac{1}{2}
{\cal G}_0 (P^2; {\bf \bar A}) + \frac{A_d}{2 (4 - d)}
\left[\tilde x P^2 \; - \; \frac{2}{d} \; P^d \right] \;, \quad
\nonumber\\
\end{eqnarray}
\begin{eqnarray}
\label{8c} P^{d-2} &=&  \tilde x \; - \; \frac{4 - d}{A_d} \; \;
{\cal G}_1 (P^2; {\bf \bar A}) \; ,
\end{eqnarray}
\begin{eqnarray}
\label{8d} {\cal G}_j (P^2; {\bf \bar A}) \; = \; (4 \pi^2)^{-j}
\int\limits^\infty_0 dy \; y^{j-1} \exp \left(- \frac{P^2 y}{4
\pi^2} \right) \nonumber\\
\times \left\{\left(\frac{\pi}{y}\right)^{d/2} \; - \; K_d (y,
{\bf \bar A}) \right\} \; .
\end{eqnarray}
Here $\tilde x \; = \; t (L' / \xi_{0+}')^{1 / \nu}$ with $\nu =
(d - 2)^{-1}$, $L' = (\det {\bf A})^{- 1/(2d)} L$, $\xi_{0+}' = (4
u'_0 n A_d a_0^{-1} / \varepsilon)^\nu$,  and $u'_0=(\det{\bf
A})^{-1/2}u_0$. We note that the geometric factor $A_d$,
(\ref{5c1}), appears in (\ref{8a}) - (\ref{8d}) in a natural way.
The reason is that only diagrammatic contributions of single-loop
structures contribute to the large - $n$ limit. The function
$(1/2){\cal G}_0(\tilde x; {\bf \bar A})$ with $\tilde x = r_0
L'^2$ is the scaling function of the excess free energy density of
the Gaussian model (see (\ref{b18}) - (\ref{b23}) of App. B). For
$T \geq T_c$ the function $P (\tilde x; {\bf \bar A})$ determines
the finite-size scaling form of the susceptibility per component
in the limit $n \to \infty$ \cite{cd2004}
\be
\label{6-39} \chi_\infty^+ (t, L; {\bf \bar A}) \; = \;
L'^{\gamma/\nu} \; g (\tilde x; {\bf \bar A}) \; , \gamma/\nu \; =
\; 2 \; ,
\ee
where $g (\tilde x; {\bf \bar A}) = [P (\tilde x; {\bf \bar
A})]^{-2}$. Below we shall present the relative anisotropy effect
\begin{eqnarray}
\label{6-42} \Delta \chi^+_{\infty,c} ({\bf \bar A}) \; = \;
\frac{g(0; {\bf \bar A}) \; -  g(0; {\bf 1})}{g(0; {\bf 1})}
\end{eqnarray}
on  $\chi_\infty^+$ at $T_c$ in three dimensions.

The result (\ref{8a}) - (\ref{6-39}) is the extension of the
result for the isotropic case (see  Eqs. (17)-(19) of
\cite{cd2002}) and corrects Eq. (44) of \cite{cd2004} where the
term $- (\ln 2)/2 $ should be dropped. The scaling function of the
excess free energy density above, at, and below $T_c$ is given by
\begin{eqnarray}
\label{8e} {\cal F}_\infty^{ex} (\tilde x; {\bf \bar A}) \; = \;
{\cal F}_\infty (\tilde x; {\bf \bar A}) \; - \; {\cal F}_{\infty,
b}^\pm (\tilde x)
\end{eqnarray}
with the bulk part
\begin{eqnarray}
\label{8f} {\cal F}_{\infty, b}^\pm (\tilde x) \; = \left\{
\begin{array}{r@{\quad \quad}l}
                         \; Y \tilde x^{d \nu}& \mbox{for} \;T > T_c\; ,\\
                         \; 0 \qquad  & \mbox{for} \;T <
                 T_c \;,
                \end{array} \right.
\end{eqnarray}
where $Y  = (d - 2) A_d/[2 d (4 - d)]$. At $T_c$ the finite-size
amplitude is given by
\begin{eqnarray}
\label{8h}{\cal F}_\infty (0; {\bf \bar A}) = \frac{1}{2} {\cal
G}_0 (P^2_c; {\bf \bar A}) - \frac{A_d}{d (4 - d)} P^d_c \;, \quad
\end{eqnarray}
where $P_c ({\bf \bar A}) \equiv P (0; {\bf \bar A})$ is
determined by
\begin{eqnarray}
\label{8i} P^{d-2}_c \; = \;  - \; \frac{4 - d}{A_d} \; \; {\cal
G}_1 (P^2_c; {\bf \bar A}) \; .
\end{eqnarray}

\setcounter{equation}{0}
\section{Other finite-size scaling functions}

The calculations of the preceding sections can be extended to
other finite-size quantities. Here we consider only those
quantities that have been studied in MC simulations of anisotropic
Ising models \cite{selke2005,schulte,stauffer}. Within our
$\varphi^4$ lattice model (\ref{2a}) for $n = 1$ at $h = 0$ on a
simple-cubic lattice with volume $V = L^d$ we shall consider the
susceptibilities $ \chi^+ = V < \Phi^2 > $, $\chi^- = V (<\Phi^2>
- <|\Phi|>^2)$, and the Binder cumulant
\begin{eqnarray}
\label{8l}  U = 1 - \frac{1}{3}<\Phi^4>/<\Phi^2>\;
\end{eqnarray}
where $\Phi = N^{-1} \sum_j \varphi_j$  (see, e.g., \cite{EDC}).
These quantities remain invariant under the transformation defined
in Sect. II \cite{dohm2006}, $\chi^\pm = (\chi^\pm)'\; , U = U'$.
As a consequence we find that, in the regime (b) defined in
Section IV.A, the finite-size scaling forms of these quantities
are
\begin{eqnarray}
\label{8o} \chi^\pm(t,L; {\bf A}) = (\chi^\pm)'(t,L'; {\bf \bar
A}) = (L'/\xi'_{0+})^{\gamma/\nu}P_\chi^\pm(\tilde x; {\bf \bar
A}) \;,
\nonumber\\
\end{eqnarray}
\begin{eqnarray}
\label{8p}  U(t,L; {\bf A}) =  U'(t,L'; {\bf \bar A}) = U(\tilde
x; {\bf \bar A})\; ,
\end{eqnarray}
where the scaling functions $P_\chi^\pm$ and $U$ are obtained from
those of \cite{EDC} by the replacements $Y \to Y (\tilde x; {\bf
\bar A})$ and $R_2 \to R_2 (1, \tilde l, {\bf \bar A})$. Note that
the functions $P^\pm_\chi$ are nonuniversal even for ${\bf \bar A}
= {\bf 1}$ since they still contain  nonuniversal overall
amplitudes $c^\pm$ proportional to the bulk amplitudes of
$\chi^\pm$ (see also \cite{CDT}). Here we consider only the {\it
relative} anisotropy effect
\begin{eqnarray}
\label{7-a} \Delta \chi^\pm_c ({\bf \bar A}) \; = \;
\frac{P_\chi^\pm (0; {\bf \bar A}) \; - P_\chi^\pm (0; {\bf
1})}{P_\chi^\pm (0; {\bf 1})}
\end{eqnarray}
on the susceptibilities $\chi^\pm (0, L; {\bf A})$ at $T = T_c$.
The analytic expressions are in $2 < d < 4$ dimensions
\be
\label{7-7} P_\chi^+ (0; {\bf \bar A}) \; = \; c^+ \left[1- 18 u^*
R_2 (1, \tilde l_c, {\bf \bar A})\right]^{-1} \vartheta_2 (Y_c
({\bf \bar A})) \; ,
\ee
\begin{eqnarray}
\label{7-c} P_\chi^- (0; {\bf \bar A}) \; = \; &c^-&
\left[1- 18 u^* R_2 (1, \tilde l_c, {\bf \bar A})\right]^{-1} \nonumber\\
&\times& \left\{ \vartheta_2 (Y_c ({\bf \bar A})) \; - \;
\left[\vartheta_1 (Y_c ({\bf \bar A}))\right]^2 \right\} \; ,\nonumber\\
\end{eqnarray}
where the constants $c^\pm$ are independent of ${\bf \bar A}$ and
drop out of the ratio (\ref{7-a}). For $Y_c ({\bf \bar A})$ and
$R_2 (1, \tilde l_c, {\bf \bar A})$ see (\ref{6-4}) and
(\ref{6-7}), for $\vartheta_m (Y)$ see (\ref{4f1}). The anisotropy
effect on the Binder cumulant $U (0; {\bf \bar A})$ at $T_c$ will
be described by the difference
\begin{eqnarray}
\label{7-b} \Delta U_c ({\bf \bar A}) \; = \; U (0;{\bf \bar A})
\; - \; U (0; {\bf 1})
\end{eqnarray}
where
\begin{eqnarray}
\label{7-d} U (0; {\bf \bar A}) \; = \; 1 - \frac{1}{3}
\vartheta_4 (Y_c ({\bf \bar A})) \left[\vartheta_2 (Y_c ({\bf \bar
A})\right]^{-2} \; .
\end{eqnarray}

\setcounter{equation}{0}
\section{Quantitative results and predictions}

For the application to three dimensions we shall employ the same
values as previously \cite{EDC,liu}, $A_3 = (4\pi)^{-1}$, $\nu=
0.6335$,  $u^*= 0.0412$, $Q^*= 0.945$, $B(u^*) = 0.50$. For
reasons of consistency, a slightly different value will be used
for $\alpha = 2 - 3 \nu = 0.0995$ in order to exactly satisfy the
hyperscaling relation (\ref{6-d}).

\subsection{Universal bulk amplitude ratios}

Evaluating our analytic expressions for the bulk amplitude ratios
(\ref{6o}) and (\ref{6p}) in three dimensions we obtain for {\it
isotropic} systems [compare (\ref{3f1})]
\be
\label{6u} f_{s,b}'^+ \;\;\xi_+'^3 = Q_1 = -0.119, \;\; A^- / A^+
= 2.04 \; .
\ee
This can be compared with the series expansion results for the
three-dimensional Ising model by Liu and Fisher \cite{liu} who
calculated the amplitude ratios $(R^+_\xi)^3 = 0.0188 \pm 0.0001$
and $ A^+ / A^- =0.523 \pm 0.009$. These calculations were carried
out for several different cubic (sc, bcc, fcc) lattice structures
in order to test bulk universality (see also \cite{tarko}). The
relation between $ R^+_\xi $ and $Q_1$ is $ (R^+_\xi)^3 = - \alpha
(1-\alpha)(2-\alpha) Q_1$. This yields the central values of the
Ising model based on series expansions

\be
\label{8-3} {Q_1}_{|Ising} \; = \; - 0.1099 \; ,\;\; {A^- /
A^+}_{|Ising} \; = \; 1.91 \;.
\ee
Considering the fact that our present theory is an effective
finite-size theory that is  not designed to produce highly
accurate bulk predictions the results (\ref{6u}) are in acceptable
agreement with (\ref{8-3}). As seen from (\ref{6m}) - (\ref{6o}),
the bulk results for the free energy are sensitive to the choice
of the geometrical factor in defining the renormalized coupling,
(\ref{5a}). The results (\ref{6u}) demonstrate the appropriateness
of the choice of $A_d$, (\ref{5c1}).

\subsection{Finite-size free energy of isotropic systems}

\subsubsection*{ 1. Test of the $d = 3$ theory : amplitude at
$T_c$}

In order to test the reliability of our finite-size theory we
first consider the isotropic case ${\bf \bar A} = {\bf 1}$,
$\xi'_{0+} =\xi_{0+}$, where accurate MC data by Mon
\cite{mon-1,mon-2,mon-3} are available.

\begin{figure}[!h]
\includegraphics[clip,width=80mm]{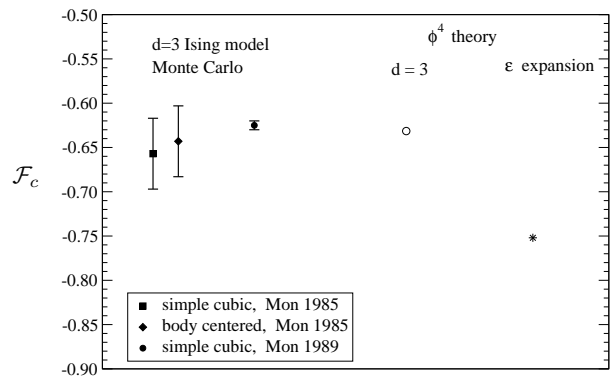}
\caption{Finite-size amplitude ${\cal F}_c({\bf 1})$ ,
(\ref{6-1}), (\ref{6w}), of the free energy density of isotropic
systems in a cubic geometry at $T_c$ for $n=1$ in three dimensions
(open circle), and at $\varepsilon=1$, (\ref{9g}) (star), of the
$\varepsilon$ expansion (\ref{9b}). MC data for the
three-dimensional Ising model on sc and bcc lattices
\cite{mon-1,mon-2}. }
\end{figure}

The first set of data was obtained for the three-dimensional Ising
model with NN couplings on sc and bcc lattices. These systems have
different values of $T_c$ and different correlation-length
amplitudes $\xi_{0+}$ but both belong to the subclass of
(asymptotically) isotropic systems with ${\bf \bar A} = {\bf 1}$.
Within the error bars, the MC results for the finite-size
amplitude ${\cal F}_c ({\bf 1})$ of the free energy density at
$T_c$ \cite{mon-1}
\begin{eqnarray}
\label{8-c}{\cal F}_c ({\bf 1})^{MC}\; = \left\{
\begin{array}{r@{\quad \quad}l}
                         \; - \;  \; 0.657 \pm 0.03 \qquad  \quad \mbox{(sc lattice)} \\
                         \; - \;  \; 0.643 \pm 0.04 \qquad \mbox{(bcc lattice)}
                \end{array} \right.
\end{eqnarray}
are consistent with the universality hypothesis. Subsequently the
more accurate MC result at $T_c$ was obtained \cite{mon-2}
\begin{eqnarray}
\label{8-d} {\cal F}_c ({\bf 1})^{MC} \; = \; - \; 0.625 \; \pm \;
0.005 \qquad \mbox{(sc lattice)}
\end{eqnarray}
which is also consistent with (\ref{8-c}).

In three dimensions the numerical values of the quantities $\tilde
l_c, J_0, I_1$, and $I_2$ in our analytical result (\ref{6-1}) are
$\tilde l_c = 2.042$, $J_0( {\tilde l}^2_c, {\bf 1}) = 1.6430$,
$I_1( {\tilde l}^2_c, {\bf 1}) = - 4.1581$, and $I_2( {\tilde
l}^2_c, {\bf 1}) = - 15.4032$. This yields the theoretical
prediction for the finite-size amplitude
\be
\label{6w} \mathcal{F}_c({\bf 1})_{d=3} \; = \; - \; 0.6315 \; ,
\ee
in excellent agreement with the MC results (\ref{8-c}) and
(\ref{8-d}). (Fig. 5)

\subsubsection*{ 2. Epsilon expansion at $T_c$}

For comparison we also evaluate the result of the $\varepsilon$
expansion (\ref{9b}). For isotropic systems $({\bf \bar A} = {\bf
1})$ the coefficients in (\ref{9b}) are well defined. The
numerical values are
\begin{equation}
\label{9e}f_0 ({\bf 1}) \; = \; - \; 0.3302 \; ,\;\;f_1 ({\bf 1})
\; = \; - \; 0.4218 \; ,
\end{equation}
where ${\bf 1}$ denotes the $4 \times 4$ unity matrix. For
$\varepsilon = 1$ the terms up to $O (\varepsilon^{1/2})$ of
(\ref{9b}) yield
\begin{equation}
\label{9g} {\cal F}_c ({\bf 1})_{\varepsilon=1} \; = \; - \;
0.7520 \; ,
\end{equation}
which is in less good agreement with the MC results (\ref{8-c})
and (\ref{8-d}). (Fig. 5)

\begin{figure}[!h]
\includegraphics[clip,width=80mm]{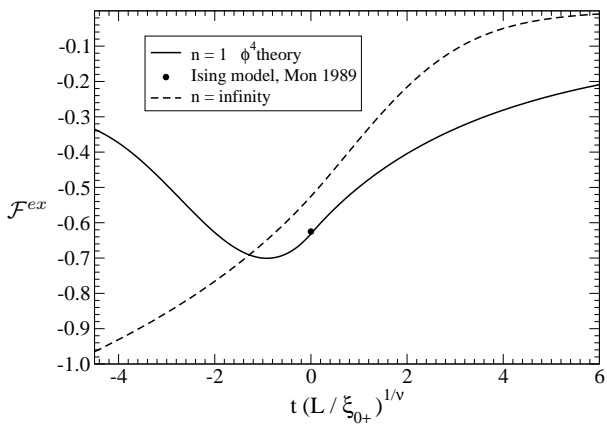}
\caption{ Scaling functions ${\cal F}^{ex} (\tilde x; {\bf 1})$,
(\ref{6t}), (\ref{6l}), (\ref{6s}) for $n=1$ (solid line) and
${\cal F}_\infty^{ex} (\tilde x; {\bf 1 })$, (\ref{8e}),
(\ref{8b}), (\ref{8f}) for $n=\infty$ (dashed line) for the excess
free energy density of isotropic systems in three dimensions as a
function of the scaling variable $\tilde x = t
(L/\xi_{0+})^{1/\nu}$. MC result (full circle) for the Ising model
on a sc lattice \cite{mon-2}. }
\end{figure}

\subsubsection*{ 3. Finite-size scaling function}

In three dimensions the numerical evaluation of the scaling
function ${\cal F}^{ex} (\tilde x, {\bf 1})_{d=3}$ as given in
(\ref{6t}), (\ref{6l}), and (\ref{6s}) yields the curve shown in
Fig. 6 in the range $ - 4.5 \leq \tilde x \leq 6 $ (solid curve).
This range corresponds to the central finite-size regime (b)
mentioned in Sect. IV A. A minimum with
\be
\label{6y}{\cal F}^{ex} (\tilde x_{min}; {\bf 1})_{d=3} = - 0.701,
\;\; \tilde x_{min} = - 0.910
\ee
exists slightly below $T_c$. For the subclass of isotropic systems
within the $(d=3, n=1)$ universality class both the position
$\tilde x_{min}$ and the value ${\cal F}^{ex} (\tilde x_{min};
{\bf 1})_{d=3}$ are predicted to be universal numbers. This can be
tested by MC simulations for families of three-dimensional Ising
models with ${\bf \bar A} = {\bf 1}$, (e.g. on $sc$, $fcc$, or
$bcc$ lattices with isotropic interactions) in a cube with
periodic boundary conditions. The nonuniversal differences of
these models are predicted to be absorbable entirely in different
values of $\xi'_{0+}$. In {\it two} dimensions such tests of
universality for the critical Binder cumulant of isotropic systems
at $T_c$ have been performed very recently by Selke
\cite{selke2007}.

Our analytical result for  ${\cal F}^{ex} (\tilde x; {\bf
1})_{d=3}$ is not applicable far outside the range of $\tilde x$
shown in Fig. 6. In the limits $\tilde x \rightarrow \pm \infty$
this result does not correctly describe the exponential decay to
zero in the regimes (a) and (c) mentioned in Sect. IV A, as
expected.

For comparison we also present the exact result for the scaling
function ${\cal F}_\infty^{ex} (\tilde x; {\bf 1 })$  in the large
- $n$ limit in three dimensions. For ${\bf \bar A} = {\bf 1}$ and
$d=3$ the numerical solutions of (\ref{8i}) and (\ref{8h}) at
$T=T_c$ are $P_c ({\bf 1}) = 1.946$ and
\begin{eqnarray}
\label{8j} {\cal F}_\infty^{ex} ( 0 ; {\bf 1 })={\cal F}_\infty
(0; {\bf 1}) = - 0.526 .
\end{eqnarray}
In Fig. 6 the scaling function ${\cal F}_\infty^{ex} (\tilde x;
{\bf 1 })$, (\ref{8b}), is shown in three dimensions (dashed
curve). Unlike the case $n=1$, ${\cal F}_ \infty^{ex} (\tilde x;
{\bf 1})$ does not have a minimum at finite $\tilde x $ below
$T_c$ but has a slow monotonic decrease towards a finite negative
constant ${\cal F}_\infty^{ex} (-\infty; {\bf 1}) = - 3.18$. Above
$T_c$ it decays exponentially to zero (but not with the correct
exponential form, see Sect. X.).

\subsection{Three-dimensional anisotropy}

In the following we present quantitative predictions for the
nonuniversal effect of a non-cubic anisotropy on the finite-size
scaling functions in three dimensions. We illustrate the
three-dimensional anisotropy effects for the example of ${ \bf A}$
and ${\bf \bar A}$  given in  (\ref{32}) and (\ref{32a}) of Sect.
II. In this example,  ${\bf \bar A}(w)$ depends only on the single
anisotropy parameter $w$, (\ref{32b}), $ - \frac{1}{2} < w < 1$.
The crucial anisotropy function is given by (\ref{4ii}) which
enters the functions $J_0, I_1,$ and $I_2$ defined in (\ref{4gg})
and (\ref{4hh}).

\begin{figure}[!h]
\includegraphics[clip,width=80mm]{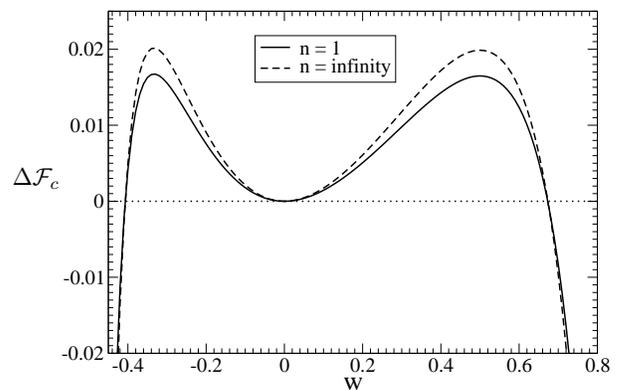}
\caption{Differences $\Delta \mathcal{F}_c ({\bf \bar A}(w))$,
(\ref{8-14}), and $\Delta \mathcal{F}_{c,\infty} ({\bf \bar
A}(w))$, (\ref{8-15}), of the finite-size amplitudes (\ref{6-1})
and (\ref{8h}) of the free energy density of anisotropic systems
with the reduced anisotropy matrix ${\bf \bar A}(w)$, (\ref{32a}),
in a cubic geometry at $T_c$ for $n=1$ (solid line) and $n=\infty$
(dashed line) in three dimensions  as a function of the anisotropy
parameter $w$, (\ref{32b}). }
\end{figure}

First we consider the anisotropy effect on the finite-size
amplitude of the free energy density for $n=1$ and $n=\infty$ at
$T=T_c$ as described by the differences

\begin{figure}[!h]
\includegraphics[clip,width=80mm]{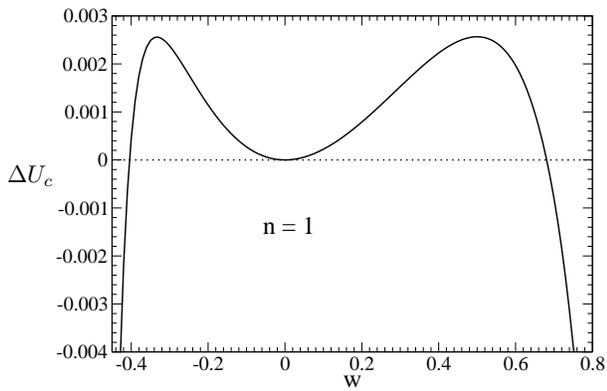}
\caption{Difference $\Delta U_c ({\bf \bar A}(w))$, (\ref{7-b}),
of the Binder cumulant (\ref{7-d}) of anisotropic systems with the
reduced anisotropy matrix ${\bf \bar A}(w)$, (\ref{32a}), in a
cubic geometry at $T_c$ for $n=1, d=3$   as a function of the
anisotropy parameter $w$, (\ref{32b}). }
\end{figure}
\be
\label{8-14} \Delta \mathcal{F}_c ({\bf \bar A}(w)) =
\mathcal{F}(0; {\bf \bar A}(w)) - \mathcal{F}(0; {\bf 1}),
\ee
\be
\label{8-15} \Delta \mathcal{F}_{c,\infty} ({\bf \bar A}(w)) =
\mathcal{F}_\infty(0; {\bf \bar A}(w)) - \mathcal{F}_\infty(0;
{\bf 1}),
\ee
where $\mathcal{F}(0; {\bf \bar A})$ and $\mathcal{F}_\infty(0;
{\bf \bar A})$ are given by (\ref{6-1}) and (\ref{8h}). This is
shown in Fig. 7  in the range $-0.45 < w < 0.80$. The anisotropy
effect is well pronounced for both positive and negative values of
$w$, with a non-negligible $n$ dependence. For both $n=1$ and
$n=\infty$ two maxima of almost equal heights exist at $w_{max} =
- 0.333$ and $w_{max} = 0.500$, with $\Delta {\cal F}_{c,max} =
0.0167$ and $\Delta {\cal F}_{c,max} = 0.0165$, respectively, for
$n=1$. [The slight difference of the two heights is presumably not
a consequence of the approximations made for $n=1$; such a
difference exists also for the exact result for $n=\infty$ where
$\Delta\mathcal{F}_{c,\infty,max} = 0.0202 $ and
$\Delta\mathcal{F}_{c,\infty,max} = 0.0199 $ at $w_{max} = -
0.333$ and $w_{max} = 0.500$, respectively.] At $w=-0.45$ and
$w=0.80$ we predict the larger negative values $\Delta {\cal F}_c
= -0.069$ and $\Delta {\cal F}_c = -0.079$, respectively (not
shown in Fig. 7).

The corresponding anisotropy effect on the Binder cumulant for
$n=1$ at $T_c$ is shown in Fig. 8 as described by the difference
$\Delta U_c ({\bf \bar A}(w))$, (\ref{7-b}) \cite{c-d2}. Figs. 7
and 8 imply that the {\it relative} anisotropy effect on the free
energy $\Delta \mathcal{F}_c ({\bf \bar A}(w))/ \mathcal{F}_c
({\bf 1})$ for $n=1$ is considerably larger than that on the
Binder cumulant. For the free energy it is predicted to be of
$O(2.5 \%) $ at the maxima which may be detectable in future MC
simulations of the three-dimensional Ising model. By contrast, the
corresponding relative effect on the Binder cumulant $\Delta U_c
({\bf \bar A}(w))/ U (0;{\bf 1})$ for $n=1$ is predicted to be
only of $O(0.6 \%) $ at the maxima. It becomes quite large,
however, in the regime $w < -0.45$, as shown in Fig. 1 of
\cite{cd2004}, and in the regime $w > 0.8$. Previous MC
simulations \cite{schulte} of the anisotropic three-dimensional
Ising model in the range $-0.48\leq w \leq 0$ are in disagreement
with our results for the Binder cumulant ( see also
\cite{stauffer}).

\begin{figure}[!h]
\includegraphics[clip,width=80mm]{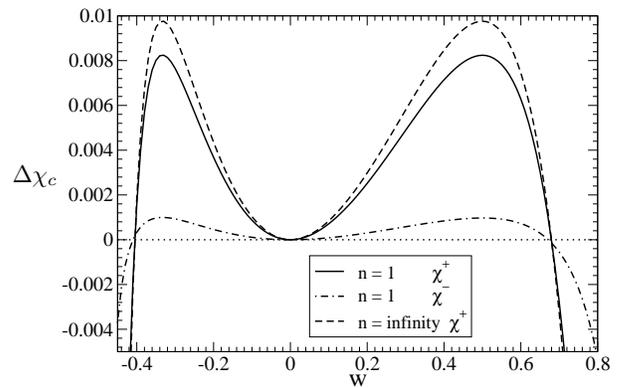}
\caption{Relative anisotropy effect on the susceptibilities
$\chi^+$ and $\chi^-$ for $n=1$, (\ref{8o}), and $\chi^+_\infty$
for $n=\infty$, (\ref{6-39}),  at $T_c$ of anisotropic systems
with the reduced anisotropy matrix ${\bf \bar A}(w)$, (\ref{32a}),
in three dimensions in a cubic geometry  as a function of the
anisotropy parameter $w$, (\ref{32b}), as described by $\Delta
\chi^\pm_c ({\bf \bar A}(w))$, (\ref{7-a}) (solid and dot-dashed
lines), and $\Delta \chi^+_{\infty,c}({\bf \bar A}(w))$,
(\ref{6-42})  (dashed line). }
\end{figure}

In Fig. 9 we also show the predicted relative anisotropy effect on
the susceptibilities $\chi^+$ and $\chi^-$ for $n=1$ and on
$\chi^+_\infty$ for $n=\infty$ at $T_c$ in the same range of $w$.
While this effect is of $O(1 \%) $ near the maxima of the
susceptibilities $\chi^+$ and $\chi^+_\infty$  the corresponding
effect on the susceptibility $\chi^-$ is only of $O(0.1 \%) $.
Previous MC simulations \cite{schulte,stauffer} on $\chi^-$ did
not resolve this small anisotropy effect.

Our prediction of the anisotropy effect on the finite-size scaling
function ${\cal F}^{ex} (\tilde x; {\bf \bar A}(w))$ for $n=1$
near the minimum below $T_c$ is shown in Fig. 10 for several $w$.
While the position of the minimum $\tilde x_{min}$ depends only
weakly on the anisotropy the value ${\cal F}^{ex} (\tilde x_{min};
{\bf 1})$ is significantly changed relative to the isotropic case
(dotted curve in Fig. 10). This effect is well outside the error
bars of the MC data by  Mon  for the isotropic case
\cite{mon-1,mon-2} and may be be detectable in future MC
simulations.

\begin{figure}[!h]
\includegraphics[clip,width=80mm]{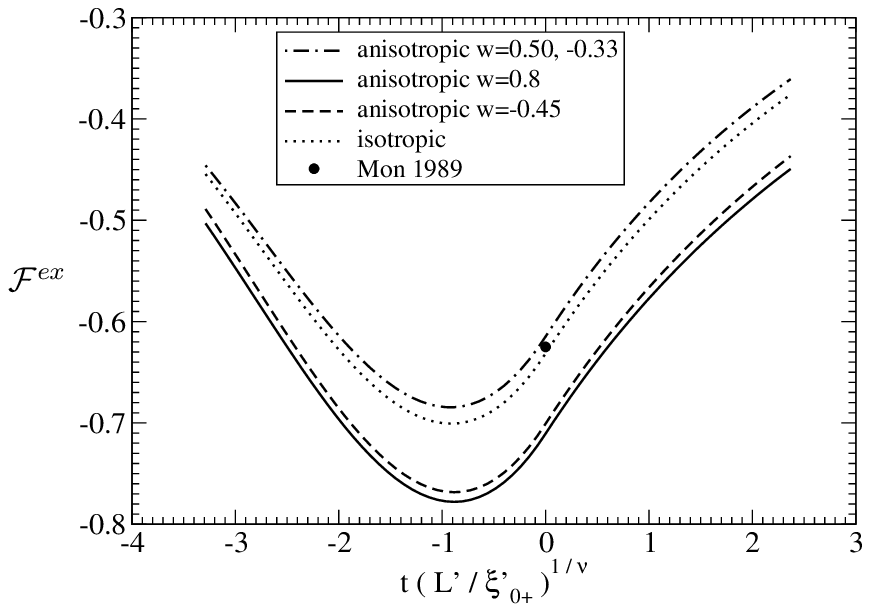}
\caption{Scaling function ${\cal F}^{ex} (\tilde x; {\bf \bar
A}(w))$, (\ref{6t}), (\ref{6l}), (\ref{6s})  of the excess free
energy density of anisotropic systems with the reduced anisotropy
matrix ${\bf \bar A}(w)$, (\ref{32a}), for $n=1$ in a cubic
geometry in three dimensions as a function of the scaling variable
$\tilde x = t (L'/\xi'_{0+})^{1/\nu}$  for several values of the
anisotropy parameter $w$, (\ref{32b}): $w = 0.50, -0.33$
(dot-dashed line), $w=0.80$ (solid line), $w= -0.45$ (dashed
line), $w=0$ (dotted line). MC result (full circle) for the
three-dimensional Ising model on a sc lattice \cite{mon-2}. }
\end{figure}

\subsection{Two-dimensional anisotropy}

Highly precise numerical information on the nonuniversal
anisotropy effect on the critical Binder cumulant $U$ of the
two-dimensional Ising model has been provided recently by MC
simulations of Selke and Shchur \cite{selke2005}. They considered
finite square lattices with isotropic ferromagnetic NN couplings
$K_x=K_y \equiv K > 0$ and an anisotropic NNN coupling J only in
the $\pm(1,1)$ directions but not in the $\pm(-1,1)$ directions
(Fig. 11 (a) ). They found a non-monotonic dependence of $U$ on
the ratio $J/K$ (as shown in Fig. 4 of Ref. \cite{selke2005}).

The anisotropy matrix of the corresponding two-dimensional
$\varphi^4$ lattice model is \cite{dohm2006}
\begin{equation}
 \label{33}
 {\bf A}_2  = \; 2 \tilde a^2 \left(\begin{array}{ccc}
  K+J & J  \\
  J & K+J  \\
\end{array}\right) \;
\end{equation}
with the reduced anisotropy matrix
\begin{equation}
  \label{33a}
 {\bf \bar A}_2(s) = {\bf A}_2/ (\det {\bf A}_2)^{1/2}  = \;(1-s^2)^{-1/2} \left(\begin{array}{ccc}
  1 & s  \\
  s & 1  \\
\end{array}\right)
\end{equation}
and with the single anisotropy parameter
\begin{equation}
\label{33b} s \; = \; \frac{J}{K \; + \; J} = (1 + K/J)^{-1}\; .
\end{equation}

By universality it is expected \cite{dohm2006} that (in some range
of ${\bf  \bar A}_2$ near ${\bf 1}$ and for sufficiently large
$L$) the two-dimensional $\varphi^4$ model has the same anisotropy
effects at $T_c$ as the two-dimensional Ising model if both models
have the same reduced anisotropy matrix ${\bf \bar A}_2$ .
Unfortunately, at the present time, it is not known how to perform
quantitative finite-size calculations for the $\varphi^4$ model in
{\it two} dimensions.

It is possible, however, to incorporate a two-dimensional
anisotropy of the type shown in Fig. 11 (a) in a three-dimensional
$\varphi^4$ (or Ising) model on a simple-cubic lattice with
isotropic NN couplings $K_x=K_y=K >0$, with an anisotropic NNN
coupling $J_1 \equiv J\neq 0$ in the $x-y$ planes, and with an
additional NN coupling $K_0> 0$ in the $z$ direction (Fig. 11
(b)). The corresponding anisotropy matrix is
\begin{equation}
 \label{33c}
 {\bf A}_3  = \; 2 \tilde a^2 \left(\begin{array}{ccc}
  K+J & J & 0 \\
  J & K+J & 0 \\
  0 & 0 & K_0 \\
\end{array}\right) \; .
\end{equation}
The eigenvalues and eigenvectors  are $ \lambda_1  =  2 \tilde a^2
(K + 2 J)$, $ \lambda_2  =  2 \tilde a^2 K $, $\lambda_3  =
2\tilde a^2 K_0 $ ,
\begin{equation}
 \label{33h}
 {\bf e}^{(1)} = \frac{1}{\sqrt{2}}\left(\begin{array}{c}
  1 \\
  1 \\
  0 \\
\end{array}\right) \; , {\bf e}^{(2)} = \frac{1}{\sqrt{2}}\left(\begin{array}{c}
  -1 \\
  1 \\
  0 \\
\end{array}\right) \; ,{\bf e}^{(3)} = \left(\begin{array}{c}
  0 \\
  0 \\
  1 \\
\end{array}\right) \; .
\end{equation}
The eigenvalues are positive in the range  $ -\frac{1}{2} < J/K <
\infty$, $K>0$, $K_0>0$. In the limit $J/K \rightarrow \infty$ and
$J/K_0 \rightarrow \infty$ (or $K \to 0_+$ and $K_0 \to 0_+$ at finite
$J > 0$) the model represents a system of variables $\varphi_i$ on
decoupled one-dimensional chains with ferromagnetic NN
interactions.

\begin{figure}[!h]
\includegraphics[clip,width=70mm]{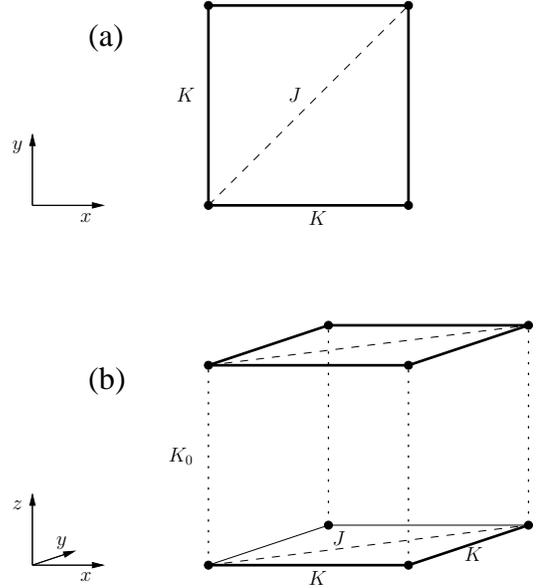}
\caption{Lattice points of the primitive cell of $(a)$ a square
lattice, $(b)$ a simple-cubic lattice, with isotropic NN couplings
$ K_x = K_y = K$ (solid lines),  with an anisotropic NNN coupling
$J$ in the $x-y$ planes (dashed lines), and a NN coupling $K_0$ in
the $z$-direction (dotted lines). The corresponding anisotropy
matrices ${\bf  A}_2$, ${\bf \bar A_2}$, and ${\bf  A}_3$, ${\bf
\bar A_3}$, are given by (\ref{33}), (\ref{33a}), and (\ref{33c}),
(\ref{33d}), respectively. }
\end{figure}

In its present form the matrix ${\bf A}_3$, (\ref{33c}), contains
both a two-dimensional anisotropy due to $J$ and  an additional
anisotropy due to the coupling $K_0$. Although ${\bf A}_3$
contains ${\bf A}_2$  as a decoupled $2 \times 2$ submatrix  it is
not expected that, for general fixed $K_0$, this three-dimensional
model exhibits the same type of anisotropy effect (as a function
of the ratio $J/K$) as the two-dimensional model with the  matrix
(\ref{33}). The reason is that it is not ${\bf A}_3$ itself but
rather the {\it reduced} anisotropy matrix ${\bf \bar A}_3 = {\bf
A}_3/ (\det {\bf A}_3)^{1/3}$ that governs the anisotropy effect
according to the results of the preceding sections. This matrix is
given by
\begin{equation}
 \label{33d}
 {\bf \bar A}_3  = \;[\overline K_0(1-s^2)]^{-1/3}\;  \left(\begin{array}{ccc}
  1 & s & 0 \\
  s & 1 & 0 \\
  0 & 0 & \overline K_0 \\
\end{array}\right) ,\;
\end{equation}
\begin{equation}
 \label{33e} \overline K_0 = \frac{K_0}{(K + J)} = \frac{K_0}{K} (1-s).
\end{equation}
We see that the $J$ dependence of  ${\bf \bar A}_3$ differs
qualitatively from that of  ${\bf \bar A}_2$ because of the
additional $s$ dependence of $\overline K_0 $ at given $K_0 / K >
0$. What is needed is a kind of {\it isotropic extension} of the
two-dimensional matrix (\ref{33a}) to three dimensions parallel to
the proposed four-dimensional extension (\ref{6-34}) of the
three-dimensional matrix ${\bf  \bar A}$, (\ref{6-33}). This is
achieved by the choice $\overline K_0 = 1$ or $K_0 = K + J $. Then
the reduced anisotropy matrix becomes
\begin{equation}
 \label{33j}
 {\bf \bar A}_3(s)  = \;(1-s^2)^{-1/3} \left(\begin{array}{ccc}
  1 & s & 0 \\
  s & 1 & 0 \\
  0 & 0 & 1 \\
\end{array}\right)
\end{equation}
with the same anisotropy parameter $s$ as in (\ref{33a}),
(\ref{33b}). [Naively one might have expected that the choice of
the additional NN coupling should have been $K_0 = K $ instead of
$K_0 = K+J$ . But in this case the third diagonal element of ${\bf
\bar A}_3$ would become ${(\bar A_3)_{zz}}= (1-s^2)^{- 1/3}
(1-s)^{2/3}$ rather than $(1-s^2)^{- 1/3}$ which would produce a
qualitatively different anisotropy effect that is not an even
function of $s$.]

We have evaluated numerically the expressions (\ref{7-d}),
(\ref{4f1}), (\ref{6-4}) in three dimensions for the Binder
cumulant $U(0;{\bf \bar A}_3(s))$ at $T_c$ using the matrix ${\bf
\bar A}_3(s)$, (\ref{33j}). The result is shown in Fig. 12 in the
range $-0.8<s<0.8$. The range $ 0 \leq s \leq 0.8$ corresponds to
the range $0 \leq J/K \leq 4.0 $ studied by Selke and Shchur
\cite{selke2005}. The anisotropy effect shows up as a
non-monotonic dependence on $s$. It is an even function of the
anisotropy parameter $s$ and exhibits two maxima of equal height
at $s_{max}=\pm 0.461$ corresponding to $J/K=0.855$ and
$J/K=-0.316$. This symmetry is a consequence of the symmetry
property $K_3(y,{\bf \bar A}_3(s)) = K_3(y,{\bf \bar A}_3(-s))$ of
the function (\ref{4ii}). The symmetry is hidden if $U$ is plotted
as a function of $J/K$ in which case the curve is asymmetrically
distorted (see our Fig. 13 and Fig. 4 of \cite{selke2005}). Our
theoretical value $U(0;{\bf 1})=U(0;{\bf \bar A}_3(0)) = 0.417$
for the isotropic three-dimensional $\varphi^4$ theory \cite{EDC}
differs somewhat from the MC result \cite{bloete} 0.465 of the
three-dimensional Ising model on a $sc$ lattice and is, of course,
far from the MC result 0.6107 \cite{kam} of the {\it
two-dimensional} Ising model on a square lattice. The magnitude of
the anisotropy effect, however, turns out to be rather insensitive
to the precise value of $U(0;{\bf 1})$ of the isotropic system.

To exhibit clearly the {\it deviations} from isotropy and for the
purpose of a comparison with the MC data \cite{selke2005} for the
anisotropic two-dimensional Ising model we have plotted in Fig. 13
our theoretical result for the {\it difference} $\Delta U_c({\bf
\bar A}_3(s))$, (\ref{7-b}), as a function of $J/K$ together with
the corresponding difference of the MC data of Fig. 4 by Selke and
Shchur \cite{selke2005}. The theoretical maximal value is $\Delta
U_{c, max} = 0.0010$ at $J/K = 0.855$ and $J/K = - 0.316$. The
isotropic value, i.e. $\Delta U_c = 0$, is found at $s=0$ and at
$s=\pm 0.6169$ corresponding to $J/K = 1.611$ and $J/K = -
0.3815$. We see that for positive $J/K$ there is remarkable
agreement between the MC data in two dimensions and the
anisotropic $\varphi^4$ theory in three dimensions, thus
confirming our expectation regarding the similarity of the
anisotropy effect in the two- and three-dimensional models. It
should be noted, of course, that no {\it exact} agreement can be
expected. Only the anisotropic {\it two-dimensional} $\varphi^4$
theory (with $n=1$) is expected to yield exactly the same
anisotropy effects (in the asymptotic critical region) as the
two-dimensional Ising model.

\begin{figure}[!h]
\includegraphics[clip,width=80mm]{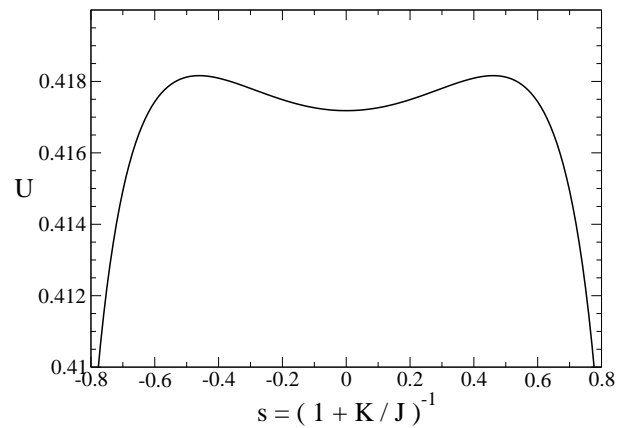}
\caption{ Binder cumulant $U(0;{\bf \bar A}_3(s))$ , (\ref{7-d}),
(\ref{4f1}), (\ref{6-4}), for anisotropic systems with the reduced
anisotropy matrix ${\bf \bar A}_3(s)$, (\ref{33j}), in a cubic
geometry at $T_c$ for $n=1$ in three dimensions  as a function of
the anisotropy parameter $s$, (\ref{33b}). }
\end{figure}

The non-monotonicity for small {\it negative} values of $J/K$ and
the maximum at $J/K = -0.316$ predicted by our theory was not
detected in the preliminary MC simulations by Selke and Shchur
\cite{selke2005} who found a {\it monotonic decrease} of $U$ when
taking a weak antiferromagnetic coupling $J$ \cite{info-1}. It
would be interesting to perform more detailed MC simulations for
the anisotropic two-dimensional Ising model in the regime of
negative $J/K$.

Very recently such MC simulations have been started by Selke
\cite{selke-neu} in order to test our prediction for the Binder
cumulant in the regime of negative values of $J/K$. The {\it
positive} value of his MC result \cite{selke-neu} $\Delta U_c^{MC}
= 0.00056 \pm 0.00015$ for $J/K = - 0.25$ indeed confirms the
predicted {\it increase} of $\Delta U_c$ for small negative $J/K$
according to Fig. 13. More quantitively, there is indeed
reasonable agreement with our theoretical result $\Delta U_c =
0.00073$ at $J/K = - 0.25$ (as shown in Fig. 13) corresponding to
$s = - 1/3$. It remains to be seen whether the predicted symmetry
with regard to $s$ is also confirmed by MC simulations.

\begin{figure}[!h]
\includegraphics[clip,width=80mm]{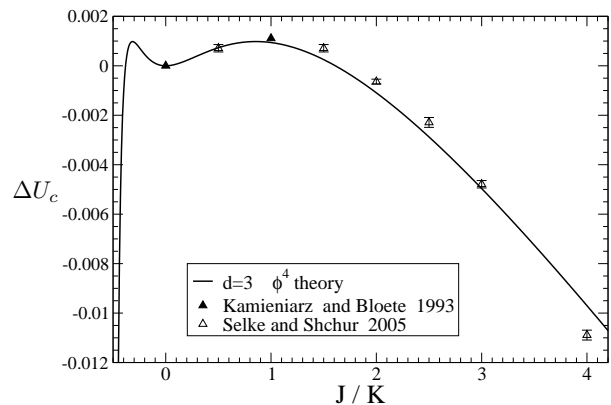}
\caption{Difference $\Delta U_c({\bf \bar A}_3(s))$, (\ref{7-b}),
of the Binder cumulant $U(0;{\bf \bar A}_3(s))$ shown in Fig. 12
but plotted as a function of $J/K$, together with the
corresponding difference of the MC data of Fig. 4 of
\cite{selke2005} for the two-dimensional Ising model. }
\end{figure}

We also apply our general result (\ref{6-1}) for the finite-size
amplitude of the free energy density at $T_c$ to the case of the
two-dimensional anisotropy determined by the matrix (\ref{33j}).
The anisotropy effect as described by the difference $\Delta {\cal
F}_c ({\bf \bar A}_3 (s))$ is shown in Fig. 14 for $n=1$ (solid
curve). The curve is an even function of $s$ and has two maxima of
equal height at $s_{max} = \pm 0.450$ corresponding to $J/K =
0.818$ and $J/K = - 0.310$. The theoretical maximal value is
$\Delta {\cal F}_{c, max} = 0.0060$ for $n = 1$. The corresponding
effect for $n=\infty$ (dashed curve) as computed from the exact
result (\ref{8h}) is slightly more pronounced than for $n=1$.
Nevertheless the anisotropy effect for $n=1$ may be detectable by
MC calculations for both the three-dimensional and two-dimensional
anisotropic Ising models. The two-dimensional model is of course a
better candidate, as noted by Selke and Shchur \cite{selke2005},
because the value of $T_c$ is known analytically as a function of
$J/K$. Although the solid curve in Fig. 14 is calculated on the
basis of the $\varphi^4$ theory in {\it three} dimensions with the
reduced anisotropy matrix (\ref{33j}) we predict that this curve
should be close to the difference $\Delta {\cal F}_c({\bf \bar
A}_2(s) )$ of the free energy density of the {\it two-dimensional}
Ising model with the reduced anisotropy matrix (\ref{33a}). It
would be interesting to test this prediction by MC simulations.

For comparison with Fig. 10 we have also computed the nonuniversal
anisotropy effect on the finite-size scaling function of ${\cal
F}^{ex}$ for the reduced anisotropy matrix (\ref{33j}) near the
minimum below $T_c$ as shown in Fig. 15 for several values of $s$.
Again this effect for $s=\pm 0.80$ is well outside the error bars
of the MC data by Mon \cite{mon-1,mon-2} for the isotropic case
and may be be detectable in future MC simulations.

\begin{figure}[!h]
\includegraphics[clip,width=80mm]{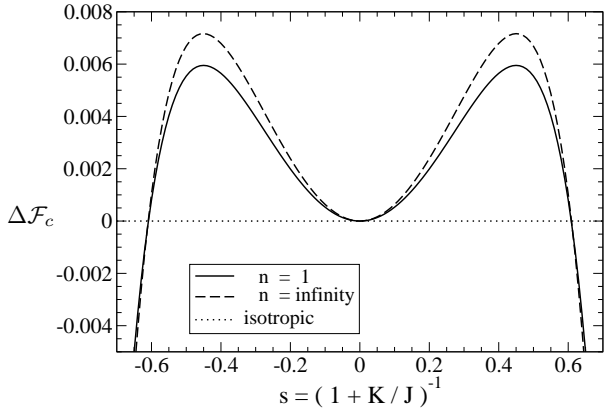}
\caption{Differences $\Delta \mathcal{F}_c ({\bf \bar A}_3(s))$,
(\ref{8-14}), and $\Delta \mathcal{F}_{c,\infty} ({\bf \bar
A}_3(s))$, (\ref{8-15}), of the finite-size amplitudes (\ref{6-1})
and (\ref{8h}) of the free energy density of anisotropic systems
with the reduced anisotropy matrix ${\bf \bar A}_3(s)$,
(\ref{33j}), in a cubic geometry at $T_c$ for $n=1$ (solid line)
and $n=\infty$ (dashed line) in three dimensions  as a function of
the anisotropy parameter $s$, (\ref{33b}). }
\end{figure}
\begin{figure}[!h]
\includegraphics[clip,width=80mm]{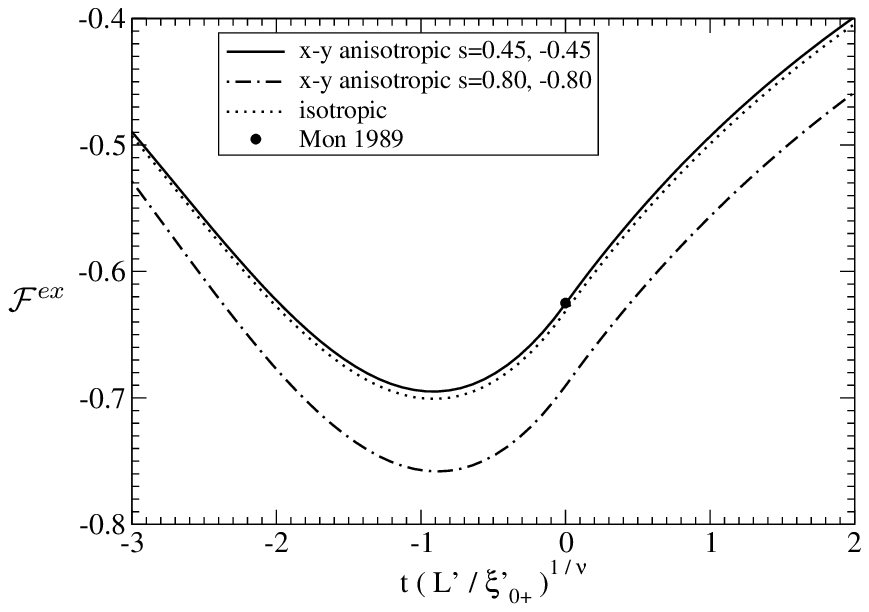}
\caption{Scaling function ${\cal F}^{ex} (\tilde x; {\bf \bar
A}_3(s))$, (\ref{6t}), (\ref{6l}), (\ref{6s})  of the excess free
energy density of anisotropic systems with the reduced anisotropy
matrix ${\bf \bar A}_3(s)$, (\ref{33j}), for $n=1$ in a cubic
geometry in three dimensions as a function of the scaling variable
$\tilde x = t (L'/\xi'_{0+})^{1/\nu}$  for several values of the
anisotropy parameter $s$, (\ref{33b}): $s = 0.45, -0.45$ (solid
line), $s=0.80, -0.80$ (dot-dashed line), $s=0$ (dotted line). MC
result (full circle) for the three-dimensional Ising model on a sc
lattice \cite{mon-2}. }
\end{figure}

\subsection{Limit $\bf{|s| \to 1}$ and Lifshitz point}

In Subsections C and D we have assumed the positivity of all
eigenvalues $\lambda_\alpha, \alpha = 1,2,3$. They vanish in the
limits $w \to 1, w \to - 1/2$ and $s \to \pm 1$ in which cases our
results for $\Delta U_c$ and $\Delta {\cal F}_c$ are not
applicable: they become singular as indicated by the curves in
Figs. 7 - 9 and 12 - 14. In the following we confine ourselves to
a brief discussion of the limit $s \to \pm 1$ of the model shown
in Fig. 11. (A similar discussion could be given for the model
shown in Fig. 3 in the limits $w \to 1, w \to - 1/2$.)

(i) According to (\ref{33b}), the limit $s \to 1$ can be performed
as  $K \to 0_+$ at fixed $J > 0$ and $K_0 > 0$ (keeping
$\lambda_1>0$ and  $\lambda_3>0$ positive while $\lambda_2 \to 0_+$). In this limit our
model is reduced to decoupled {\it two-dimensional} lattices which
have ferromagnetic NN couplings $J$ and $K_0$ in the $\pm (1,1,0)$
directions and in the $\pm (0,0,1)$ directions, respectively. Such
a model has a ferromagnetic critical point of the $(d = 2, n=1)$
universality class. Therefore it is expected at the outset that
the results of our $\varphi^4$ theory at fixed $d = 3$ must break
down for $s \to 1$.

(ii) To discuss the limit $s \to - 1$ we first perform a rotation
in wave-vector space, ${\bf q} = {\bf U}_3 {\bf k}$, by means of
the orthogonal matrix determined by the eigenvectors (\ref{33h}),

\begin{equation}
 \label{33k}
 {\bf  U}_3  = \; \frac{1}{\sqrt 2}\left(\begin{array}{ccc}
  1 & 1 & 0 \\
  -1 & 1 & 0 \\
  0 & 0 & \sqrt 2 \\
\end{array}\right) \;.
\end{equation}
Correspondingly the inverse propagator $r_0 + \delta \widehat K
({\bf k})$ of the Hamiltonian (\ref{2g}) is transformed to $r_0 +
\delta \widetilde K ({\bf q})$, where $\delta \widetilde K ({\bf
q}) \equiv \delta \widehat K ({\bf U}_3^{-1} {\bf q})$ with the
interaction part
\begin{eqnarray}
\label{33l} \delta \widetilde K ({\bf q}) \;=\; \sum_{\alpha=1}^3
\lambda_\alpha q_\alpha^2 \; + \; \sum^d_{\alpha,
\beta, \gamma, \delta} \widetilde B_{\alpha \beta \gamma \delta} \; q_\alpha
q_\beta q_\gamma q_\delta \;+ \; O (q^6) \nonumber\\
\end{eqnarray}
where $\lambda_\alpha$ is given after (\ref{33c}).

On the level of Landau theory a Lifshitz point exists at
$\lambda_1 = 0$ corresponding to $s = - 1$ or $J/K = - 1/2$ with a
wave-vector instability in the $(1,1,0)$ direction. It is
expected, however, that, due to fluctuations, the Lifshitz point
occurs at a shifted value $\lambda_1 = \lambda_{1 LP}$ that
depends on nonuniversal details of the model. [Similarly, $r_{0c}$
depends on all details of the interaction, see (\ref{4zz}).] This
corresponds to a shifted coupling ratio $(J/K)_{LP} = - 1/2 +
\lambda_{1 LP} / (2 \tilde a^2)$, presumably with $\lambda_{1 LP}
< 0$. To describe the critical behavior near the Lifshitz point
would require to introduce a renormalized shifted eigenvalue
according to $\lambda_R = Z_\lambda (\lambda_1 - \lambda_{1 LP} )$
\cite{diehl-2}. It would be interesting to locate this Lifshitz
point by MC simulations and to detect the nonuniversal change of
finite-size effects [critical Binder cumulant and free energy at
$T_c(J/K)$] upon approaching this point along the "$\lambda$-line"
$T=T_c (J/K)$ as $J/K \to (J/K)_{LP}$. This change can be compared
with our predictions shown in the curves of Figs. 12 - 14 for
negative $J/K$ and negative $s$.

\setcounter{equation}{0}
\section{Hypothesis of restricted universality }

The results for the finite-size scaling functions (\ref{6l}) -
(\ref{6-7}), (\ref{6t}), (\ref{8b}), (\ref{8p}), and (\ref{7-d})
depend on the nonuniversal anisotropy matrix ${\bf \bar A}$ but
are independent of the bare coupling $u_0$, of the lattice spacing
$\tilde a$, of the cutoff of $\varphi^4$ field theory, and of the
fourth-order moments $B_{\alpha
\beta \gamma \delta}$ etc.. We anticipate that the finite-size scaling functions would also
remain independent of higher-order couplings, such as those of
$\varphi^6$ terms and of higher-order gradient terms etc., if they
were included in the Hamiltonian.  A special matrix ${\bf \bar A}$
with given matrix elements $\bar A_{\alpha\beta}$ can be obtained
from various different lattice structures with a large variety of
different couplings, both in $O(n)$ symmetric $\varphi^4$ lattice
models and in  $O(n)$ symmetric fixed-length spin models. We
expect that ${\cal F} (\tilde x, {\bf \bar A})$ and $ U(\tilde x,
{\bf \bar A})$ are  the same for all those systems whose geometry
and boundary conditions are the same and whose reduced anisotropy
matrix ${\bf \bar A}$ is the same. We consider this feature as a
kind of {\it restricted universality within a $(d, n)$
universality class}. A nontrivial aspect of this feature is that
it is governed by the {\it bare} anisotropy matrix ${\bf \bar A}$
containing the {\it unrenormalized microscopic} couplings. Our
approximate results do not yet provide a rigorous proof for the
validity of this hypothesis. It would be interesting to test this
hypothesis by MC simulations for microscopic spin models with such
anisotropy matrices.

For concreteness consider the three-dimensional anisotropic model
(i) of Sect. II A with the three couplings $K$, $J$, and
$\overline K$ as described by the matrix
\begin{equation}
 \label{9-a}
 {\bf  A} = 2\tilde a^2 \left(\begin{array}{ccc}
  D & J+\overline K & J+\overline K \\
  J+\overline K & D & J+\overline K \\
  J+\overline K & J+\overline K & D \\
\end{array}\right) \; ,
\end{equation}
with $D= K + 2J + \overline K$ and with the reduced matrix ${\bf
\bar A}$, (\ref{32a}). For a given fixed value of the anisotropy
parameter $w$, (\ref{32b}), a family of anisotropic spin models
with different couplings $K, J$, and $\overline K$ are predicted
to have the same finite-size scaling functions if the third-NN
coupling $\overline K$ is chosen as
\begin{eqnarray}
\label{9aa}\overline K =\frac{1}{1-w}\big[w K - (1- 2w) J\big]
\end{eqnarray}
in the range where $\lambda_\alpha > 0$. Equation (\ref{9aa})
represents a surface in the space of the three couplings $K, J$,
and $\overline K$. At $\overline K=0$, this surface becomes a
"$\lambda$ - line"
\begin{eqnarray}
\label{9bb} J = \frac{w}{1-2w}K
\end{eqnarray}
along which, at a given fixed value of $w$, $-1/2<w<1/2$, all
finite-size scaling functions of models with the anisotropy matrix
(\ref{9-a}) are predicted to remain unchanged when changing $K$
and $J$ simultaneously according to (\ref{9bb}) (in the range
$K+J>0, K+4J>0$).

A non-trivial test of our hypothesis applied to {\it two}
dimensions can  be performed for the following example. Consider a
triangular-lattice model with a shape of a rhombus and with three
NN couplings $K_1$, $K_2$, $K_3$, and a NNN coupling $J$ only in
the $\pm (3/2, \sqrt3/2)$ directions (Fig. 16). The anisotropy
matrix of this system with lattice constant $\tilde a = 1 $ is
\begin{eqnarray}
\label{9cc} {\bf A} \; &= \; \frac{1}{2} \left(\begin{array}{cc}
  4 K_1+K_2+K_3+9J \;\; & \;\; \sqrt{3}\; (K_2-K_3+3J) \\ \\
  \sqrt{3} \;(K_2-K_3+3J) \; \;& \;\; 3 (K_2+K_3+J)
\end{array}\right). \nonumber\\
\end{eqnarray}
In the absence of the NNN coupling $J$, isotropy is possible only
for the symmetric case $K_1=K_2=K_3$. In this case the critical
Binder cumulant for the ($n=1,d=2$) universality class for
periodic boundary conditions is  known to very high accuracy
\cite{kam}: $U = 0.6118277 \pm 0.0000001$. Apart from this case,
the system can become isotropic even for $K_3\neq K\equiv K_1=K_2$
if the NNN coupling $J$ is chosen as
\begin{eqnarray}
\label{9cx} J \;=\;\frac{1}{3} \;(K_3-K)
\end{eqnarray}
according to (\ref{9cc}) in which case
\begin{eqnarray}
\label{9dd} {\bf A} \; &= \; \frac{1}{2} \left(\begin{array}{cc}
  2 K + 4 K_3 \;\; & \;\;  0 \\ \\
  0 \; \;& \;\; 2 K + 4 K_3)
\end{array}\right). \nonumber\\
\end{eqnarray}
and ${\bf \bar A} = {\bf 1}$. Then, on the basis of our hypothesis
of restricted universality, the critical Binder cumulant for $K_3
\neq K, J \neq 0$ is predicted to have exactly the same value as
found by Kamieniarz and Bl\"{o}te \cite{kam} for $J=0, K_3=K$. A
corresponding prediction should hold also for other boundary
conditions (e.g., free boundary conditions).

\begin{figure}[!h]
\includegraphics[width=70mm]
{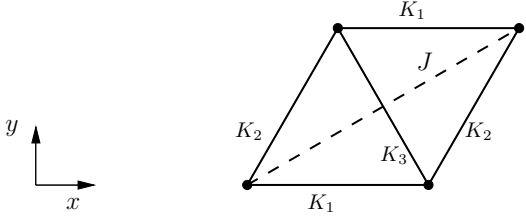} \caption{Lattice points of a triangular
lattice with the shape of a rhombus. The solid lines indicate the
anisotropic NN couplings $ K_1$, $K_2$, $K_3$, the dashed line
indicates the anisotropic NNN coupling $J$. The anisotropy matrix
${\bf  A}$  is given by (\ref{9cc}).
 }
\end{figure}

\setcounter{equation}{0}
\section{Finite-size effects outside the central finite-size regime}

The result ${\cal F}^{ex}(\tilde x; {\bf \bar A})$,  (\ref{6t}),
(\ref{6l}), needs to be complemented outside the central
finite-size regime where this result does not have the correct
exponential structure for large $|\tilde x|$. This is the regime
below the dashed lines in Fig. 1. It turns out that it is
necessary to further distinguish between a scaling and a
nonscaling regime (the latter is the shaded region in Fig. 1).
Both regimes belong to the asymptotic critical region. In these
regimes ordinary perturbation theory is appropriate. Here we
perform the corresponding analysis above and below $T_c$ at the
one-loop level. In order to distinguish the perturbation results
of this section from those of the preceding sections we explicitly
include the indices + and $-$ in the notation of ${\cal F}^{ex, +}
$, $f'^+$, ${\cal F}^{ex, -}$, and $f'^-$.

\subsection{Scaling regime}

The starting point of ordinary perturbation theory for the bare
free energy density (\ref{2s}) for $n=1$ at $h'=0$ is in one-loop
order [i.e., up to $O (1)$] (see App. B)
\begin{eqnarray}
\label{VIIa} f'^+ \; = \;f'_0(r_0, L', K_{i,j}, v') + O (u'_0)\;,
\end{eqnarray}
\begin{eqnarray}
\label{VIIb} f'^- \;= \frac{1}{2} r_0 M_{mf}'^2 + u'_0 M_{mf}'^4 +
f'_0(-2r_0, L', K_{i,j}, v')  + O(u'_0) \nonumber\\
\end{eqnarray}
above and below $T_c$, respectively, where $M_{mf}'^2$ is given by
(\ref{4n}) and
\begin{eqnarray}
\label{VIIaa} f'_0(r_0, L', K_{i,j}, v')\; = \; - \; \frac{\ln (2
\pi)} {2 v'} \nonumber\\+ \;\frac{1}{2 L'^d} \sum_{\bf k'} \ln
\{[r_0 + \delta \widehat K' (\mathbf k')] (v')^{2/d} \} .
\end{eqnarray}
Because of the ${\mathbf k'} = {\bf0}$ term, the sum exists only
for $r_0 > 0$. Rewriting these expressions in terms of $r_0 -
r_{0c}$ with $r_{0c}$ given by (\ref{4zz}) yields up to $O (1)$
\begin{eqnarray}
\label{VIIc} f'^+ \; =  f'_0(r_0 - r_{0c}, L', K_{i,j}, v'),
\end{eqnarray}
\begin{eqnarray}
\label{VIId} f'^- \;&=&\;  - \frac{1}{64 u'_0} [-2(r_0 -
r_{0c})]^2 + \frac{3}{2} (r_0 - r_{0c}) \int\limits_{{\bf k}'}
\frac{1}{\delta \widehat K'
 ({\bf k}')}\nonumber\\
&+& f'_0(-2(r_0 - r_{0c}), L', K_{i,j}, v').
\end{eqnarray}
We define the finite-size parts $\delta f'^\pm$ of $ f'^\pm$ in
the same way as $\delta f'$ in (\ref{4z5}) and (\ref{4z6}).
Calculating the sum in the continuum limit $v' \rightarrow 0$ at
fixed $|r_0 - r_{0c}| \neq 0$ (see App. B and C) one obtains for
$2<d<4$
\begin{eqnarray}
\label{VIIe}&&\delta f'^+(r_0 - r_{0c}, u'_0, L', {\bf \bar A}) =
\; - \frac{A_d}{d\varepsilon}\; (r_0 - r_{0c})^{d/2} \;
 \nonumber\\ && \;+ \; \frac{1}{2L'^d} \;
 {\cal G}_0((r_0
- r_{0c}) L'^2; {\bf \bar A}) + O (u_0') \;,
\end{eqnarray}
\begin{eqnarray}
\label{VIIf}&&\delta f'^-(r_0 - r_{0c}, u'_0, L', {\bf \bar A}) =
- \frac{1}{64 u'_0} [-2(r_0 - r_{0c})]^2 \;\nonumber\\ && -
\frac{A_d}{d\varepsilon} \;[-2 (r_0 - r_{0c})]^{d/2} \;+ \;
\frac{1}{2L'^d} \;
 {\cal G}_0(-2 (r_0
- r_{0c}) L'^2; {\bf \bar A}) \nonumber\\ && + \; O(u_0')
\end{eqnarray}
where ${\cal G}_0$ is given by (\ref{8d}). Eqs. (\ref{VIIe}) and
(\ref{VIIf}) correspond to (\ref{4aa}). The renormalized
counterparts $ f'^\pm_R$ of $\delta f'^\pm$ are defined in the
same way as $ f'_R$ in (\ref{5d}), (\ref{5aa}). The explicit form
of the functions $f'^\pm_R$ depends on the choice of the flow
parameter. For the application to the regime $|\tilde x| \gg 1$,
we make the bulk choice $\mu^2 l_+^2 = r(l_+)$ for $T > T_c$  and
$ \mu^2 l_-^2 = -2r(l_-)$ for $T < T_c$ , with $\mu^{-1} =
\xi'_{0+}$, (\ref{5y}) (the same choice will be made for the
calculations in subsection B below). Then the functions $f'^\pm_R$
are given by
\begin{eqnarray}
\label{VIIi} && f'^+_R \big(r(l_+),u'(l_+),l_+\mu,L',{\bf \bar
A}\big) = \; -\; A_d (l_+\mu)^d / (4d) \nonumber\\ &&+ \frac{1}{2
L'^d} {\cal G}_0(l_+^2 \mu^2 L'^2;{\bf \bar A}) + O (u'(l_+)) \; ,
\end{eqnarray}
\begin{eqnarray}
\label{VIIj} && f'^-_R \big(r(l_-),u'(l_-),l_-\mu,L',{\bf \bar
A}\big) = \nonumber\\ &&  -\; A_d (l_-\mu)^d \Big\{\frac{1}{64
u'(l_-)} + \frac{1}{4d} \Big\} + \frac{1}{2 L'^d} {\cal G}_0(l_-^2
\mu^2 L'^2;{\bf \bar A})\;\nonumber\\ && + \;O(u'(l_-)) \; .
\end{eqnarray}
For $l_\pm \to 0 $, this leads to the finite-size scaling function
of the excess free energy density in one-loop order in the limit
of zero lattice spacing
\begin{eqnarray}
\label{VIIk}
 {\cal F}^{ex, \pm}_{1-loop}(\tilde x; {\bf \bar A}) = \frac{1}{2}
 \; {\cal G}_0 (L'^2 / \xi'^2_\pm ; {\bf \bar A}) \;
 \; + \; O (u^*)\;  ,
\end{eqnarray}
where $\xi'_+ = \xi'_{0 +} t^{- \nu}$ and
\begin{equation}
\label{VIIp} \xi'_- = \xi'_{0 -} | t |^{- \nu},\; \xi'_{0 - } /
\xi'_{0 +} = 2^{- \nu} + O (u^*)
\end{equation}
are the bulk second-moment correlation lengths above and below
$T_c$, respectively (for $\xi'_{0 +}$ see (\ref{5y})). Here we
have confined ourselves to the simplest form of perturbation
theory for the regime $|\tilde x| \gg 1, \tilde x = t (L' /
\xi'_{0+})^{1/\nu} $. As a shortcoming of this approach, ${\cal
F}^{ex, \pm}_{1-loop}$ diverges for $\tilde x \to 0$ at fixed
finite $L'$ which originates from the ${\bf k}' = 0$ term of
(\ref{VIIaa}). [This divergence could formally be suppressed by an
$L'$ dependent choice of $l_\pm$ but this would not avoid a
structurally incorrect nonanalytic $t$ dependence at $t = 0$ for
finite $L'$.]

${\cal F}^{ex, +}_{1-loop}(\tilde x; {\bf \bar A})$ serves the
purpose of complementing ${\cal F}^{ex}(\tilde x; {\bf \bar A})$,
(\ref{6t}), (\ref{6l}) in the large  $\tilde x$ regime. This is
illustrated by the thin solid line in Fig. 17 for the example of
three-dimensional {\it isotropic} systems (and for systems with
cubic symmetry) with ${\bf \bar A} = {\bf 1}$, $L' = L$,
$\xi'_{0+} = \xi_{0+} \; , \xi'_{0 -} = \xi_{0 -}$ . The curves
match reasonably well above $T_c$. [No perfect matching can be
expected because of the missing $O(u^*)$ terms in (\ref{VIIk}) and
because ${\cal F}^{ex}(\tilde x; {\bf 1})$ is not applicable to
the region $ \tilde x \gg 1$ where it has an {\it algebraic}
approach to a {\it finite} limit ${\cal F}^{ex}(\infty; {\bf 1}) =
-2u^* = -0.082$ for $\tilde x \rightarrow \infty$.]

By contrast, ${\cal F}^{ex, -}_{1-loop}(\tilde x; {\bf \bar A})$
and ${\cal F}^{ex}(\tilde x; {\bf \bar A})$ do not match well
below $T_c$ for two reasons: (i) The two-loop terms of $O(u^*)$ in
(\ref{VIIk}) are non-negligible, (ii) our approximate result
${\cal F}^{ex}(\tilde x; {\bf \bar A})$ as represented by
(\ref{6t}), (\ref{6l}), (\ref{6s}) is not applicable to the region
$ \tilde x < -5$. [In this region this result for ${\cal
F}^{ex}(\tilde x; {\bf 1})$  has an unphysical maximum ${\cal
F}^{ex} (\tilde x_{max}; {\bf 1}) = - 0.303$ at $\tilde x_{max} =
- 5.61$ and has an algebraic approach to a finite limit ${\cal
F}^{ex}(-\infty; {\bf 1}) = -0.49$ for $\tilde x \rightarrow
-\infty$.] Thus substantial further work is needed for a
satisfactory description of the region well below $T_c$.

Nevertheless, ${\cal F}^{ex, \pm}_{1-loop}$ has the advantage of
displaying the expected {\it exponential} large $|\tilde x|$
behavior. The leading large $\tilde x$ behavior of ${\cal G}_0$
for the isotropic case is (see App. B)
\begin{eqnarray}
\label{VIIm} {\cal G}_0 (\tilde x^{2 \nu} ; {\bf 1}) &=&-2d
\Big(\frac{\tilde x^\nu}{2\pi}\Big)^{(d-1)/2} \exp (-\tilde x^\nu
) + O(\exp (-2\tilde x^\nu )). \nonumber\\
\end{eqnarray}
For $d=3$, Eqs. (\ref{VIIm}) and (\ref{VIIk}) yield for $ |\tilde
x|^\nu\gg 1$
\begin{eqnarray}
\label{VIIn}{\cal F}^{ex, \pm}_{1-loop}(\tilde x; {\bf 1 })=
-\frac{3}{2\pi} (L/\xi_\pm) \exp (- L/\xi_\pm) + O(u^*),
\nonumber\\
\end{eqnarray}
as shown by the dashed lines in Fig. 17 (with $L/\xi_+ = \tilde
x^\nu \; , L/\xi_- = 2^\nu | \tilde x |^\nu$).

In case of noncubic anisotropy, all curves  in Fig. 17 including
the thin solid lines and dashed lines are, of course, affected by
the anisotropy matrix ${\bf \bar A} \neq 1$ in a way similar to
that shown in Figs. 10 and 15. It would be straightforward to
illustrate this effect by complementing Fig. 17 accordingly by
means of curves representing ${\cal F}^{ex} (\tilde x; {\bf \bar
A}_3(s))$ and ${\cal F}^{ex, \pm}_{1-loop}(\tilde x; {\bf \bar
A}_3(s))$, with   ${\bf \bar A}_3(s)$ given by (\ref{33j}), for
several examples of $s$. In this case the scaling argument
(horizontal axis of Fig. 17) needs to be replaced by $ t
(L'/\xi'_{0+})^{1/\nu}$.

\begin{figure}[!h]
\includegraphics[clip,width=80mm]{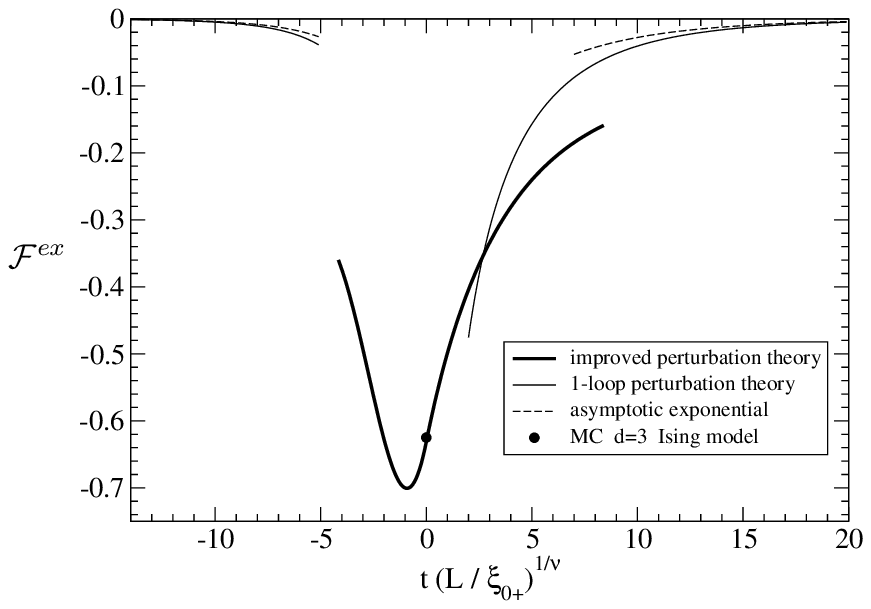}
\caption{Scaling functions ${\cal F}^{ex} (\tilde x; {\bf 1})$,
(\ref{6t}), (\ref{6l}), (\ref{6s}) for $d=3$ (thick solid line),
${\cal F}^{ex, \pm}_{1-loop} (\tilde x; {\bf 1 })$, (\ref{VIIk})
for $d=3$ (thin solid lines), and (\ref{VIIn}) (dashed lines) for
the excess free energy density of isotropic systems as a function
of the scaling variable $\tilde x = t (L/\xi_{0+})^{1/\nu}$. MC
result (full circle) for the Ising model on a sc lattice
\cite{mon-2}.  No scaling function exists in the large - $| \tilde
x |$ regions above and below $T_c$ which are sensitive to all nonuniversal details of the
model according to ${\cal F}^{ex, \pm}(L/\xi_\pm ; {\bf 1 };
\tilde a/\xi_\pm )$, (\ref{VIIpp}), with (\ref{Xh}). }
\end{figure}

\subsection{Non-scaling regime}

\subsubsection*{ 1. Anisotropic $\varphi^4$ lattice model with finite lattice constant}

So far we have taken the continuum limit which is well justified
in the range shown in Figs. 10, 15, and 17 provided that $\xi'_\pm
/ \tilde a \gg 1$ is sufficiently large. In earlier work
\cite{cd2000-1} it was pointed out for the example of the
susceptibility that the finite lattice constant $\tilde a$ becomes
non-negligible in the limit of large $L/\tilde a$ at fixed $T \neq
T_c $ in the regime where the finite-size scaling function has an
exponential form. Here we further discuss this issue in the
context of the excess free energy of the model (\ref{2a}) with
$V=L^d$ and cubic anisotropy, i.e., on a simple-cubic lattice with
lattice constant $\tilde a$ and only NN couplings $K$. In this
case we have ${\bf A} = 2 \tilde a^2 K {\bf 1}$, ${\bf \bar A} =
{\bf 1}$, $L = (2 \tilde a^2 K)^{1/2} L'$, $\tilde a = (2 \tilde
a^2 K)^{1/2} \tilde a'$, and there exist well defined bulk
second-moment correlation lengths $\xi_\pm = (2 \tilde a^2
K)^{1/2} \xi'_\pm$ above and below $T_c$ (for $n=1$). As shown in
App. B, the excess free energy density in one-loop order attains
the following form in the limit of large $L/\tilde a = L'/\tilde
a' $ at fixed arbitrary $\tilde a/\xi_\pm = \tilde a'/\xi'_\pm >
0$
\begin{eqnarray}
\label{VIIo} f^{ex, \pm}(t,L)\;\;
\mathop{\longrightarrow}_{L/\tilde a \gg 1}\;\; L^{-d} {\cal
F}^{ex, \pm} (L/\xi_\pm ; {\bf 1 }; \tilde a / \xi_\pm) \; ,
\end{eqnarray}
\begin{eqnarray}
\label{VIIpp}{\cal F}^{ex, \pm}(L/\xi_\pm ; {\bf 1 }; \tilde
a/\xi_\pm )= -d  \left[1 + \left(\frac{\tilde a}{2
\xi_\pm}\right)^2 \right]^{d-1} \nonumber\\ \times
\left(\frac{L}{2\pi \xi_\pm}\right)^{(d-1)/2} \exp \left\{-
\frac{L}{\xi_\pm} \left[\frac{2 \xi_\pm}{\tilde a} {\rm arsinh}
\left(\frac{\tilde a}{2 \xi_\pm} \right)\right] \right\} \; .
\nonumber\\
\end{eqnarray}
This result applies to the shaded region of Fig. 1. The
exponential part of (\ref{VIIpp}) can be rewritten as $\exp (- L /
\xi_{{\bf e}\pm})$ with the {\it exponential correlation lengths}
\begin{eqnarray}
\label{VIIr} \xi_{{\bf e} \pm} \; = \; \frac{\tilde a}{2}
\left[{\rm arsinh} \left(\frac{\tilde a}{2
\xi_\pm}\right)\right]^{-1}
\end{eqnarray}
above and below $T_c$, respectively. As a nontrivial relation
between bulk properties and finite-size effects \cite{cd2000-2},
the lengths $\xi_{{\bf e}\pm}$ describe the exponential part of
the {\it bulk} order-parameter correlation function \cite{fish-2}
in the large-distance limit in the direction of one of the cubic
axes at arbitrary fixed $T \neq T_c$ above and below $T_c$ (for $n
= 1$), respectively. This relation is exact in the large-$n$ limit
above $T_c$ \cite{cd2000-2}.

It has been shown \cite{cd2000-1} that, because of the exponential
structure of the finite-size part of the susceptibility, the
$\tilde a$ dependence of $\xi_{{\bf e}\pm}$ cannot be neglected
even for small $\tilde a / \xi_\pm \ll 1$ if $L/\xi_\pm > [24 \ln
2] (\xi_\pm / \tilde a)^2$ is sufficiently large (see Fig. 1 of
Ref. \cite{cd2000-1}). The same argument now applies to the
$\tilde a$ dependence of the exponential part of ${\cal F}^{ex,
\pm}(L/\xi_\pm ; {\bf 1 }; \tilde a/\xi_\pm )$, (\ref{VIIpp}).
This implies that finite-size scaling and universality are
violated in the large - $| \tilde x |$ tails of ${\cal F}^{ex,
\pm}$ at any $\tilde a / \xi_\pm > 0$ even arbitrarily close to
$T_c$ because ultimately, for $|x| \to \infty$ (i.e., for large
$L$ at fixed $|t| \neq 0$), the tails of ${\cal F}^{ex, \pm}$
become explicitly dependent on $\tilde a$. (Below we shall show
that the tails depend also on the bare four-point coupling $u_0$.)
Thus no finite-scaling form (\ref{1b}), (\ref{1c}), or (\ref{1d})
with a single scaling argument $\propto t L^{1/\nu}$ and with a
single nonuniversal amplitude $C_1$ can be defined in this large -
$| \tilde x |$ region \cite{scaling}.  Higher-loop contributions
cannot remedy this violation. It is obvious that an even larger
variety of different nonscaling effects exist in the exponential
finite-size region of systems with {\it non-cubic} anisotropies
(${\bf \bar A} \neq {\bf 1}$).

\subsubsection*{ 2. Isotropic $\varphi^4$ field theory}

The diversity of nonuniversal non-scaling effects in the region $
L/\xi_\pm \gg 1$ discussed above exists not only in anisotropic
lattice models but also in fully isotropic systems. We demonstrate
this point for the isotropic $\varphi^4$ field theory based on the
standard Hamiltonian
\begin{eqnarray}
\label{Xa}H_{field}  &=& \int\limits_{V} d^d x \big[\frac{r_0} {2}
\varphi^2 +  \frac{1} {2} (\nabla \varphi)^2 +
 u_0 \varphi^4  \big]
\end{eqnarray}
in a cube with $V=L^d$ and periodic b.c. and with some cutoff
$\Lambda$ in {\bf k} space. Keeping the cutoff finite may be a
valuable tool for testing universality as has been convincingly
demonstrated by Nicoll and Albright \cite{nicollalbright} in the
context of {\it bulk} universality \cite{priv}. In a similar
spirit this was done in \cite{cd2000-1,cd2002} with regard to {\it
finite-size} universality in the large-$n$ limit at and above
$T_c$. We shall show that ${\cal F}^{ex, \pm}$ depends on the {\it
bare} coupling $u_0$ and on the cutoff procedure for large $L$
above and below $T_c$ of the $n=1$ universality classes. For the
case of a sharp cutoff we shall also correct a previous
misinterpretation \cite{cd2002,dan-8} of the {\it singular} part
of the excess free energy density at $T_c$.

Since $f^{ex, \pm}$ has a finite limit for
$\Lambda\rightarrow\infty$ we first we calculate $f^{ex, \pm}$ at
infinite cutoff $\Lambda = \infty$ within the minimal
renormalization scheme at fixed dimension $2<d<4$
\cite{dohm1985,schl,str}. In one-loop order we obtain
\begin{eqnarray}
\label{Xb} f^{ex, \pm}_{\Lambda = \infty}(t,L)\;\;=\;\; L^{-d}
{\cal F}^{ex, \pm}_{\Lambda = \infty} (L/\xi_\pm ) \;
\end{eqnarray}
where for large $ L/\xi_\pm $
\begin{eqnarray}
\label{Xc}{\cal F}^{ex, \pm}_{\Lambda = \infty} (L/\xi_\pm )= -d
\left(\frac{L}{2\pi \xi_\pm}\right)^{(d-1)/2} \exp \left\{-
\frac{L}{\xi_\pm} \right\}\;\;
\end{eqnarray}
with the bulk second-moment correlation lengths
\begin{eqnarray}
\label{Xd} \xi_\pm (t; u) = \xi_{0 \pm}(u) |t|^{- \nu} \left\{1 +
C_\pm (t, u) \right\}.
\end{eqnarray}
(There is no difference between exponential and second-moment
correlation lengths at infinite cutoff at the one-loop level.) The
function $C_\pm (t, u)$ represents the Wegner series
\begin{eqnarray}
\label{Xex} C_\pm (t, u)= \sum_{m=1}^\infty  a^{(m)}_\pm(u)\;
|t|^{\Delta m}
\end{eqnarray}
with the universal Wegner exponent $ \Delta = \omega \nu$, $
\omega = \partial \beta_u(u,\varepsilon) / \partial u|_{u=u*}$,
and the Wegner amplitudes $a^{(m)}_\pm(u)$ depending in the
nonuniversal renormalized coupling $u$. The latter is defined by
\begin{eqnarray}
\label{Xe}u =  A_d Z_{u}(u,\varepsilon)^{-1}
Z_{\varphi}(u,\varepsilon)^2 u_0 \xi_{0 +}^{ \varepsilon}
\end{eqnarray}
(with the choice $\mu=\xi_{0 +}^{-1}$) where $Z_u(u,\varepsilon)$
and $Z_\varphi(u,\varepsilon)$ are the standard $Z$ factors
\cite{larin}. Equation (\ref{Xe}) determines $u$ as an implicit
function of $u_0 \xi_{0 +}^{ \varepsilon} $. Although $C_\pm$ is
an negligible additive correction in (\ref{Xd}) for sufficiently
small $|t|$ this is not the case in the exponential part of
(\ref{Xc}) which, for small $C_\pm $, can be rewritten as
\begin{eqnarray}
\label{Xf}  \exp \left\{- \frac{L}{\xi_\pm} \right\} = A(L, t, u)
\exp \left\{- \frac{L}{\xi_{0\pm} |t|^{-\nu}} \right\},
\end{eqnarray}
\begin{eqnarray}
\label{Xg} A(L,t, u) = \exp \left\{ C_\pm(t,u) \frac{L}{\xi_{0\pm}
|t|^{-\nu}} + O(C_\pm ^2)\right\},
\end{eqnarray}
with the nonuniversal nonscaling prefactor $A(L, t, u)$ that
cannot simply be replaced by 1 for small $|t|$. Even for
arbitrarily small $|t| \neq 0$ the prefactor becomes nonnegligible
if $L$ is sufficiently large, $L \gg |C_\pm |^{-1}{\xi_{0\pm}
|t|^{-\nu}}$. Thus the tails of the large $L$ dependence of ${\cal
F}^{ex, \pm}_{\Lambda = \infty}$ become nonuniversal and have a
nonscaling $L$ dependence through the prefactor $A(L, t, u)$. This
applies to the shaded area of the asymptotic critical region above
and below $T_c$ shown in Fig. 1. The same argument applies to  the
preceding subsection: it is necessary to keep the complete
non-asymptotic form of the second-moment bulk correlation lengths
at finite $\tilde a$
\begin{eqnarray}
\label{Xh} \xi_\pm (t; u_0 \tilde a^\varepsilon) = \xi_{0 \pm}
|t|^{- \nu} \left\{1 +  C_\pm (t, u_0
\tilde a^\varepsilon) \right\} \nonumber\\
\end{eqnarray}
in (\ref{VIIpp}) and (\ref{VIIr}) and to include all correction
terms in $ C_\pm (t, u_0 \tilde a^\varepsilon)$. \cite{scaling}

The reasoning described above must also be extended to the case
when a smooth cutoff $\Lambda$ in ${\bf k}$ space is taken into
account. This can be done by including an isotropic (Pauli-Villars
type) term $\frac{1}{2}(\nabla^2\varphi)^2/\Lambda^2$ in the
Hamiltonian (\ref{Xa}) \cite{cd2000-1,parisi}. In this case the
structure of ${\cal F}^{ex, \pm}$ still remains exponential
$\propto \exp [- L / \xi_{{\bf e}\pm}(\Lambda)]$ for large
$L/\xi_\pm$ but the exponential correlation lengths
\begin{eqnarray}
\label{Xi} \xi_{{\bf e \pm}}(\Lambda) = \xi_{\pm} [1 -
\frac{1}{2}\Lambda^{-2}\xi_{\pm}^{-2} +
O(\Lambda^{-4}\xi_\pm^{-4})]
\end{eqnarray}
become cutoff dependent. This causes a cutoff
dependent prefactor  $A(L,t,u, \Lambda)$ in (\ref{Xf}).

As pointed out in \cite{cd2000-2} there exists a close relation
between the $L$ dependence of finite-size effects and the ${\bf
x}$ dependence of the bulk order-parameter correlation function
$G_b$ discussed in Sect. III. In retrospect, the arguments
presented above apply also to the exponential part  of $ G_b
\propto |{\bf x}|^{-d+2} \exp(-|{\bf x}|/\xi_{{\bf e \pm}})$ even
if it is isotropic because here the same correlation lengths
$\xi_{{\bf e \pm}}$ appear as in the large $L$ decay of the
finite-size quantities. No scaling functions $\Phi_\pm$,
(\ref{3c}), (\ref{3d}), (\ref{3n}) can be defined in the
exponential large-distance regime $|{\bf x}|/ \xi_{\pm} \gg 1$
(shaded region in Fig. 2). Thus the exponential  tails of
$G_b({\bf x}; t)$ of the $\varphi^4$ theory have a nonscaling form
that depends on $u_0$ and the (smooth) cutoff even for arbitrarily
small $t \neq 0$, $h=0$ and $t= 0$, $h \neq 0$. In addition, for
anisotropic systems, it depends on the anisotropy matrix $\bf A$
and the higher order tensors $\bf B$ etc..

Although the nonscaling effect on the {\it relative} quantity
${\cal F}^{ex, \pm} / {\cal F}^\pm_b$ becomes arbitrarily large
for sufficiently large $L/\xi_\pm$  this happens in a region where
the magnitude of ${\cal F}^{ex, \pm}$ itself is exponentially
small. Thus, from a purely quantitative point of view, this is
only a very small effect for systems with short-range interactions
and periodic boundary conditions.

This is in contrast to the corresponding non-scaling finite-size
effects in the presence of (effective) long-range correlations
caused by a {\it sharp} momentum cutoff $ -\Lambda \leq k_\alpha <
\Lambda$ used in \cite{cd2002,cd2000-1}. Such a cutoff has often
been used in the formulation and application of the RG theory
based on the $\varphi^4$ Hamiltonian (\ref{Xa}) (see, e.g.,
\cite{wilson,nicollalbright,amit}). As far as thermodynamic bulk
properties are concerned this is well justified as the sharp
cutoff does not affect the critical exponents and the
thermodynamic bulk scaling functions.  Thus the $\varphi^4$ model
(\ref{Xa}) with a sharp cutoff is a legitimate model of
statistical mechanics that belongs to the same $(d,n)$
universality class as systems with short-range interactions or
with subleading long-range interactions. This implies the validity
of thermodynamic two-scale factor universality in the presence of
a sharp cutoff. Chen and the present author \cite{cd2000-1,cd2002}
have raised  the question whether this remains true also for
confined systems. It was found, for the susceptibility and for the
excess free energy in the large-$n$ limit above $T_c$, that a
sharp cutoff is not compatible with an exponential size dependence
and violates finite-size scaling in the large - $L$ regime above
$T_c$. This behavior was traced back to the well known
\cite{cd2002,wilson} artifact that the sharp cutoff in {\bf k}
space  causes long-range correlations in real space as can be
demonstrated in the bulk order-parameter correlation function
$G_b({\bf x}; t; \Lambda) $ \cite{cd2002,cd2003} whose
algebraically decaying non-scaling part dominates the
exponentially decaying scaling part. By means of a RG one-loop
calculation for $n=1$ we find that this property holds both above
and below $T_c$ for sufficiently large $L$.

In contrast to \cite{cd2002}, however, we do not obtain a
violation of finite-size scaling in the central finite-size region
including $T=T_c$. Our present analysis is based on an appropriate
decomposition of the excess free energy into singular and
nonsingular parts in the sense of (\ref{2dd}) whereas in
\cite{cd2002} no $L$ dependent {\it nonsingular} part was defined.
We find that, in the presence of a sharp cutoff $\Lambda$ and for
large $L\Lambda$, Eq. (\ref{Xb}) with (\ref{Xc}) is to be replaced
by
\begin{eqnarray}
\label{Xk} f^{ex, \pm}_{\Lambda}(t,L)\;\;= f^{ex,
\pm}_{\Lambda,s}(t,L) + f^{ex}_{\Lambda,ns}(L),
\end{eqnarray}
with the singular part
\begin{eqnarray}
\label{Xl} f^{ex, \pm}_{\Lambda,s}(t,L)\;\;=
L^{-2}\Lambda^{d-2}\tilde \Phi_d(\xi^{-1}_\pm\Lambda^{-1}) +
f^{ex, \pm}_{\Lambda= \infty}(t,L),\;\;\;\;\;\;\;\;
\end{eqnarray}
\begin{eqnarray}
\label{Xm} \tilde \Phi_d(z)\;\;=\;\;\int_{0}^{\infty}
 d y \;\big[e^{-(1+z^2)y} - e^{-y}\big] E_d(y),\;\;\;\;
\end{eqnarray}
\begin{eqnarray}
\label{Xo} E_d(y)\;\;= \frac{d }{6(2\pi)^{d-2}}\Big[\int_{-1}^{1}
 d q \;e^{-q^2y}\Big]^{d-1} ,\;\;
\end{eqnarray}
and the $L$ dependent nonsingular part
\begin{eqnarray}
\label{Xn} f^{ex}_{\Lambda,ns}(L)\;\;=
L^{-2}\Lambda^{d-2}\int_{0}^{\infty}
 d y \;e^{-y} E_d(y) .\;\;\;\;\;\;
\end{eqnarray}
Although our one-loop result (\ref{Xk}) - (\ref{Xn}) for the {\it
total} excess free energy density $f^{ex, \pm}_{\Lambda}(t,L)$ is
equivalent to equations (8) and (16) of \cite{cd2002}, there is a
crucial difference with regard to singular part. In contrast to
the non-vanishing function $ \Phi_{d,d'}(z)$ of \cite{cd2002}, our
function $\tilde \Phi_d(z)$ vanishes at criticality,  $\tilde
\Phi_d(0)=0$.  The temperature independent part (\ref{Xn})
$\propto L^{-2}$ should not be attributed to the singular part as
was done in \cite{cd2002}. Our definition of the nonsingular part
$f^{ex}_{\Lambda,ns}(L) \propto L^{-2}$ is parallel to the
standard analysis of bulk systems with a specific heat $C_\pm =
A_\pm |t|^{-\alpha} + C_B$ with a negative critical exponent
$\alpha$ whose finite value $C_B$ at the finite cusp must {\it
not} be included in the singular scaling part $\sim |t|^{-\alpha}$
but rather in the nonsingular "background" contribution of the
specific heat. The nonuniversal power-law term $\propto
L^{-2}\Lambda^{d-2}$ in (\ref{Xl}) dominates in the shaded region
of Fig. 1 compared to the scaling part $f^{ex, \pm}_{\Lambda=\infty}
\propto L^{-d}$ but vanishes at $T=T_c$ and is subleading in the
central finite-size regime.  Thus the {\it leading} finite-size contributions in the $\varphi^4$ model with a
sharp cutoff are in agreement with universal finite-size scaling
{\it in the central finite-size regime} if the singular part of
the free energy is identified correctly. Consequently, the
leading singular part of the Casimir force (in film geometry) at
bulk $T_c$ \cite{cd2002} remains universal within the subclass of
isotropic systems even in the presence of a sharp cutoff but an
additional regular part $\propto L^{-2} $ exists that is
nonuniversal and is dominant compared to the singular part
$\propto L^{-d}$. This unusual behavior is due to the long-range
correlations caused by the sharp cutoff \cite{cd2001}, as noted
already in \cite{cd2002}, which is of course a mathematical
artifact and not generic for systems with purely short-range
interactions.  As pointed out by Dantchev et al. \cite{dan-8}, the
sharp cutoff implies an unphysical discontinuity of the slope of
the interaction $\delta \widehat K({\bf k}) = k^2$ at the boundary
of the Brillouin zone which is the mathematical origin of the
$L^{-2}$ terms. For the reasons given above, however, we disagree
with the opinion expressed in \cite{dan-8} that the concept of
finite-size scaling as developed for systems with short range
interactions does not apply to the $\varphi^4$ model (\ref{Xa})
with a sharp cutoff. The authors of \cite{dan-8} did not perform
an analysis based on a decomposition of the type (\ref{2dd}) and
(\ref{Xk}). Our analysis shows that, in spite of the mathematical
artifact of $\widehat K({\bf k})$ at the Brillouin-zone boundary,
the concept of finite-size scaling is well applicable to the
central finite-size regime including $T=T_c$ and that a violation
of finite-size scaling occurs only in the large-$L$ regime at $T
\neq T_c $ (shaded region in Fig.1), as in the other cases
discussed above in the presence of a lattice cutoff or a smooth
cutoff.

Finally  we discuss the case of an additional subleading
long-range interaction of the van der Waals type   as defined in
(\ref{2g}) and (\ref{2hh}). It was pointed out by Dantchev and
Rudnick \cite{dan-1} that it affects the finite-size
susceptibility in the regime $L/\xi_+ \gg 1$, similar to the
effect caused by a sharp cutoff \cite{cd2000-1}. The effect of
this interaction on the excess free energy $f_s^{ex}$ and on the
critical Casimir force in the case of film geometry was first
studied by Chen and the present author \cite{cd2002,cd-2003}.  The
asymptotic structure for $L/\xi_+ \gg 1$ in one-loop order above
$T_c$ at $h=0$ is
\begin{eqnarray}
\label{Xp} f_s^{ex}(t,L)\;\;= L^{-d} \Big[{\cal F}^{ex}(L/\xi_+ )
+ b L^{2-\sigma} \Psi (L/\xi_+)\Big]. \;\;\;
\end{eqnarray}
which is similar to (\ref{3xx}). We have verified that, for $n=1$,
the same structure is valid also for cubic geometry with periodic
b.c. above and below $T_c$ where the function $\Psi_{cube}$ has an
algebraic large-$L$ behavior $\sim (L/\xi_\pm )^{-2}$. The  latter
dominates the exponentially decaying scaling part ${\cal F}^{ex,
\pm} \sim \exp(-L/\xi_{{\bf e}\pm})$ in the shaded region of Fig.
1. This implies that, in this region, {\it two} nonuniversal
length scales $b^{1/(\sigma-2)}$ and $\xi_{\pm}$ at $h=0 $ govern
the {\it leading} singular part of the excess free energy density
\begin{eqnarray}
\label{Xq} f_s^{ex,\pm}(t,L)\;\;\sim \;\;L^{-d} \;\; \Big[\frac{b^{1/(\sigma-2)}}{ L}\Big]^{\sigma - 2}\;\;
\Big[\frac{\xi_\pm}{L}\Big]^{2}, \;\;\;
\end{eqnarray}
even arbitrarily close to criticality. In addition, there is the
nonuniversal $u_0$ - dependent exponential tail of ${\cal F}^{ex,
\pm}$. In (\ref{Xq}), both the amplitude $\sim b$ and the power $-d-\sigma$ of the $L$ dependence are nonuniversal. Thus the universal scaling form (\ref{1b}), with only {\it
one} length scale $C_1^{-\nu}$ at $h=0$, is not valid in the
entire range of its scaling arguments for isotropic systems with
van der Waals type interactions although such systems are members
of the same ($d$, $n=1$) universality class as, e.g., Ising models
with short-range interactions. The structure of (\ref{Xp}) and
(\ref{Xq}) and the corresponding structure for the critical
Casimir force in film geometry has been confirmed in
\cite{dan-8,dantchev2006,dantchev2007}.

\renewcommand{\thesection}{\Roman{section}}
\setcounter{equation}{0} \setcounter{section}{1}
\renewcommand{\theequation}{\Alph{section}.\arabic{equation}}
\section*{ACKNOWLEDGMENT}

I thank W. Selke for informing me about his new MC data prior to publication.
I also thank B. Kastening and W. Selke for useful discussions.
Support by DLR under Grant No. 50WM0443 is gratefully acknowledged.

\renewcommand{\thesection}{\Roman{section}}
\setcounter{equation}{0} \setcounter{section}{1}
\renewcommand{\theequation}{\Alph{section}.\arabic{equation}}
\section*{Appendix A : Universal bulk amplitude relations}

In this Appendix we present explicit expressions for the universal
constants $Q_i$, $\widetilde Q_3$, and $P_i$, (\ref{3f}) -
(\ref{3l1}), in terms of universal scaling functions. Near $T_c$
the sum rule (\ref{3-n}) yields
\begin{eqnarray}
\label{a1} \chi_b' (t, h') &=& - A'_1 | t |^{d \nu}
\partial^2 W_\pm (A'_2 h' | t |^{- \beta \delta}) / \partial h'^2
\nonumber\\ &=& D_1' \; \xi'_\pm (t, h')^{2 - \eta} \; \widetilde
\Phi_\pm (D'_2 h' | t |^{- \beta \delta})
\end{eqnarray}
with the universal function
\begin{equation}
\label{a2} \widetilde \Phi_\pm (y) = 2 \pi^{d/2} \Gamma (d/2)^{-1}
\int\limits_0^\infty ds s^{1 - \eta} \Phi_\pm (s, y) \;.
\end{equation}
At $t > 0, h' = 0$, (\ref{a1}) yields $\chi_b' (t, 0) = \Gamma_+'
| t |^{- \gamma}$ with
\begin{equation}
\label{a3} \Gamma_+' = - A_1' \; A_2'^2 \; W_2 \; = \; D_1' \;
(\xi'_{0 +})^{2 - \eta} \; \widetilde \Phi_+ (0)
\end{equation}
where $W_2 = \lim_{y \to 0} \partial^2 W_+ (y) /
\partial y^2$. At $t = 0 \; , h' \neq 0$ we have from (\ref{3d}) $\xi'_\pm (0, h')
\equiv \xi'_h = \xi_c' | h' |^{- \nu / (\beta \delta)}$ with
\begin{equation}
\label{aa3}\xi_c' = \xi_{0 +}' (D'_2)^{- \nu / (\beta \delta)}
\widehat X \;,
\end{equation}
thus (\ref{a1}) yields $\chi_b' (0, h') = \Gamma'_c | h' |^{-
\gamma / (\beta \delta)}$ with
\begin{eqnarray}
\label{a4} \Gamma'_c &=& - A_1' \; A_2'^{1 + 1/\delta} \; \widehat
W \nonumber\\ &=& D_1' \left[\xi_{0 +}' \;D_2'^{- \nu / (\beta
\delta)} \widehat X \right]^{2 - \eta} \; \widetilde \Phi (\infty)
\end{eqnarray}
where $\widetilde \Phi (\infty) \equiv \widetilde \Phi_\pm
(\infty)$ and
\begin{equation}
\label{a5} \widehat W \; = \; \lim_{y \to \infty} \left\{ | y
|^{\gamma / (\beta \delta)} \; \partial^2 W_\pm (y) / \partial y^2
\right\} \; ,
\end{equation}
\begin{equation}
\label{a6} \widehat X \; = \; \lim_{y \to \infty} \left\{ | y
|^{\nu / (\beta \delta)} \; X_\pm (y) \right\} \; .
\end{equation}
(\ref{a3}) and (\ref{a4}) yield
\begin{eqnarray}
\label{a7}\Gamma_+' / \Gamma_c' &=& A'^{1 - 1/\delta} _2 \; W_2 \;
\widehat W^{-1} \nonumber\\ &=& D_2'^{1 - 1 / \delta} \;
\widetilde \Phi_+ (0) \left[\widehat X^{2 - \eta} \; \widetilde
\Phi (\infty) \right]^{-1} \; ,
\end{eqnarray}
thus we obtain the universal ratio
\begin{equation}
\label{a8} P_2 \; = \; \frac{A_2'}{D_2'} \; = \;
\left[\frac{\widehat W \; \widetilde \Phi_+ (0)} {W_2 \; \widehat
X^{2 - \eta} \; \widetilde \Phi (\infty)}\right]^{\delta / (\delta
- 1)} \;.
\end{equation}
Eqs. (\ref{aa3}) and (\ref{a7}) yield a universal ratio $Q_2$
different from $P_2$,
\begin{equation}
\label{a9} Q_2 \; = \; \frac{\Gamma_+'}{\Gamma_c'} \;
\left(\frac{\xi_c'} {\xi_{0 +}'}\right)^{2 - \eta} \; = \;
\frac{\widetilde \Phi_+ (0)} {\widetilde \Phi (\infty)}
\end{equation}
where we have used $1 - 1/\delta = \gamma / (\beta \delta)$ and
$(2 - \eta) \nu = \gamma$.

Following Privman and Fisher \cite{pri} we assume that the {\it
unsubtracted} bulk correlation function $\widetilde G_b' ({\bf
x}'_i - {\bf x}_j' ; t, h') = \lim_{V' \to \infty} \langle
\varphi' ({\bf x}'_i) \varphi' ({\bf x}'_j)\rangle'$ has the
asymptotic scaling form
\begin{equation}
\label{a10} \widetilde G_b' ({\bf x}' ; t, h') = D_1' | {\bf x}'
|^{- d + 2 - \eta} \; Z_\pm \left( | {\bf x}' | / \xi' \; , D_2'
h' | t |^{- \beta \delta} \right)
\end{equation}
with the same constants $D_1'$ and $D_2'$ as in Eq. (\ref{3c}) and
with a universal scaling function $Z_\pm (x, y)$. From
(\ref{a10}), (\ref{2w}), and (\ref{3a}) we obtain the square of
the bulk order parameter below $T_c$
\begin{eqnarray}
\label{a11} \left[m'_b (t)\right]^2 = \lim_{h' \to 0} \; \;
\lim_{|{\bf x}'| \to \infty} \; \widetilde G'_b ({\bf x}' ; t, h')
\qquad \nonumber\\ = D_1' \; (\xi'_{0 -})^{- d + 2 - \eta} | t
|^{\nu (d - 2 + \eta)} \; \widehat Z \nonumber\\ = \lim_{h' \to 0}
\left[\partial f'_b (t, h') / \partial h'\right]^2  = (A_1' \;
A_2')^2 | t |^{2 \beta} \; W_1
\end{eqnarray}
with $\widehat Z = \lim_{y \to 0} \lim_{x \to \infty} \left\{x^{-
d + 2 - \eta} Z_- (x, y) \right\}$ and $W_1 = \lim_{y \to 0}
\partial W_- (y) / \partial y$. In order to derive $Q_1$ and $P_3$
we use (\ref{a3}) and (\ref{a11}) and obtain
\begin{eqnarray}
\label{a12} D_1' &=& - A_1' \; A'^2_2 \; (\xi'_{0 +})^{- 2 + \eta}
\; W_2 / \widetilde \Phi_+ (0) \nonumber\\ &=& (A_1' \; A_2')^2 \;
(\xi'_{0 -})^{d - 2 + \eta} \; W_1 / \widehat Z \; .
\end{eqnarray}
Together with (\ref{3e}) this yields the universal quantities
\begin{equation}
\label{a13} Q_1 = A_1' \; (\xi'_{0 +})^d  = -  \widehat Z \; W_2
\left[X_- (0) \right]^{- d + 2 - \eta} /\left[W_1 \; \widetilde
\Phi_+ (0)\right]
\end{equation}
and
\begin{equation}
\label{a14} P_3 \; = \; D_1' \; A'^{- 1 - \gamma / (d \nu)}_1
(A_2')^{-2} \; = \; - \; Q_1^{- \gamma / (d \nu)} \; W_2 /
\widetilde \Phi_+ (0) \;.
\end{equation}
Finally we consider the universal ratio (\ref{3l}). The amplitude
$\widehat D'_\infty$ is given by $\widehat D'_\infty \; = \; D'_1
\; \Phi_\pm (0,0) \; \widehat C$ with the universal constant
\cite{dohm2006,Kastening}
\begin{eqnarray}
\label{a16} \widehat C &=& \frac{\widehat D'_\infty}{D'_\infty} =
\frac{\widehat \Phi_\pm (0,0)}{\Phi_\pm (0,0)} = \frac{(4
\pi)^{d/2} \Gamma (\frac{2-\eta}{2})}
{2^{d-2+\eta}\Gamma(\frac{d-2+\eta}{2})}\;.
\end{eqnarray}
Together with (\ref{a3}) this yields a universal ratio $Q_3$
different from $P_3$ ,
\begin{equation}
\label{a17} Q_3 \; = \; \widehat D'_\infty (\xi'_{0 +})^{2 - \eta}
/ \Gamma_+' \; = \; \Phi_\pm (0,0) \; \widehat C / \widetilde
\Phi_+ (0) \; .
\end{equation}
The universal constant $\widetilde Q_3$ in (\ref{3l1}) is
\be
\label{a19} \widetilde Q_3 \; = \; \Phi_\pm (0,0) / \widetilde
\Phi_+ (0) \; .
\ee

\renewcommand{\thesection}{\Roman{section}}
\setcounter{equation}{0} \setcounter{section}{2}
\renewcommand{\theequation}{\Alph{section}.\arabic{equation}}
\section*{Appendix B : Gaussian model with lattice anisotropy}

In order to derive the Gaussian part of (\ref{4z}) and the results
of Sect. X we consider the Hamiltonian (\ref{2a}) and (\ref{2g})
for $r_0 = a_0 t > 0$, $u_0 = 0$ and $h = 0$ with $N$ scalar
variables $\varphi_j$ on a simple-cubic lattice with lattice
constant $\tilde a$ in a cubic volume $V = L^d = N \tilde a^d$
with periodic boundary conditions. This Hamiltonian will be
denoted by $H^G$. The Jacobian of the linear transformation
$\varphi_j \rightarrow \hat \varphi({\bf k})$  is $ \left|
\partial \varphi_j/\partial \hat \varphi ({\bf k}) \right| =
({\tilde a}L)^{-dN/2}$. The dimensionless partition function is
\begin{eqnarray}
\label{b6} Z^G &=& \left[\prod_{j=1}^N \; \int\limits_{-
\infty}^\infty \frac{d \varphi_j}{\tilde a^{1 - d/2}} \right] \exp
(- H^G) \nonumber\\
&=&  \left[\prod_{\bf k}\;\frac{1}{\tilde a L^{d/2}} \int d \hat
\varphi({\bf k}) \right] \exp (- H^G) \nonumber\\ &=& \prod_{\bf
k} \left(\frac{2 \pi} {[r_0 + \delta \widehat K (\mathbf k)]
\tilde a^2}\right)^{1/2} \;.
\end{eqnarray}
For the transformed system one obtains
\begin{eqnarray}
\label{bb6} Z'^G &=& \left[\prod_{\bf k'}\;\frac{1}{v'^{1/d}
L'^{d/2}} \int d \hat \varphi'({\bf k'}) \right] \exp (- H'^G)
\nonumber\\ &=& \prod_{\bf k'} \left(\frac{2 \pi} {[r_0 + \delta
\widehat K' (\mathbf k')] v'^{2/d}}\right)^{1/2} \;
\end{eqnarray}
with $v'=(\det {\bf A})^{-1/2} \tilde a^d$. Eqs. (\ref{b6}) and
(\ref{bb6}) define the integration measure $\int d \hat
\varphi({\bf k})$ and $\int d \hat \varphi'({\bf k'})$. The
Gaussian free energy densities divided by $k_B T$ are
\begin{eqnarray}
\label{b7} f^G =   - \frac{\ln (2 \pi)} {2 \tilde a^d} &+&
\frac{1}{2 L^d} \sum_{\bf k} \ln \{[r_0 + \delta \widehat K
(\mathbf k)] \tilde a^2 \} \;,
\end{eqnarray}
\begin{eqnarray}
\label{bb11} f'^G =  - \frac{\ln (2 \pi)} {2 v'} &+& \frac{1}{2
L'^d} \sum_{\bf k'} \ln \{[r_0 + \delta \widehat K' (\mathbf k')]
v'^{2/d} \} \; . \nonumber\\
\end{eqnarray}
The correctness of the additive constant of $f^G$ can be checked
by performing the integrations of $Z^G$ in real space for $K_{i,j}
= 0, \delta \widehat K (\mathbf k)=0$,
\be
\label{b8} \prod_{j=1}^N \;  \int\limits_{- \infty}^\infty \frac{d
\varphi_j}{\tilde a^{1 - d/2}} \; \exp \left[- \tilde a^d
\sum_{j=1}^N \frac{r_0}{2} \varphi^2_j \right] = \left(\frac{2
\pi} {r_0 \tilde a^2} \right)^{N/2} \;.
\ee
This is valid also for free boundary conditions. The additive
constant of $f^G$ was not correct in previous work \cite{text,aa}.
In order to calculate (\ref{b7}) we consider
\begin{eqnarray}
\label{b9} \Delta(r_0, L, K_{i,j}, \tilde a) = L^{-d} {\sum_{\bf
k}} \ln
\{[r_0 + \delta \widehat K (\mathbf k)] \tilde a^2\}\nonumber\\
- \int\limits_{\bf k} \ln \{[r_0 + \delta \widehat K (\mathbf k)]
\tilde a^2\}
\end{eqnarray}
where the sum $\sum_{\bf k}$ and the integral  $\int_{\bf k}$ have
finite cutoffs $\pm \pi / \tilde a$ for each $k_\alpha$ [see Eq.
(\ref{4z5a})]. Using the integral representation
\be
\label{b10} \ln w = \; \int \limits_0^\infty dy y^{-1} \left[ \exp
{\left(-y\right)} - \exp {\left(- w y\right)}\right]
\ee
and interchanging the integration $\int dy$ with ${\sum_{\bf k}}$
and $\int _{\bf k}$ we obtain, because of $L^{-d} \sum_{\bf k} 1 =
\int_{\bf k} 1$,
\begin{eqnarray}
\label{b11} \Delta(r_0, L, K_{i,j}, \tilde a) \; = \;  \int
\limits_0^\infty dy y^{-1} e^{-r_0 \tilde a^2
y}\Big[\int\limits_{\bf k} \exp \{-  \delta \widehat K (\mathbf k)
\tilde a^2 y \} \nonumber\\ - \; L^{-d} \sum_{\bf k} \exp \{-
\delta \widehat K (\mathbf k) \tilde a^2 y \} \Big] \; . \qquad
\qquad
\end{eqnarray}
Since $\delta \widehat K (\mathbf k)$ is a periodic function of
each component $k_\alpha$ of $\bf k$ the sum in (\ref{b11})
satisfies the Poisson identity \cite{cd2000-2,morse}
\begin{eqnarray}
\label{b12} L^{-d} \sum_{\bf k} \exp \{- \delta \widehat K
(\mathbf k) \tilde a^2 y \} \nonumber\\ = {\sum_{\bf n}}
\int\limits_{\bf k} \exp \{ -  \delta \widehat K (\mathbf k)
\tilde a^2 y + i {\bf k} \cdot {\bf n} L\}
\end{eqnarray}
where ${\bf n} = (n_1, n_2, ..., n_d)$ and ${\bf k} \cdot {\bf n}
= \sum^d_{\alpha = 1} k_\alpha n_\alpha$. The sum $\sum_{\bf n}$
runs over all integers $n_\alpha \; , \alpha = 1, 2, ..., d$ in
the range $- \infty \leq n_\alpha \leq \infty$. This leads to the
exact representation
\begin{eqnarray}
\label{b13} \Delta(r_0, L, K_{i,j}, \tilde a) \; = \;-\int
\limits_0^\infty dy y^{-1} e^{-r_0 \tilde a^2 y} \nonumber \\
\times \sum_{{\bf n} \neq 0} \int\limits_{\bf k} \exp \{- \delta
\widehat K (\mathbf k) \tilde a^2 y + i {\bf k} \cdot {\bf n} L\}.
\end{eqnarray}
Note that here the integral $\int _{\bf k}$ still has finite
lattice cutoffs $\pm \pi/\tilde a$. We shall evaluate $\Delta(r_0,
L, K_{i,j}, \tilde a)$ for large $L/\tilde a \gg 1$ and
distinguish two regimes: (i) $L r_0^{1/2} \lesssim O (1)$,
$r_0^{1/2} \tilde a \ll 1$, and (ii) $L r_0^{1/2} \gg 1$ for
arbitrary  fixed $r_0^{1/2} \tilde a > 0$.

 In the regime (i), the large - ${\bf k}$ dependence of $\delta
\widehat K (\mathbf k)$ does not matter. Therefore we may replace
$\delta \widehat K (\mathbf k)$ by its small - ${\bf k}$ form
${\bf k} \cdot {\bf A k}$ and let the integration limits of
$\int_{\bf k}$ go to $\infty$. Furthermore it is useful to
substitute the integration variable $z = 4 \pi^2 \tilde a^2 y /
L^2$. Then we obtain
\begin{eqnarray}
\label{b14} \Delta (r_0, L, K_{i,j}, \tilde a) \to \Delta (r_0, L;
{\bf A}) = - \int\limits_0^\infty dz z^{-1} e^{- r_0 L^2 z/(4
\pi^2)} \nonumber\\ \times \; \sum_{{\bf n} \neq {\bf 0}}
\int\limits_{\bf k}^\infty \exp [- {\bf k} \cdot {\bf Ak} L^2 z/(4
\pi^2) + i {\bf k} \cdot {\bf n} L] \; .\qquad \qquad
\end{eqnarray}
The Gaussian integral over ${\bf k}$ yields
\begin{eqnarray}
\label{b15} \int\limits_{\bf k}^\infty \exp [- {\bf k} \cdot {\bf
Ak} L^2 z/(4 \pi^2) + i {\bf k} \cdot {\bf n} L] \nonumber\\
= (\det {\bf A})^{- 1/2} \left(\frac{\pi}{L^2 z} \right)^{d/2}
\exp (- {\bf n} \cdot {\bf A}^{-1} {\bf n} \pi^2 / z) \; .
\end{eqnarray}
Eqs. (\ref{b14}) and (\ref{b15}) lead to
\begin{eqnarray}
\label{b18} \Delta (r_0, L; {\bf A}) \; = \;  L^{-d} {\cal G}_0
(r_0 L'^2, {\bf \bar A}) \;
\end{eqnarray}
where ${\cal G}_0$ and $K_d (y, {\bf \bar A})$ are given by
(\ref{8d}) and (\ref{4ii}). Thus, in the regime (i), we derive
from Eqs. (\ref{b7}), (\ref{b9}), (\ref{b11}), and (\ref{b18})
\begin{eqnarray}
\label{b20} f^G &=&  f_b^G + \frac{1}{2} L^{-d} {\cal G}_0(r_0
L'^2; {\bf \bar A})  \;,
\end{eqnarray}
where the bulk part $f_b^G$ is obtained from (\ref{b7}) by the
replacement  $L^{-d}\sum_{\bf k} \to  \int_{\bf k}$. We note that
within the {\it anisotropic} Gaussian model there exists no unique
correlation length. This exists only for the transformed system
with the (asymptotically isotropic) Hamiltonian $H'^G$ [Eqs.
(\ref{2p}) and (\ref{2q}) for $u'_0 = 0, h' = 0]$ for which the
corresponding result reads
\begin{eqnarray}
\label{b-20} f'^G =  f_b'^G + \frac{1}{2} L'^{-d} {\cal G}_0(r_0
L'^2; {\bf \bar A})  \;,
\end{eqnarray}
with the bulk part  $f'^G$ given in (\ref{4z3}) for $u'_0=0$. Now
the parameter $r_0$ is related to the second-moment bulk
correlation length of $H'^G$ [see (\ref{3b})]
\begin{equation}
\label{b-22} \xi'^G_+ = r_0^{-1/2} = \xi_{0+}'^G t^{-1/2} \; , \;
\xi_{0+}'^G = a_0^{-1/2} \;.
\end{equation}
This leads to the identification of the scaling function of the
Gaussian excess free energy density in the regime (i)
\begin{eqnarray}
\label{b23}{\cal F}^{G, ex}(\tilde x; {\bf \bar A}) = (1/2) {\cal
G}_0 (\tilde x; {\bf \bar A})
\end{eqnarray}
with $\tilde x = t (L' / \xi_{0+}'^G)^{1/\nu}$, $\nu = 1/2$.

In the regime (ii), $\Delta(r_0, L, K_{i,j}, \tilde a)$ will
attain an exponential $L$ dependence and the complete ${\bf k}$
dependence of $\delta \widehat K (\mathbf k)$ does matter. For
simplicity we consider only nearest-neighbor couplings $K_{i,j} =
K$ on a $sc$ lattice, $\delta \widehat K ({\bf k})  =  4 K
\sum_{\alpha = 1}^d  \left[1 - \cos (\tilde a  k_\alpha) \right]$,
with the second-moment bulk correlation length of the Gaussian
model
\begin{equation}
\label{b-25} \xi_+^G = \tilde a \left(\frac{2K}{r_0}\right)^{1/2}
= \xi_{0 +}^G t^{- 1/2} \; , \; \xi_{0 +}^G = \tilde a
\left(\frac{2K}{a_0}\right)^{1/2} \; .
\end{equation}
At the level of the Gaussian model there exist no Wegner
corrections to (\ref{b-25}). Using (\ref{b10}) we obtain, similar
to App. A of  \cite{cd2003}, the exact representation for
(\ref{b9})
\begin{eqnarray}
\label{b22} \Delta (r_0, L, K, \tilde a) = \tilde a^{-d}
\int\limits_0^\infty dy \; y^{-1} \; e^{- \tilde r_0 y}
\nonumber\\ \times\big\{\left[S (\infty, y)\right]^d - \left[S
(L/\tilde a, y)\right]^d \big\}\; ,
\end{eqnarray}
\begin{eqnarray}
\label{b25} S (L /\tilde a, y) = S (\infty, y) + 2 \; e^{-2y} \;
\sum_{m=1}^\infty \; I_{mL/\tilde a} (2y) \; ,
\end{eqnarray}
with $S (\infty, y)  =  e^{- 2 y} I_0 (2 y)$ and $\tilde r_0
\equiv r_0 / (2K) = (\tilde a/\xi_+^G)^2$ where
\begin{eqnarray}
\label{b26} I_M (2y) \; = \; (2 \pi)^{-1} \; \int\limits_{-
\pi}^\pi d \theta \cos (M \theta) \; e^{2y \cos \theta} \;\;
\end{eqnarray}
are the modified Bessel functions with integer $M$ (see (9.6.19)
of \cite{abra-1}). In the limit of large $L/\tilde a$ for fixed
$\tilde r_0 > 0$ only the range of $y \sim O (L/\tilde a)$ is
relevant and only the $m=1$ term of (\ref{b25}) suffices to obtain
the leading exponential behavior. Thus we substitute $2 y = z
L/\tilde a$ in (\ref{b22}) and use the asymptotic formulae for
large $L/\tilde a$ (see (9.7.7) and (9.7.1) of \cite{abra-1} and
App. A of \cite{cd2000-2})
\begin{eqnarray}
\label{b27} I_{L/\tilde a} (z L/\tilde a) \sim (2 \pi L q/\tilde
a)^{- 1/2} \exp \left\{ \frac{L}{\tilde a} \left[q + \ln\left(
\frac{z}{1+q}\right)\right]\right\}, \nonumber\\
\end{eqnarray}
\begin{equation}
\label{b28} I_0 (z L/\tilde a) \sim \; e^{z L/\tilde a} (2 \pi z
L/\tilde a)^{- 1/2} \; ,
\end{equation}
where $q = (1+z^2)^{1/2}$. The maximum of the resulting
exponential part of the integrand of (\ref{b22}) is at $z = \bar
z$ where $\bar z \; = \; \left[\tilde r_0 (1 + \tilde r_0 /
4)\right]^{- 1/2}$. Expanding around $z = \bar z$ and performing
the integration over $z$ yields for large $L/\tilde a$ at
arbitrary fixed $\tilde r_0 > 0$
\begin{eqnarray}
\label{b30} \Delta (r_0, L, K, \tilde a) = - \frac{2d}{L^d}
\left(\frac{L/\tilde a}{2 \pi \bar z}\right)^{(d-1)/2} e^{-
L/\xi_{\bf e}^G}
\end{eqnarray}
with the exponential correlation length
\begin{eqnarray}
\label{b31} \xi_{\bf e}^G \; = \; \frac{\tilde a}{2} \left[{\rm
arsinh} \left(\frac{\tilde a}{2 \xi_+^G}\right)\right]^{-1} \; .
\end{eqnarray}
No universal finite-size scaling function of the Gaussian model
can be defined in the region $L/\xi_+^G\gg1$ because of the
explicit $\tilde a$ dependence of (\ref{b30}) and (\ref{b31}).

Within a RG treatment of the $ \varphi^4$ lattice model the
Gaussian results can be considered as the bare one-loop
contribution. By means of such a RG treatment at finite lattice
constant $\tilde a$ parallel to Sect. 2 and App. A of
\cite{cd2000-1}, the results derived above  acquire the
correct critical exponents of the $n=1$ universality class
including corrections to scaling. This leads to the one-loop
results (\ref{VIIo}) - (\ref{VIIr}) which are valid  for arbitrary $\tilde a /
\xi_\pm > 0$.

For field theory with  $\delta \widehat K (\mathbf k)= k^2$  and a
sharp cutoff $-\Lambda\leq k_\alpha <\Lambda$ the Gaussian excess
free energy density is given by $(1/2)\Delta_\Lambda$ where
$\Delta_\Lambda$ is given by (\ref{b9})  with $\tilde a$ replaced
by $\Lambda^{-1}$. From  (A.31) of App. A of \cite{cd1998} and a
RG treatment at finite $\Lambda$ we obtain (\ref{Xk}) -
(\ref{Xn}).

\renewcommand{\thesection}{\Roman{section}}
\setcounter{equation}{0} \setcounter{section}{3}
\renewcommand{\theequation}{\Alph{section}.\arabic{equation}}
\section*{Appendix C : Sums over higher modes}

Using (\ref{b9}) and the integral representation (\ref{b10}) with
$w = r L'^2 / (4 \pi^2)$ we obtain from (\ref{b18}) and (\ref{8d})
\begin{eqnarray}
\label{c1}\frac{1}{2L'^d} {\sum_{\bf k'\neq0}} \ln \{[r + \delta
\widehat K' (\mathbf k')] v'^{2/d}\} & = &
\frac{1}{2}\int\limits_{\bf k'} \ln \{[r + \delta \widehat K'
(\mathbf k')] v'^{2/d}\} \nonumber\\+ \frac{1}{2L'^d}\ln
\left(\frac{L'^2}{ v'^{2/d} 4\pi^2}\right) & & + \;\frac{1}{2L'^d}
J_0(r L'^2, {\bf \bar A}),
\end{eqnarray}
\begin{eqnarray}
\label{c2} J_0(rL'^2, {\bf \bar A})= \int\limits_0^\infty
\frac{dy}{y}
  \Big[\exp {\left(-\frac{rL'^2y}{4\pi^2}\right)}
  \nonumber\\ \times \left\{
  (\pi / y)^{d/2}
  - K_d (y, {\bf \bar A}) + 1 \right\}  - \exp (-y)\Big] .\;\;
\end{eqnarray}
The $v'$ dependent finite-size part in (\ref{c1}) comes from the
absence of the  ${\bf k'}={\bf 0}$ term and is exactly cancelled by the
corresponding logarithmic term in (\ref{4z}). For $d>0$ the
function $J_0(r L'^2, {\bf A})$ is finite for all $0 \leq r L'^2 <
\infty$ and diverges for large $rL'^2$ as $J_0 (r L'^2, {\bf A})
\sim - \ln [r L'^2 / (4 \pi^2)]$. By means of differentiation with
respect to $r$ we obtain
\begin{eqnarray}
\label{c3} S_m(r) = L'^{-d} \sum_{{\bf k'} \neq {\bf 0}} [r +
\delta \widehat K' (\mathbf k')] ^{-m} = \nonumber\\
\int\limits_{\bf k'}[r + \delta \widehat K' (\mathbf k')]^{-m} +
\frac{(L')^{2m-d}}{(2 \pi)^{2m}} I_m (r L'^2, {\bf \bar A}) ,
\end{eqnarray}
\begin{eqnarray}
\label{c4} I_m (rL'^2, {\bf \bar A}) = \int\limits_0^\infty
{\rm{d}}y \;y^{m-1} \exp[- r L'^2 y / (4 \pi^2)] \nonumber\\
\times \{K_d (y, {\bf \bar A}) - (\pi / y)^{d/2} - 1\} \qquad.
\end{eqnarray}
For $2 < d < 4$ the behavior of these functions for $r \to 0$ is
$I_1 (r, {\bf \bar A}) \to I_1 (0, {\bf \bar A}) = {\rm finite}$,
$r^2 I_2 (r, {\bf \bar A}) \sim O (r^{d/2})$,
\begin{eqnarray}
\label{c7} J_0 (r, {\bf \bar A}) &-& \frac{r}{4 \pi^2} \; I_1 (r,
{\bf \bar A}) \; - \; \frac{r^2}{32 \pi^4} \; I_2 (r, {\bf \bar
A}) \nonumber\\ &=& J_0 (0, {\bf \bar A}) \; + \; O (r^{d/2}) \; .
\end{eqnarray}
For $d>0$ the behavior for $r \to \infty $ is $r I_1 (r, {\bf \bar
A}) \to - 4 \pi^2 $, $r^2 I_2 (r, {\bf \bar A}) \to - 16 \pi^4$.
For $2<d<4$ and $r > 0$ the bulk integral in (\ref{c1}) is
\begin{eqnarray}
\label{c8}  \int\limits_{\bf k'} \ln \{[r + \delta \widehat K'
(\mathbf k')] v'^{2/d}\} = \int\limits_{\bf k'} \ln \{[ \delta
\widehat K' (\mathbf k')] v'^{2/d}\} \nonumber\\+ \;\; r \;
\int\limits_{\bf k'} [\delta \widehat K' (\mathbf k')]^{-1} -
2A_d\; r^{d / 2}/(d\varepsilon),
\end{eqnarray}
apart from nonasymptotic corrections that depend on  $r v'^{2/d}$
and vanish for $ r \rightarrow 0$ at fixed finite $\tilde v'$. The
bulk integrals in (\ref{c3}) follow  by differentiation with
respect to $r$.


\begin{thebibliography}{99}
\bibitem{text}
In Eq. (52) of Ref. \cite{cd2003} the  term $[\ln \pi + (\widetilde L / \tilde
a)^{1-d} \ln 2]$ should be replaced by $\ln (2 \pi)$. I thank B.
Kastening for pointing out this correction to me.
\end{thebibliography}
\end{document}